\documentclass[iop,revtex4]{emulateapj}
\usepackage{amsmath}
\usepackage{ulem}
\usepackage[dvips, usenames]{color}

\shorttitle{Near-infrared DIB}
\shortauthors{Hamano et al.}

\begin{document}

\title{Near-infrared diffuse interstellar bands in $0.91-1.32$ micron\footnote{Based on data collected at Araki 1.3m telescope, which is operated by the Koyama Astronomical Observatory, Kyoto Sangyo University.}}

\author{Satoshi Hamano, Naoto Kobayashi\altaffilmark{1}}
\affil{Institute of Astronomy, School of Science, University of Tokyo, 2-21-1 Osawa, Mitaka, Tokyo 181-0015, Japan}

\author{Sohei Kondo}
\affil{Laboratory of Infrared High-resolution spectroscopy (LIH), Koyama Astronomical Observatory, Kyoto Sangyo University, Motoyama, Kamigamo, Kita-ku, Kyoto 603-8555, Japan}

\author{Yuji Ikeda\altaffilmark{1}}
\affil{Photocoding, 460-102 Iwakura-Nakamachi, Sakyo-ku, Kyoto, 606-0025, Japan}

\author{Kenshi Nakanishi}
\affil{Koyama Astronomical Observatory, Kyoto Sangyo University, Motoyama, Kamigamo, Kita-ku, Kyoto 603-8555, Japan}

\author{Chikako Yasui\altaffilmark{1}, Misaki Mizumoto\altaffilmark{2}, Noriyuki Matsunaga\altaffilmark{1}, Kei Fukue\altaffilmark{1}}
\affil{Department of Astronomy, Graduate School of Science, University of Tokyo, Bunkyo-ku, Tokyo 113-0033, Japan}

\author{Hiroyuki Mito\altaffilmark{1}}
\affil{Kiso Observatory, Institute of Astronomy, School of Science, The University of Tokyo, 10762-30 Mitake, Kiso-machi, Kiso-gun, Nagano, 397-0101, Japan}

\author{Ryo Yamamoto, Natsuko Izumi}
\affil{Institute of Astronomy, School of Science, University of Tokyo, 2-21-1 Osawa, Mitaka, Tokyo 181-0015, Japan}

\author{Tetsuya Nakaoka\altaffilmark{1}, Takafumi Kawanishi\altaffilmark{1}, Ayaka Kitano\altaffilmark{1}, Shogo Otsubo\altaffilmark{1}}
\affil{Department of Physics, Faculty of Sciences, Kyoto Sangyo University, Motoyama, Kamigamo, Kita-ku, Kyoto 603-8555, Japan}

\author{Masaomi Kinoshita}
\affil{Solar-Terrestrial Environment Laboratory, Nagoya University, Furo-cho, Chikusa-ku, Nagoya, Aichi, 464-8601, Japan}

\author{Hitomi Kobayashi}
\affil{Kyoto Nijikoubou, LLP, 17-203, Iwakura-Minamiosagi-cho, Sakyo-ku, Kyoto 606-0003, Japan}

\and

\author{Hideyo Kawakita\altaffilmark{3}}
\affil{Laboratory of Infrared High-resolution spectroscopy (LIH), Koyama Astronomical Observatory, Kyoto Sangyo University, Motoyama, Kamigamo, Kita-ku, Kyoto 603-8555, Japan}

\email{hamano@ioa.s.u-tokyo.ac.jp}

\altaffiltext{1}{Laboratory of Infrared High-resolution spectroscopy (LIH), Koyama Astronomical Observatory, Kyoto Sangyo University, Motoyama, Kamigamo, Kita-ku, Kyoto 603-8555, Japan}
\altaffiltext{2}{Institute of Space and Astronautical Science (ISAS), Japan Aerospace Exploration Agency (JAXA), 3-1-1, Yoshinodai, Chuo-ku, Sagamihara, Kanagawa, 252-5210, Japan}
\altaffiltext{3}{Department of Physics, Faculty of Sciences, Kyoto Sangyo University, Motoyama, Kamigamo, Kita-ku, Kyoto 603-8555, Japan}

\begin{abstract}

We present a comprehensive survey of diffuse interstellar bands (DIBs) in $0.91-1.32 \mu$m with the newly developed near-infrared (NIR) spectrograph WINERED, mounted on the Araki 1.3 m Telescope in Japan. We obtained high-resolution ($R=28,300$) spectra of 25 early-type stars with color excesses of $0.07<E(B-V)<3.4$. In addition to the five DIBs previously detected in this wavelength range, we identified 15 new DIBs, 7 of which were reported as DIB ``candidates'' by Cox. We analyze the correlations among NIR DIBs, strong optical DIBs, and the reddening of the stars. Consequently, we found that all NIR DIBs show weaker correlations with the reddening rather than the optical DIBs, suggesting that the equivalent widths of NIR DIBs depend on some physical conditions of the interstellar clouds, such as UV flux. Three NIR DIBs, $\lambda\lambda$10780, 10792, and 11797, are found to be classifiable as a ``family,'' in which the DIBs are well correlated with each other, suggesting that the carriers of these DIBs are connected with some chemical reactions and/or have similar physical properties such as ionization potential. We also found that three strongest NIR DIBs $\lambda\lambda$10780, 11797, and 13175 are well correlated with the optical DIB $\lambda 5780.5$, whose carrier is proposed to be a cation molecule with high ionization potential, indicating that the carriers of the NIR DIBs could be cation molecules. 

\end{abstract}


\section{Introduction}

Diffuse interstellar bands (DIBs) are ubiquitous absorption lines detected in the reddened stars, which originate from foreground interstellar clouds. 
Recent surveys have discovered over 500 DIBs in the near-UV to near-infrared (NIR) wavelength range \citep{jen94,hob08,hob09}. DIBs have also been found in extragalactic objects such as the Magellanic Clouds \citep{wel06, cox07}, M31 \citep{cor11}, and high-$z$ damped Lyα systems \citep{law08}. 

Although no DIB carriers have been identified successfully, organic compounds based on carbon atoms mostly contribute to  absorption features \citep{sar08}. 
In particular, polycyclic aromatic hydrocarbons (PAHs) are expected to be possible carriers of DIBs because they are rich and stable even in harsh interstellar environments, where the molecules are exposed to UV photons \citep{sar08}. 
However, despite recent progress in spectroscopic studies of such molecules in the laboratory, no PAHs have been assigned to any DIBs yet. \citet{cla03}, \citet{sal11}, and \citet{gre11} could not detect the absorption lines of some neutral PAHs in the near-UV spectra of stars; they derived the upper limits of the abundances of some specific neutral PAHs.

DIBs in the NIR wavelength range are essential for testing the PAH-DIB hypothesis because
the electronic transitions of ionized large-sized PAHs are thought to fall in the NIR wavelength range \citep{mat05,rui05}. If the DIB carriers are identified as PAHs, we will be able to better understand the variety and abundance of specific PAHs in interstellar clouds, which will eventually allow us to study the formation and destruction processes of such organic material in the universe. Owing to the recent progress in NIR high-resolution spectroscopy, the number of discovered NIR DIBs has recently increased rapidly. \citet{geb11} detected 13 new DIBs in the $H-$band toward the Galactic center.
\citet{cox14} reported 11 NIR DIB candidates in $0.9 < \lambda < 2.5\mu$m detected in their spectral survey of NIR DIBs with the X-Shooter spectrograph on the Very Large Telescope (VLT). They showed that some of these candidates are detected near the wavelength of the electronic transitions of ionized PAHs from the matrix isolation experiments of \citet{mat05}. However, as \citet{cox14} noted, the NIR spectra of gas-phase PAHs in the laboratory are needed for identification.
The NIR wavelength range is also useful in exploring the DIBs toward heavily reddened stars because of its higher transmittance in interstellar clouds compared to the optical wavelength range \citep{geb11, zas14}. 

In addition to the spectroscopic studies of molecules in the laboratory, for the identification of the carriers it is also important to constrain the physical properties of the DIB carriers in the astronomical environment. For this purpose, the correlations among the strength of DIBs and the amount of other atomic or molecule species in the gas clouds have been extensively investigated for the DIBs in the optical wavelength range \citep{kre87,cam97,mcc10,fri11}. Regarding the NIR DIBs, although earlier studies succeeded in detecting new DIBs, their properties in the interstellar clouds have not been investigated owing to small sample size \citep{job90, cox14}. 
 
In this study, we present the first comprehensive survey of NIR DIBs in the $z-$, $Y-$ and $J-$bands with the newly developed NIR Echelle WINERED spectrograph \citep{ike06,yas06,yas08,kon15} mounted on the Araki 1.3 m Telescope in Japan. WINERED offers a resolution of $R = 28,300$ with high sensitivity in wide wavelength coverage of $0.91-1.36 \mu$m, which is critical for detecting and surveying weak DIBs. DIBs in this wavelength range have not been searched systematically for many lines of sight \citep{job90,foi94,cox14}. We constructed the largest data set so far of NIR high-resolution spectra for 25 reddened early-type stars, and we investigated the correlations among the NIR DIBs detected herein, strong optical DIBs in the literature, and the reddening of stars. The rest of the paper is organized as follows. In \textsection{2}, we describe our observations, targets, and the data reduction procedures. In \textsection{3}, the search for NIR DIBs is described. In \textsection{4}, the properties of the identified DIBs are described. In \textsection{5}, we discuss the correlations among reddening, NIR DIBs, and optical DIBs. In \textsection{6}, we discuss the carrier properties of several DIBs. We present a summary of this study in \textsection{7}. 

\section{Observation, Targets, and Data Reduction}

We used the newly developed high-resolution NIR Echelle spectrograph, WINERED \citep{ike06,yas06,yas08,kon15}, mounted on the F/10 Nasmyth focus of the Araki 1.3 m telescope at Koyama Astronomical Observatory, Kyoto-Sangyo University, Japan \citep{yos12}. WINERED uses a 1.7 $\mu$m cutoff 2048 $\times$ 2048 HAWAII-2RG array with a pixel scale of $0''.8$ pixel$^{-1}$, simultaneously covering a wavelength range $0.91-1.36\mu$m. We used a slit $48''$ in length and $1''.6$ (2 pix) in width, which corresponds to a spectral resolution of $R=28,300$ or $\Delta v = 11$ km s$^{-1}$. 

Bright early-type (O, B, A) stars with $J < 6.2$ mag were selected as targets 
 (Table \ref{targets}). The color excesses of the targets are in the range $0.07 < E(B-V) < 0.85$ except for Cyg OB2 No.12 ($E(B-V) = 3.4$). 
 We also obtained spectra of Rigel ($E(B-V)=0.0$) as a reference, which should not show any absorption features due to interstellar dust. 
Observations of almost all targets were conducted from 2013 November to 2014 January. Only two stars (Cyg OB2 No.12 and HD190603) were observed in 2012 July when we used an engineering array with worse sensitivity, in particular at shorter wavelength ($\lambda < 1\mu$m), compared with that of the currently used scientific array. Therefore, the signal-to-noise ratio (S/N) and quality of the spectra of these stars are not as good as the other data. The telescope was nodded by about 30 arcsec along the direction of the slit length between alternating frames to offset the sky emission. Only for HD 37043, which is embedded in the Orion Nebulae, were the frames obtained without nodding, and corresponding sky frames were obtained in a field far from the star to avoid the emission from the Orion Nebulae.

All the data were reduced using standard procedures with IRAF, including sky subtraction, flat fielding, the extraction of spectra, and wavelength calibration with Argon lamp spectra. For the correction of telluric absorption lines, we obtained the spectra of the A0V standard stars without reddening at an airmass similar to those of the targets (Table \ref{targets}). The spectra of the targets were divided by those of the standard stars to remove telluric absorption lines. Note that we did not remove the hydrogen and metal absorption lines in standard stars, because the profile fitting and removal of those features is difficult in many cases owing to contamination of atmospheric absorption lines. Therefore, emission-like features appear in the divided spectra. The telluric absorption lines by water are stronger for two stars, HD190603 and Cyg OB2 No.12, than for the other stars because these two stars were observed in July, when the humidity is very high in Japan. The spectra are normalized by a low-order Legendre function using the IRAF task, \textit{continuum}, for each order of Echelle spectra.

\section{The Search for DIBS}

In our wavelength range, four major DIBs are known at $\lambda = 9577, 9633, 11797, \text{and }13175$\r{A} \citep{job90,foi94}. In addition, \citet{gro07} found a DIB at $\lambda =$10780\r{A}, which is correlated with $E(B-V)$. We first searched for these five DIBs in the target spectra. Although $\lambda \lambda$9577 and 9633, which are proposed to be the electronic transitions of the fullerene cation \citep[C$_{60}^+$;][]{foi94,jen97,gal00}, are detected toward some reddened stars, their EWs cannot be estimated confidently because of heavy telluric absorption lines. Conversely, the fairly strong DIBs, $\lambda \lambda$10780, 11797 and 13175, are detected toward many lines of sight. In addition to these five major DIBs, three weak DIBs at $\lambda = 9017, 9210, \text{and }9258$\r{A} were reported in the Orion Nebulae by \citet{mis09}, who suggested that these DIBs are additional features of C$_{60}^+$. Because the DIBs, $\lambda \lambda 9210 \text{and } 9258$, which fall in our wavelength range, are blended with stellar absorption lines, it was difficult to evaluate their EWs without modeling the stellar spectra. The ``C$_{60}^+$'' DIBs in the wavelength range of WINERED, $\lambda \lambda 9210, 9258, 9577, \text{and }9633$, will be discussed in a separate study after a complete evaluation of the contamination by the overlapped telluric and stellar absorption lines.

Subsequently, we searched for new DIBs in the wavelength range of $0.91 < \lambda < 1.32 \mu$m. Note that the wavelength ranges $0.93 < \lambda < 0.96 \mu$m,  $1.11< \lambda < 1.16\mu$m, and $\lambda > 1.32 \mu$m were not usable in searching for DIBs owing to heavy telluric absorption lines. We searched for the absorption lines satisfying the following two criteria: 1) the absorption lines are detected with a 5$\sigma$ level in the spectra of  reddened stars, and 2) the absorption lines do not exist in the spectra of the reference star Rigel or the model stellar spectra of the A0Ia, A0V, B5Ia, and B5III stars. We synthesized model spectra using the SPTOOL program \citep{tak95}, which employs ATLAS9 model atmospheres \citep{kur93}.\footnote{The atomic line data was adopted from the Vienna Atomic
Line Database \citep[VALD; ][]{kup99}. 
The effective temperatures and surface gravities are as follows : $(T_\text{eff} , \log
g)=(9900,2.1)$ for A0Ia, (9727,4.27) for A0V, (13600,2.5) for B5Ia, and (15000,4.22) for B5III. These values are adopted from \citet{gra92} ($T_\text{eff}$ of  A0V and $\log g$ of A0V and B5V), \citet{gra09} ($T_\text{eff}$ of A0Ia and B5V), and \citet{cox00} ($\log g$ of A0Ia and B5Ia).}
We first searched for such absorption lines in the spectrum of HD20041, because even weak DIBs are expected to be detectable toward HD20041 in view of the high S/N of the spectrum and the large EWs of $\lambda \lambda$11797 and 13175.
Then, we searched for absorption lines in the spectra of other stars that are detected near the wavelength of the absorption lines detected toward HD20041. 
Consequently, 15 absorption features were detected toward multiple lines of sight as new DIB candidates.

To confirm that the candidate absorption features are DIBs originating from the interstellar clouds, we checked the consistency between the velocity of the intervening gas clouds and those of the detected absorption features. The rest-frame wavelengths of the absorption features were determined by fitting Gaussian profiles to the absorption features detected toward HD20041 (see the next section for details). As for the stars toward which the interstellar atomic absorption lines (\ion{K}{1}, \ion{Na}{1}, and \ion{Ca}{1}) have been reported in the literature (see the Table \ref{targets} footnote for references), we compare the velocities of the detected absorption lines we detected with those of the atomic absorption lines. Regarding the other stars without known interstellar atomic lines, we compare the velocities of the candidate absorption lines with those of the confirmed DIB $\lambda 11797$. The absorption lines whose velocities are inconsistent with those of the interstellar clouds should be categorized as non-DIBs. 
However, no absorption features were rejected by checking their velocities. Because their EWs do not show any dependence on the stellar spectral types but exhibit positive correlations with $E(B-V)$ (see \textsection{5.1}), we conclude that all of the absorption features are truly DIBs. 

\begin{figure*}
 \centering
 \includegraphics[width=17cm,clip]{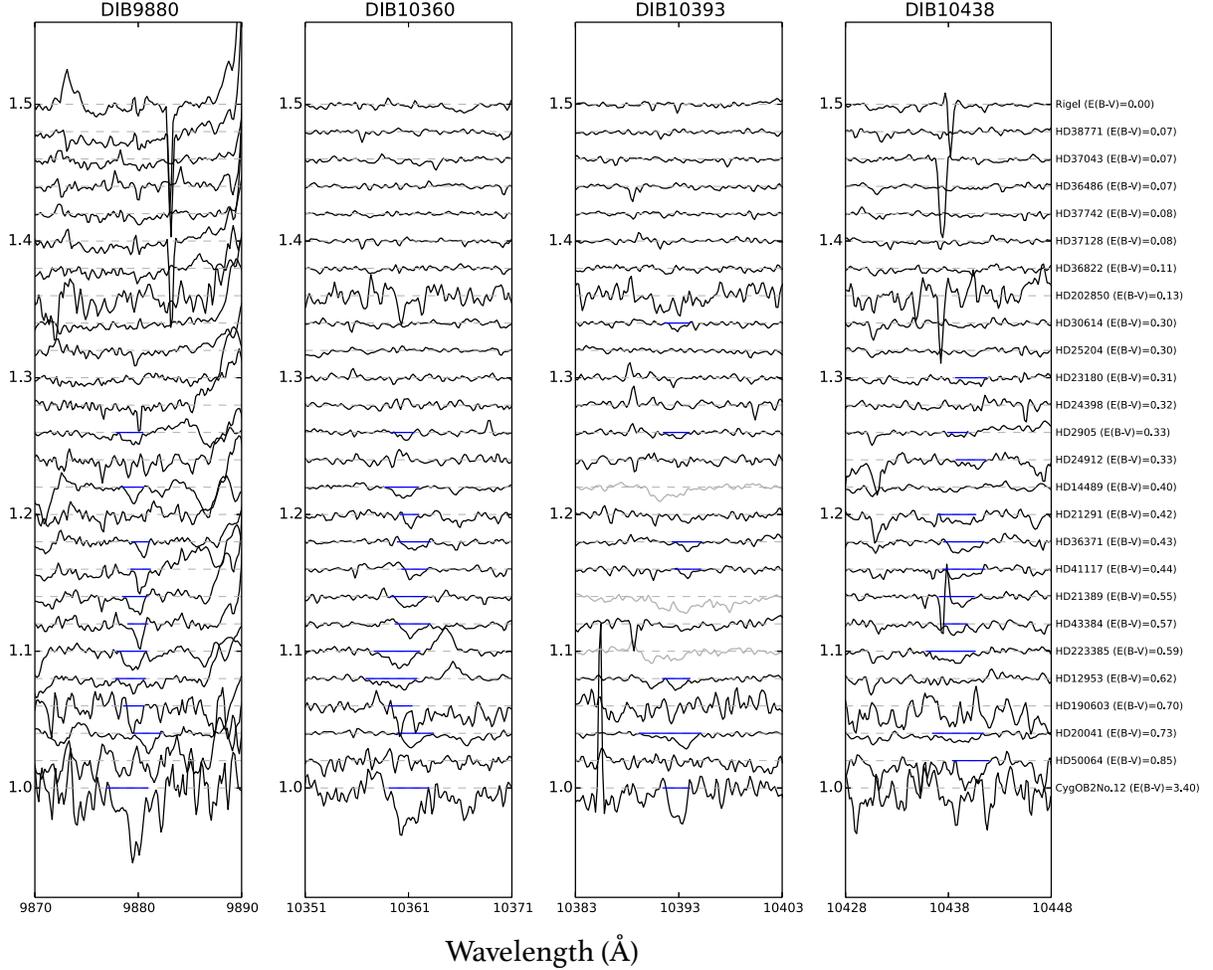}
  \caption{Spectra of four DIBs, $\lambda \lambda 9880, 10360, 10393, \text{and }10438$, for all targets plotted in increasing order of $E(B-V)$ from top to bottom. The spectra are normalized and plotted with arbitrary offsets. The continuum level of each star is shown by a dashed thin line. The spectra of the stars, in which the EWs or upper limits of the DIBs cannot be evaluated due to overlapped stellar and/or telluric absorption lines, are plotted with gray lines. The integrated ranges used in the calculation of the EWs are shown as thick blue lines.}
 \label{DIBspec1}
\end{figure*}

\begin{figure*}
 \centering
 \includegraphics[width=17cm,clip]{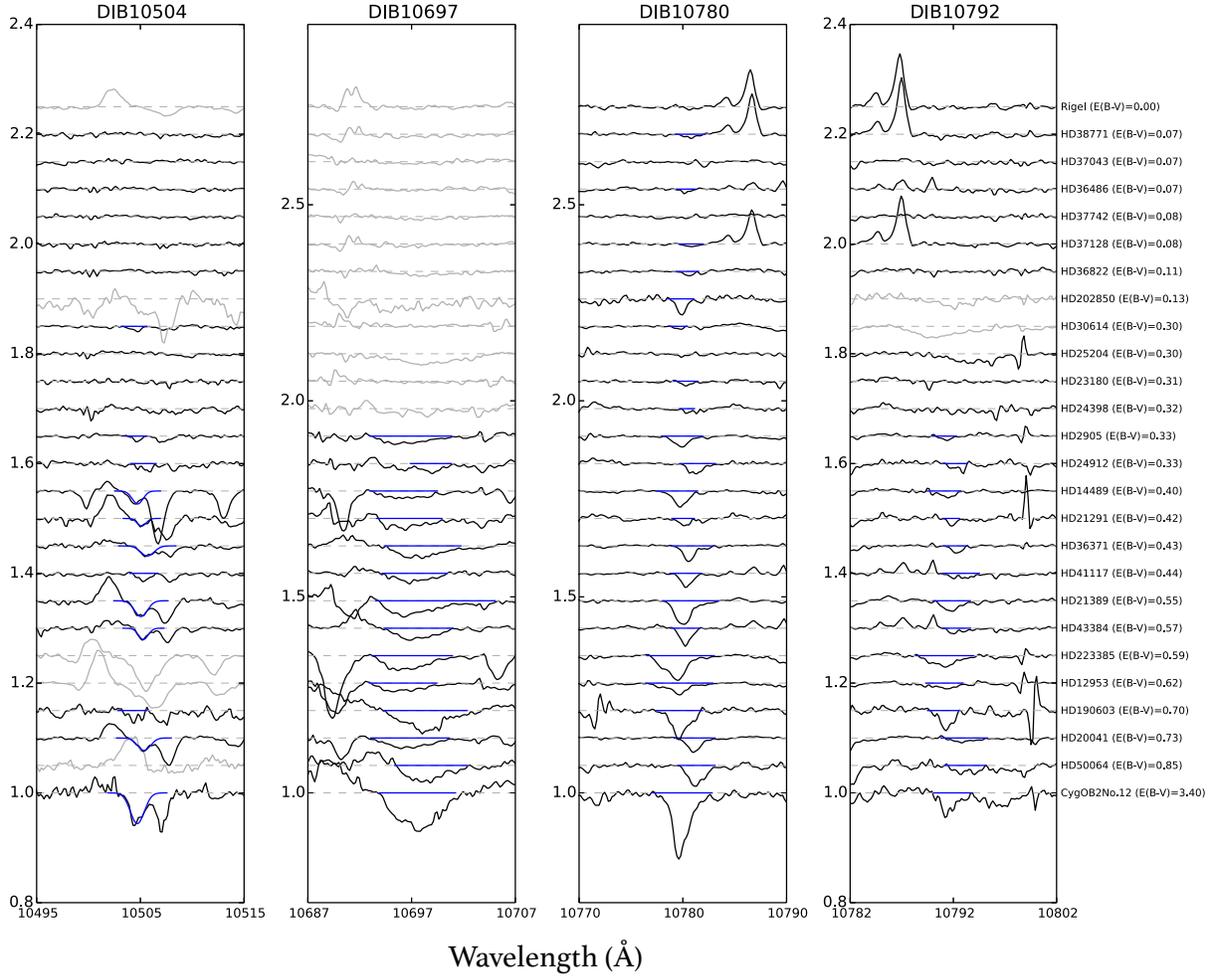}
  \caption{Spectra of four DIBs $\lambda \lambda 10504, 10697, 10780, \text{and }10792$ for all targets.  The notations are the same as those in Figure \ref{DIBspec1} except for DIB $\lambda 10504$, for which the EWs are estimated by fitting Gaussian profiles (shown as blue lines) to avoid the blending of the stellar absorption lines (see \textsection 4.3.1 for detail). The spectra of $\lambda 10697$ were divided by the spectra of telluric standard stars, from which overlapped \ion{C}{1} and \ion{Si}{1} stellar absorption lines were removed (see \textsection{4.3.2} for details).}
 \label{DIBspec2}
\end{figure*}

\begin{figure*}
 \centering
 \includegraphics[width=17cm,clip]{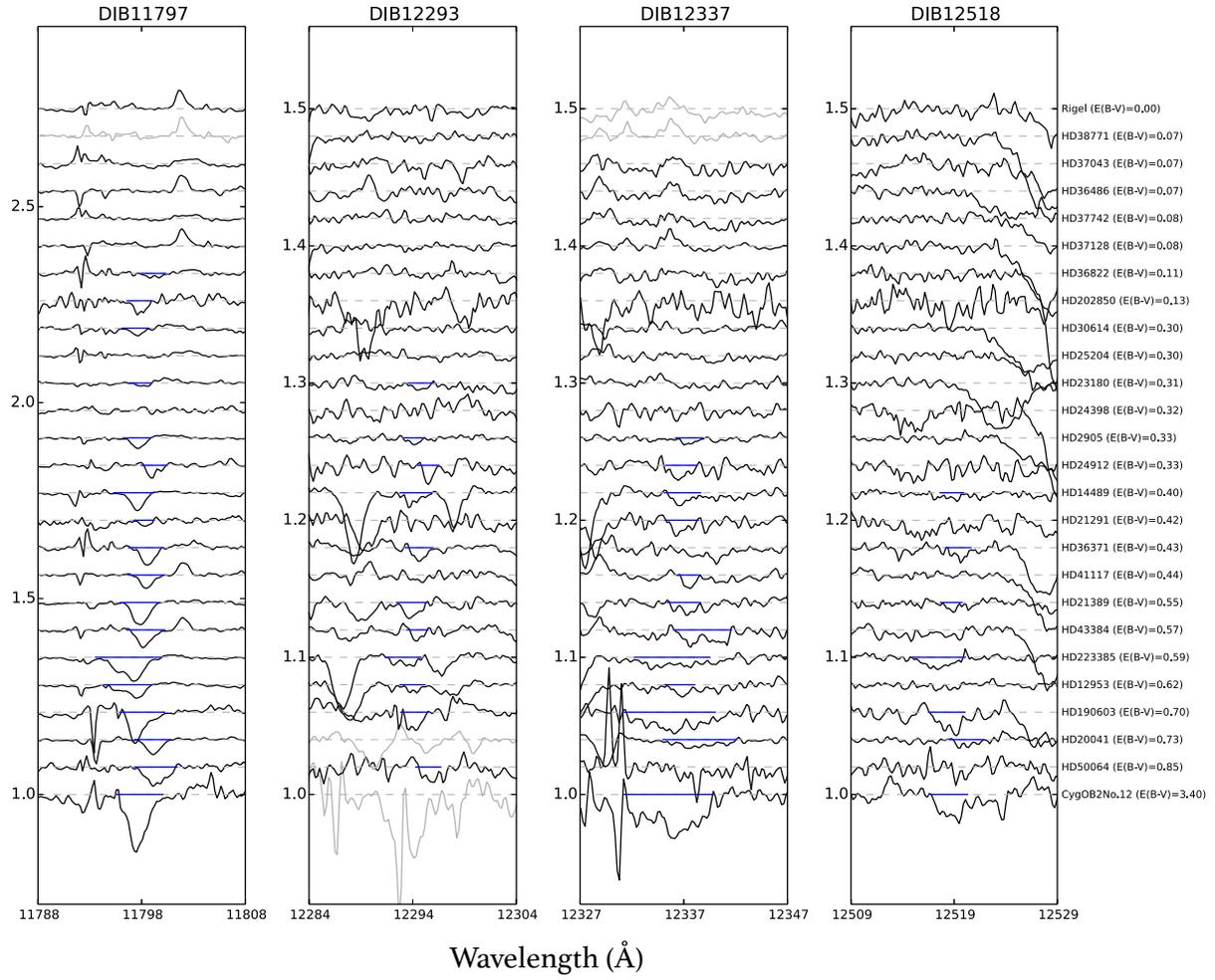}
  \caption{Spectra of DIBs $\lambda \lambda 11797, 12293, 12337, \text{and }12518$ for all targets. The notations are the same as those in Figure \ref{DIBspec1}. }
 \label{DIBspec3}
\end{figure*}

\begin{figure*}
 \centering
 \includegraphics[width=17cm,clip]{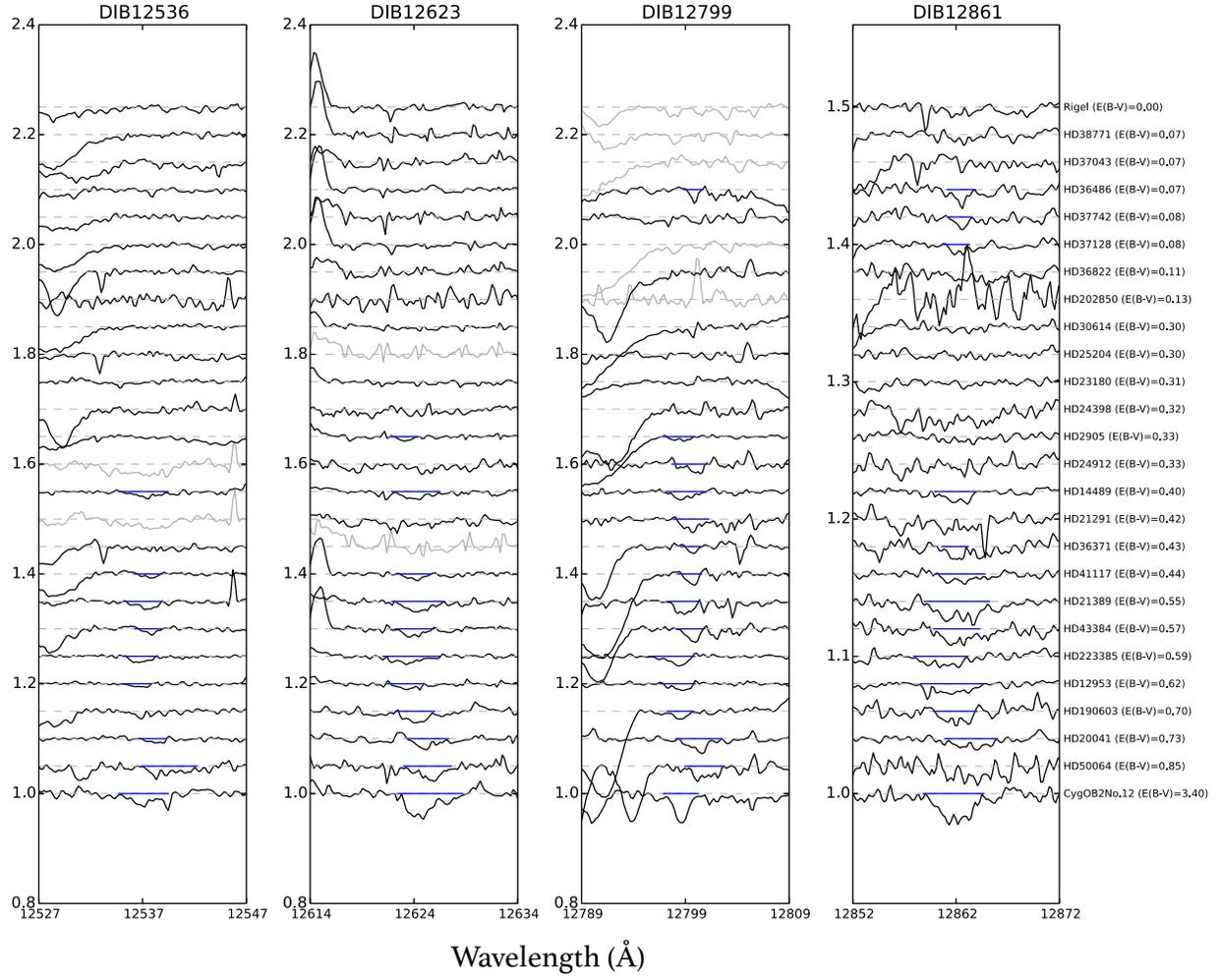}
 \caption{Spectra of DIBs $\lambda \lambda 12536, 12623, 12799, \text{and }12861$ for all targets. The notations are the same as those in Figure \ref{DIBspec1}. The spectra of $\lambda 12799$ were normalized with high-order polynomial function to remove the overlapped \ion{He}{1} and \ion{H}{1} stellar absorption lines (see \textsection{4.3.5} for details)).}
 \label{DIBspec3.5}
\end{figure*}

\begin{figure}
 \centering
 \includegraphics[width=9cm,clip]{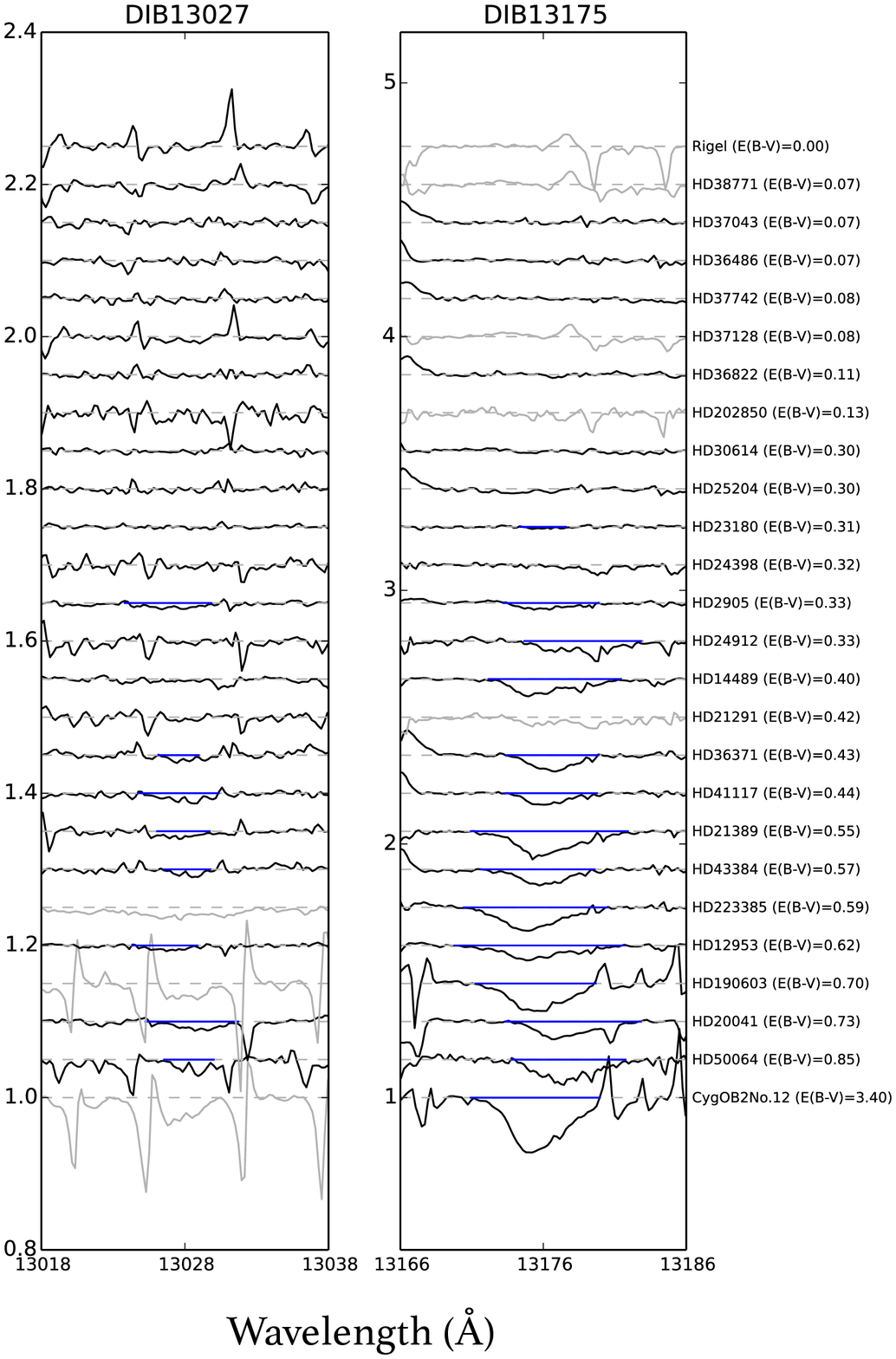}
 \caption{Spectra of DIBs $\lambda \lambda 13027 \text{and } 13175$ for all targets. The notations are the same as  those inFigure \ref{DIBspec1}. }
 \label{DIBspec4}
\end{figure}

In summary, we identified 15 new DIBs in addition to five DIBs previously identified in this wavelength range. Table \ref{DIBs} summarizes all the DIBs detected. The table also shows the effect of the stellar and telluric absorption lines on the DIBs. Figures \ref{DIBspec1}- \ref{DIBspec4} show the DIB spectra for all observed stars. 
The EWs of all DIBs toward all observed stars are summarized in Table \ref{DIBew1}. We calculated the EWs by a simple integration because the DIB profiles are not well established. Only for DIB $\lambda 10504$, whose EWs cannot be evaluated by simple integration because of blending with stellar lines (see \textsection{4.3.1} for details), we fit Gaussian profiles to the DIB profile to estimate the EWs. The wavelength range used for the integration is arbitrarily set to include the whole DIB region (see Figures \ref{DIBspec1}-\ref{DIBspec4}). The uncertainties of the EWs ($\sigma _{\text{err}}$) or 5$\sigma$ upper limits (EW$_{\text{upp}}^{5\sigma}$) are estimated from the S/N of the neighboring continuum using the following equations: 
\begin{align}
\sigma _{\text{err}}  &= \Delta \lambda \sqrt{w} (\text{S/N})^{-1}, \\
EW _{\text{upp}}^{5\sigma} &= 5 \Delta \lambda \sqrt{w_{\text{inst}}} (\text{S/N})^{-1}, 
\end{align}
where $\Delta \lambda$ is the wavelength interval per pixel ($ \sim 0.2$ \r{A}), $w$, and $w_{\text{inst}}$ are the widths of the DIB and the instrumental profile in pixels, respectively. Here, $w_{\text{inst}}$ is defined as $2J + 1$ pixel, where $J(=2 \text{pixels})$ is the slit width used in our observation.
We did not consider additional systematic uncertainties due to continuum fitting, telluric absorption lines, and stellar lines. Several concerns on the systematic uncertainties in the EWs of each DIB will be discussed in \textsection{4.3}.  

Recently, \citet{cox14} reported 11 new NIR DIB candidates with their NIR DIB survey with VLT/X-shooter, 8 of which fall in our wavelength range, although they could not confirm that the candidates are truly DIBs because they detected the candidates only toward 3 stars at most. Of their candidates, 7 are included in our newly identified 15 DIBs for which 
we confirmed that they are truly DIBs by detecting them toward many lines of sight. Note that a line at $\lambda \sim$10507\r{A}, which is reported as a DIB candidate by Cox, is identified as a stellar absorption line of \ion{N}{1} in this study. The DIB λ10792 was originally reported as an unidentified line by \citet{gro07}.

\section{Properties of NIR DIBs}

\subsection{EW Distribution}

\begin{figure}[!ht]
 \centering
 \includegraphics[width=9cm,clip]{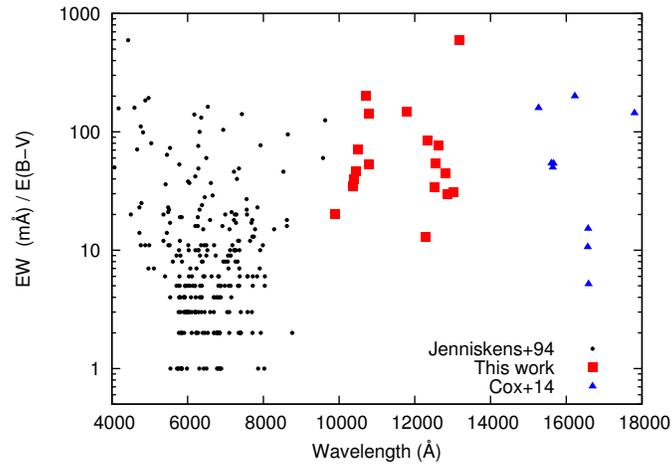}
 \caption{Distribution of the ratios of the EWs to $E(B-V)$ as a function of the wavelength for all DIBs in the optical and near-infrared. Black circles, red squares, and blue triangles represent the points of optical DIBs from \citet{jen94}, NIR DIBs found in this study, and DIBs in the $H-$band from \citet{cox14}, respectively.}
 \label{ewdist}
\end{figure}

Figure \ref{ewdist} shows the distribution of the ratios of the EWs to $E(B-V)$ for DIBs in 4000 \r{A} $< \lambda < 18000$ \r{A}. The data for optical DIBs in the wavelength range 4000 \r{A}$< \lambda < 10000$\r{A} are adopted from the catalog published on the Web\footnote{http://leonid.arc.nasa.gov/DIBcatalog.html} by P. Jenniskens. The catalog is primarily based on the studies by \citet{jen94} and \citet{kre95} (see the Web site for other references). We do not include the recent surveys of DIBs in the optical wavelength range such as \citet{tua00,hob08,hob09} in this figure, because most of the DIBs found in these surveys are weak. The data for $H-$band DIBs are adopted from \citet{cox14}. Because the authors do not show the ratios of EWs to $E(B-V)$, we roughly estimated them by dividing the EWs toward 4U 1907$+$09 by the color excess, $E(B-V) = 3.48$. Regarding the DIBs identified in this study, we determine the ratios of EWs to $E(B-V)$ by fitting linear functions ($EW = a \times E(B-V) + b$) to EW$- E(B-V)$ plots (see \textsection{5.1} and Table \ref{ebvcc}). From Figure \ref{ewdist}, we found that the NIR DIB $\lambda 13175$ is one of the strongest DIBs ever detected. We also found that there are many DIBs in the range $0.91\mu\text{m}< \lambda < 1.32\mu$m, in which only five DIBs were previously confirmed. All of the DIBs detected herein are stronger than 10 m\r{A}, below which the majority of optical DIBs are distributed. Assuming a similar distribution to that of the optical DIBs, we expect that the number of weak NIR DIBs increases by an order of magnitude through deeper spectroscopic observations.

\subsection{Wavelength and FWHM}

\begin{figure*}[!ht]
 \centering
 \includegraphics[width=16cm,clip]{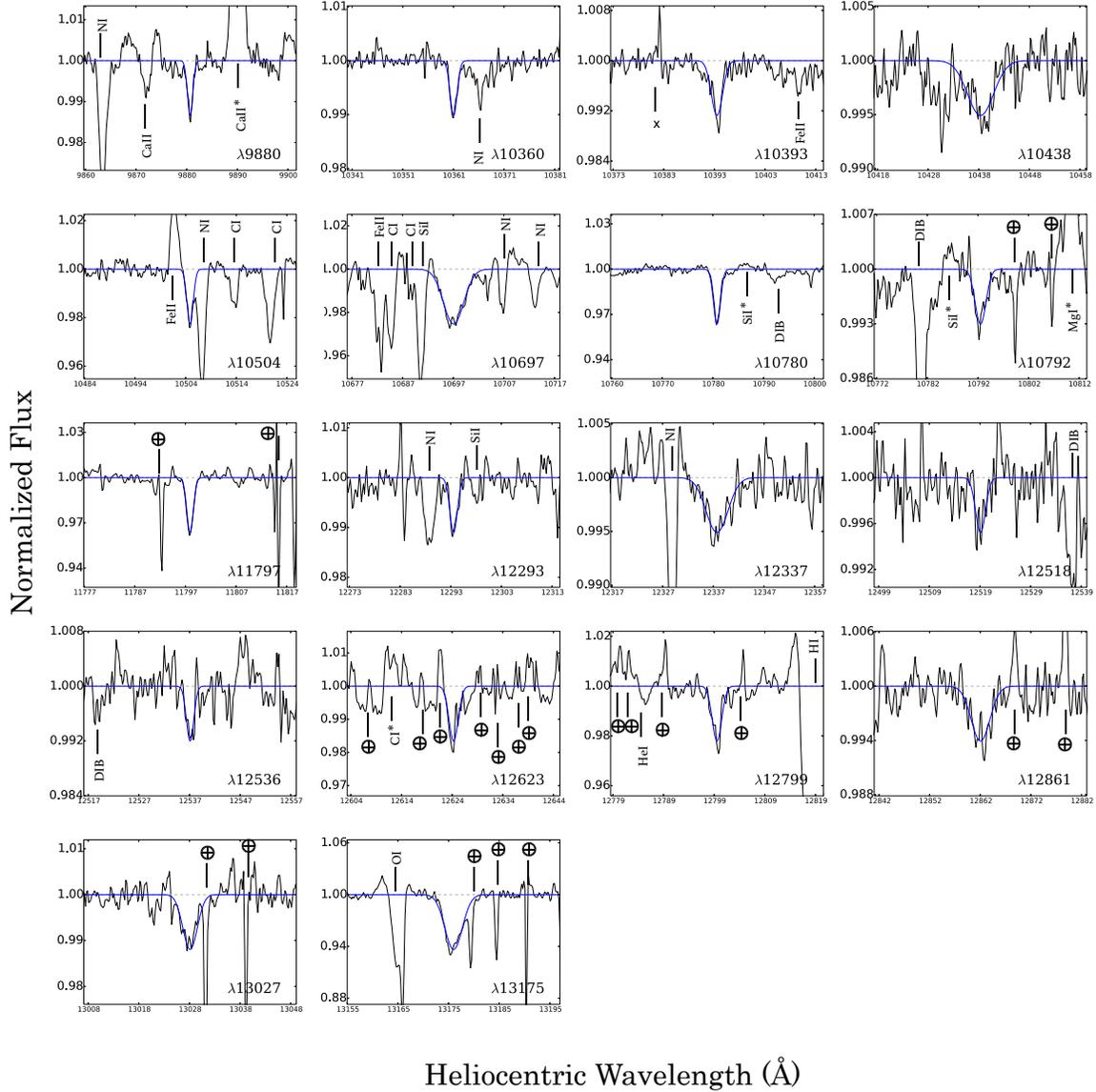}
 \caption{Profiles of identified DIBs toward HD20041 ($E(B-V) = 0.73$). The spectrum of HD21389 ($E(B-V) = 0.55$) is shown only for $\lambda 12293$ because $\lambda 12293$ toward HD20041 is blended by the stellar absorption lines (see the main text for details). The black line shows the observed spectrum, while the blue line shows the Gaussian profile fitted to the DIB. The stellar absorption and emission lines are marked with vertical lines and the species names (emission lines with asterisk: artificial emission-like features from the absorption lines of telluric standard stars). The residuals of corrected telluric absorption lines are marked with ``$\bigoplus$''. The spurious features are marked with x's.}
 \label{profilefitting}
\end{figure*}

We measured the rest-frame wavelengths in air ($\lambda _0$) and FWHM of all identified DIBs by fitting Gaussian functions to profiles. We fitted the Gaussian functions to the profiles of HD20041, toward which almost all DIBs are detected with high S/N. Because the DIB $\lambda 12293$ toward HD20041 is blended with two \ion{N}{1} stellar absorption lines, we used the spectrum of HD21389 only for this DIB. Table \ref{DIBs} summarizes the fitting results, and Figure \ref{profilefitting} shows the spectra of the DIBs and fitted Gaussian profiles. We shift the measured wavelength to the rest-frame wavelength using the heliocentric velocities of the strongest velocity component of the interstellar clouds toward these stars (Table \ref{targets}). 
The systematic uncertainties may remain in the measured $\lambda _0$ and FWHMs of the DIBs because of the possible velocity dispersion of multiple velocity components toward these two stars.

\subsection{Comments on Individual DIBs}

\subsubsection{$\lambda \lambda$10360, 10393, 10438, and 10504}

These DIBs fall in the wavelength ranges without any telluric absorption lines. Because the S/N of raw spectra is higher compared with the S/N of spectra divided by that of telluric standard stars, their EWs are calculated from the undivided spectra. DIB $\lambda 10504$ is contaminated by two stellar lines: \ion{Fe}{2} $\lambda 10501.5$ emission line of supergiant stars of early A-type and late B-type and \ion{N}{1} $\lambda 10507.0$ absorption line of stars earlier than late B-type (Figure \ref{profilefitting}). In the case when the relative velocity between the background star and the interstellar clouds is large, the observed DIB is blended by the lines. A relative velocity of only $\sim 30$ km s$^{-1}$, which produces a wavelength shift of $\sim 1$\r{A}, can affect the calculation of EWs of the DIB. In this study, we do not calculate the EWs of the DIB toward HD50064 and HD223385, which are clearly detected but significantly blended by the stellar lines (see Fig \ref{DIBspec2}). When the DIB is separately detected with slight blending, we calculate the EWs by fitting two Gaussian profiles to the blended DIB profile and \ion{N}{1} absorption line (Figure \ref{DIBspec2}). We do not estimate systematic uncertainties from the fitting procedure. For a more precise analysis of this DIB, a model spectrum that reproduces both the absorption and emission lines for each star is required.

\subsubsection{$\lambda 10697$}

The DIB $\lambda 10697$ is surrounded by strong stellar absorption lines of \ion{C}{1} and \ion{Si}{1} as well as telluric absorption lines (Figure \ref{profilefitting}). Because these stellar metal absorption lines are also detected in the A0V telluric standard stars, many false emission-like features appear in all telluric-corrected spectra. These artificial features significantly affect the continuum fitting and evaluation of EWs or upper limits. Therefore, we attempted to remove from the telluric standard spectra the \ion{C}{1} lines at $10689.72, 10683.03, 10691.24, \text{and }10707.32$\r{A} and the \ion{Si}{1} lines at 10685.34 and 10694.25\r{A}.

First, we define the instrumental profile of stellar metal absorption lines using the absorption profile of \ion{S}{1} 10459.41\r{A} and \ion{Si}{1} 10827.09\r{A} lines, which are free from any blending of telluric absorption lines and show a clear line profile. The profiles of \ion{S}{1} 10459.41\r{A} and \ion{Si}{1} 10827.09\r{A} lines are used for the removal of \ion{C}{1} and \ion{Si}{1} lines, respectively. 
We then fit the instrumental profiles to the \ion{C}{1} and \ion{Si}{1} lines by shifting the wavelength and scaling the intensity on the basis of the intensity ratios of the metal absorption lines in the model spectra of A0V stars. 
We construct a spectrum containing only \ion{C}{1} and \ion{Si}{1} absorption lines by synthesizing the profiles of each absorption line. 
These stellar metal lines are removed from the telluric standard spectra by dividing them by the synthetic spectra of \ion{C}{1} and \ion{Si}{1} lines. Note that the slight residuals appear to remain in the resultant spectra for some telluric standard stars probably because the \ion{S}{1} line profile is slightly different from the \ion{C}{1} lines. 
Finally, telluric absorptions are corrected using the \ion{C}{1} and \ion{Si}{1} corrected spectra of telluric standard stars. Consequently, the continuum fitting is improved except for Cyg OB2 No.12 and HD190603, for which the data were obtained in 2012 July (when the climate is humid in Japan), and the metal absorption lines of the telluric standard star HIP87108 could not be removed well because of the low S/N and quite strong telluric absorption lines. Figure \ref{DIBspec2} shows the spectra of $\lambda 10697$ divided by the spectra of telluric standard stars, from which the \ion{C}{1} and \ion{Si}{1} absorption lines were removed. The EWs of $\lambda 10697$ are estimated, but the upper limit of the EW is not evaluated  because of the possible residual of stellar metal lines when the DIB is not detected.



\subsubsection{$\lambda 12293$}

The DIB is located between two \ion{N}{1} absorption lines, $\lambda \lambda$12289.18 and 12298.54 (Figure \ref{profilefitting}). Because of the relatively large wavelength offset $\sim5$\r{A} ($\sim120$ km s$^{-1}$), it is unusual that they are blended significantly. However, these stellar lines affect the determination of the continuum level around the DIB. In particular, the DIB of HD20041 is surrounded by both the \ion{N}{1} absorption lines and false emission-like features from the \ion{N}{1} absorption lines in the telluric standard star. Although the DIB itself is clearly detected toward HD20041, we cannot estimate its EW. 

\subsubsection{$\lambda 12799$}

Because the DIB is located in the wings of the strong and broad \ion{He}{1} and \ion{H}{1} (Pa $\beta$) absorption lines, we remove these broad lines by fitting a high-order polynomial function with an IRAF task (i.e., \textit{continuum}) for telluric-corrected spectra (Figures \ref{DIBspec3.5} and \ref{profilefitting}). Although we do not consider various possible systematic uncertainties, we can identify this absorption as a DIB because of a clear positive correlation with $E(B-V)$ (see \textsection{5.1}). Note that the upper limit of the EWs for the DIB cannot be evaluated for Rigel, HD38771, and HD37043 because the telluric absorption lines cannot be removed sufficiently well.

\section{Correlations}

To reveal the properties of the DIB carriers in an astronomical environment, it is a first step to investigate the correlations among DIBs as well as the correlations of DIBs with the parameters of interstellar clouds, such as reddening \citep{her93, cam97, fri11}. The correlations of the NIR DIBs at $\lambda > 1\mu$m have been discussed for a limited number of lines of sight \citep{job90, geb11, cox14} except for the strongest DIB in the $H-$band, $\lambda 15272$, which was examined by \citet{zas14} for about 60,000 lines of sight in the SDSS-III/APOGEE survey. Here, we analyze the correlations among the NIR DIBs that we identified at $0.91< \lambda < 1.32 \mu$m, $E(B-V)$, and optical DIBs, with a significant number of lines of sight (i.e., 25) for the first time. The EWs of the optical DIBs are adopted from \citet{fri11}, who investigated the correlations among eight strong optical DIBs ($\lambda \lambda$5487.7, 5705.1, 5780.5, 5797.1, 6196.0, 6204.5, 6283.8, and 6613.6). Among the 25 stars that we observed, the optical spectra of 19 stars were also obtained in \citet{fri11}. Therefore, it is possible that the correlation between optical and NIR DIBs can be investigated using a significant number of lines of sight.

\subsection{Correlations with $E(B-V)$}

Figure \ref{ebvcc1} show the EW$-E(B-V)$ plot. We also plot the data of \citet{cox14} for the DIBs that they detected. However, we do not include their data in the calculation of the correlation coefficients in order to preserve the uniformity of data. We fit the linear functions ($EW = a \times E(B-V) + b$) to each EW$-E(B-V)$ plot. The results are summarized in Table \ref{ebvcc}. 
In the calculation of the correlation coefficients, we exclude the data point of Cyg OB2 No.12, for which the EWs of the optical DIBs are known to be much lower than that expected from the correlation with $E(B-V)$ \citep{chl86}. To compare the correlation coefficients with those of the optical DIBs, we also show the correlation coefficients between $E(B-V)$ and the EWs of the eight optical DIBs taken from \citet[][; Table \ref{ebvcc}]{fri11}. To eliminate the bias effect and the difference in sample size, we calculate the correlation coefficients using only the 19 stars observed in both NIR (our data) and optical wavelength ranges \citep{fri11}. As an example, Figure \ref{ebvcorr_rep} shows the correlations of the optical DIB $\lambda 5780.5$ and the NIR DIB $\lambda 11797$ with $E(B-V)$. 

All NIR DIBs are found to have a positive correlation with $E(B-V)$, which is expected for DIBs. 
We focus on the DIBs detected toward more than 10 lines of sight, whose correlation coefficients appear to be reliable. 
The correlation coefficients of all the NIR DIBs are found to be lower than those of the eight strong optical DIBs calculated for the same sample ($r \sim 0.9$). When discussing the difference between two correlation coefficients, we must consider the attenuation of the correlations due to measurement errors and the statistical significance of the difference between correlation coefficients. Here, we will consider whether the differences in the correlation coefficients between optical and NIR DIBs are statistically significant. We focus on the correlation coefficients of three NIR DIBs, $\lambda \lambda 10780, 11797, \text{and }13175$, because they are detected toward many lines of sight ($>10$) and the uncertainties of their EWs are expected to be low owing to their large EWs and the smaller effect of contamination by stellar lines. Regarding the other DIBs, statistical tests are not meaningful because of the limited sample size and/or large systematic uncertainties. 

We conduct a statistical test for the pairs of the three NIR DIBs and eight optical DIBs to clarify the difference between the correlation coefficients using the Fisher $z$-transformation. Consequently, the correlation coefficients of the five optical DIBs with $r>0.91$ are found to be different from those of the three NIR DIBs, with a 95\% significance level. Therefore, the difference in the correlation coefficients with $E(B-V)$ between the optical and NIR DIBs are suggested to be statistically significant for some pairs. As for the other pairs and NIR DIBs, further observations are necessary.

The high correlation coefficients of optical DIBs show that the abundances of their carriers are simply in proportion to the amount of gas in the lines of sight. Considering the lower correlation coefficients of NIR DIBs, 
we infer that the abundance of the NIR DIB carriers is more sensitive to other physical parameters of the interstellar clouds (such as ionization UV flux) as well as to the amount of gas in the line of sight.

\subsection{Correlations among NIR DIBs: A Family of NIR DIBs?}

We also examine the correlations between all pairs of NIR DIBs except for 5 weak DIBs, which are detected toward fewer than 10 stars, and 2 DIBs $\lambda \lambda 12293 \text{ and }12799$, whose EWs have large systematic uncertainties owing to a blending of neighboring stellar absorption lines. From the NIR DIB set identified, we search for so-called families, in which DIBs are well correlated with each other. DIBs in a family are expected to arise from molecules having similar physical and/or chemical properties \citep{mou99,mcc10}.

The correlation coefficients for all 55 pairs consisting of 11 DIBs are shown in Table \ref{nirdibcc}. 
As in the previous section, we do not include the EWs of DIBs toward Cyg OB2 No.12 in the calculation. All pairs are found to have positive correlation coefficients. In particular, three pairs consisting of $\lambda \lambda 10780, 10792, \text{and }11797$ have the highest correlation coefficients with $r\sim0.95$, suggesting that these DIBs can be classified as a family (Figure \ref{NIRcorrelation}). Because all three DIBs are fairly strong and detected toward over 10 lines of sight, the correlation coefficients should be reliable. Note that $\lambda 13175$ is also well correlated with the three DIBs $\lambda \lambda 10780, 10792, \text{and }11797$, but the correlation coefficients ($r\sim 0.9$) are slightly lower than those among the three DIBs. 
Because there are several points that do not quite fit the correlation, carriers of these three DIBs would not be the same molecule. Note that $\lambda 10792$ toward Cyg OB2 No.12 is considerably weaker than that expected from the correlations (Figure \ref{NIRcorrelation}), suggesting that the carrier of this DIB may become less abundant due to some chemical processes, which would be efficient in a dense cloud with $n_\text{H} \sim 10^3$cm$^{-3}$ \citep{cas05}. Alternatively, the high X-ray flux from many massive stars in the Cygnus OB2 association including Cyg OB2 No.12 may affect the abundance of the $\lambda 10792$ carrier \citep{gre01}.

Two pairs, $\lambda 10504 - \lambda 12623$ and $\lambda 10504 - \lambda 13175$, are also found to have high correlation coefficients of $r>0.95$ (Figure \ref{NIRcorrelation2}). However, the correlation coefficient of the remaining pair, $\lambda 12623 - \lambda 13175$, is lower ($r=0.84$). This is because the EWs of $\lambda \lambda 12623 \text{ and }13175$ toward HD50064, whose EW of $\lambda 10504$ cannot be measured owing to the blending of the stellar absorption lines (see \textsection{4.3.1}), do not match the correlation. As the number of lines of sight used for the calculation of the correlation coefficients is small for these pairs (e.g., seven stars for the $\lambda 10504 - \lambda 12623$ pair), the correlation coefficients could be affected by the sample selection. Although these DIBs may form another family in view of the high correlation coefficients, it is necessary to increase the data points to confirm this from more solid correlations.

\subsection{Correlations of NIR DIBs with Optical DIBs}

Finally, we examine the correlations between the optical and NIR DIBs. We focus on four NIR DIBs, $\lambda \lambda 10438, 10780, 11797, \text{and }13175$, whose correlation coefficients with optical DIBs can be calculated with over 10 data points. Regarding the other DIBs, the number of data points available for the calculation of the correlation coefficients is too small for a statistical discussion. As in the previous two sections, we exclude Cyg OB2 No.12 from the calculation of the correlation coefficients. Table \ref{optcc} shows the correlation coefficients of 32 pairs consisting of 8 optical DIBs and 4 NIR DIBs. As a reference, we also calculate the correlation coefficients among optical DIBs using only the stars observed by us and \citet{fri11}.

\citet{fri11} showed that the correlation coefficients of all pairs consisting of eight optical DIBs have $r>0.9$, except for the two pairs $\lambda5487.7 - \lambda 5797.1$ ($r=0.87$) and $\lambda 5797.1 - \lambda 6283.8$ ($r=0.86$). The correlation coefficients calculated in this study for the optical DIBs are as high as theirs for all pairs. However, none of the correlation coefficients between NIR and optical DIBs are found to be as high as those among optical DIBs, suggesting that these NIR DIBs do not form a family with the optical DIBs. This is consistent with the discussion in \textsection{5.1}, in which a clear difference in the correlation coefficients with $E(B-V)$ between optical and NIR DIBs is shown.

In addition, we find that the three strongest NIR DIBs, $\lambda \lambda 10780, 11797, \text{and }13175$ are better correlated with $\lambda \lambda 5705.1, 5780.5, 6204.5, \text{and }6283.8$ ($r=0.8 - 0.9$) than with $\lambda \lambda 5797.1, 6196.0, \text{and }6613.6$ ($r=0.6-0.7$) although there are a few exceptions such as $\lambda 5705.1 - \lambda 13175$ ($r=0.75, N=10$), whose correlation coefficient is lower compared to those of $\lambda 5705.1 - \lambda 10780$ and $\lambda 5705.1 - \lambda 11797$. Figure \ref{NIROPTcorrelation} shows the correlations between four NIR DIBs and two representative optical DIBs in the two groups, $\lambda \lambda 5780.5 \text{ and } 5797.1$. The DIB $\lambda 10438$ is moderately well correlated with the eight optical DIBs (correlation coefficients are $\sim$0.7-0.8). Because the DIB $\lambda 10438$ is weaker than the other three DIBs, the correlation coefficients could be affected by the relatively large uncertainty of the EWs.

These two groups of optical DIBs classified by their correlation with NIR DIBs are consistent with the previous studies on the correlations among optical DIBs. \citet{cam97} showed that two pairs of $\lambda \lambda 5780.5, 6204.5$ and $\lambda \lambda 5797.1, 6613.6$, which are members of the groups, are well correlated. $\lambda \lambda 6196.0, 6613.6$ is the only pair that is proposed to have a perfect correlation \citep{mou99,mcc10}. 
However, the differences in the correlation coefficients between the two groups are not statistically significant. For all pairs, we cannot exclude the null hypothesis that the correlation coefficients are not truly different. For example, the probability that the two correlation coefficients of $\lambda 5780.5 - \lambda 10780$ ($r=0.85, N=17$) and $\lambda 5797.1 - \lambda 10780$ ($r=0.70, N=15$) are not different is estimated to be $p=0.32$, which is too high to reject the null hypothesis. 
Therefore, it is necessary to increase the number of samples to clarify the difference in correlation coefficients.

\section{The carriers of NIR DIBs}

\subsection{Cation Molecules}

In \textsection{5.3}, we suggested that $\lambda \lambda 10780, 11797, \text{and }13175$ are better correlated with $\lambda \lambda 5705.1, 5780.5, 6204.5, \text{and }6283.8$ ($r=0.8 - 0.9$) than with $\lambda \lambda 5797.1, 6196.0, \text{and }6613.6$ ($r=0.6-0.7$). 
Among the optical DIBs, the environmental dependencies of two representative DIBs in the two groups, $\lambda \lambda$5780.5 and 5797.1, have been extensively investigated \citep{kre87, cam97, vos11, fri11}. The ratio $\lambda 5797.1 / \lambda 5780.5$ depends on the local UV intensities of the clouds because of the different ionization potentials of the DIB carriers \citep{cam97, vos11}. \citet{cam97} suggested that the ionization potential of the $\lambda 5791.1$ carrier is lower than that of the $\lambda 5780.5$ carrier. On the basis of modeling of the photoionization equilibrium, \citet{son97} estimated the ionization potential of the $\lambda 5797.1$ carrier as $\sim$ 11 eV, whereas they suggested that the $\lambda 5780.5$ carrier is a cation molecule according to its high ionization potential. The stronger correlations of the three NIR DIBs with $\lambda 5780.5$ rather than with $\lambda 5797.1$ may show that their carriers are also cation molecules with ionization potentials as high as the $\lambda 5780.5$ carrier, probably ionized PAHs. It is also suggested that the other weaker NIR DIBs that are well correlated with the three strongest DIBs $\lambda \lambda 10780, 11797, \text{and }13175$, such as $\lambda \lambda 10504, 10792, \text{and }12623$ are also from cation molecules (see \textsection{5.2}).

In our sample, HD23180 ($E(B-V)=0.31$) and HD24398 ($E(B-V)=0.32$) have high $\lambda 5797.1 / \lambda 5780.5$ ratios of $\sim$0.7-0.8, suggesting that the clouds toward these stars are $\zeta -$type, meaning that the UV intensity is weak relative to the cloud density \citep{vos11}. We compare the EWs of NIR DIBs in the $\zeta -$type clouds with those toward two stars with similar $E(B-V)$, HD2905 ($E(B-V) = 0.33$) and HD24912 ($E(B-V)=0.33$), whose $\lambda 5797.1 / \lambda 5780.5$ ratios are $\sim 0.3$, which is near the threshold for the classification of $\zeta-, \sigma-$type clouds. 
$\lambda 5780.5$ in $\zeta -$type clouds is weaker than those toward HD2905 and HD24912 by a factor of two to three although their color excesses are almost the same. This can be interpreted as the results of the $\lambda 5780.5$ carrier decreasing in $\zeta -$type clouds, where the molecules are shielded from the UV flux. From our results, the NIR DIBs, $\lambda \lambda 10780, 11797, \text{and }13175$, in $\zeta -$type clouds are also found to be weaker than those toward HD2905 and HD24912 by a factor of three to four, which is similar to the factor of $\lambda 5780.5$. This similar dependence on the UV intensity also illustrates that the carriers of $\lambda \lambda 10780, 11797, \text{and }13175$ and $\lambda 5780.5$ may have similar ionization potentials.

\subsection{Carrier Candidates}

We showed that a considerable number of DIBs are widely distributed in the NIR wavelength range as in the optical wavelength range. A number of ionized large-sized carbon molecules are proposed to have the transitions in the NIR wavelength range, which is consistent with our suggestion that the carriers of some NIR DIBs are cation molecules. First, ionized buckminsterfullerenes (C$_{60}^+$ and C$_{60}^-$) are also suggested to have transitions in the NIR wavelength range from the laboratory experiment \citep{ful93}. In fact, DIBs $\lambda \lambda 9577$ and 9633 are expected to be the transitions of C$_{60}^+$ \citep{foi94}. Next, ionized PAHs (cation and anion) are theoretically suggested to have lower-energy transitions in the NIR wavelength range \citep{sal96,rui05}. \citet{mat05} obtained the absorption spectra of some ionized PAH molecules in the NIR wavelength range. Recently, \citet{jon13} proposed that heteroatom-doped (e.g., Si, O, N, and S) hydrogenated amorphous carbons (a-C: H: X) can be the DIB carriers. However, there is no strong evidence of any of those candidates of NIR DIB carriers. Further observational constraints on DIB properties as well as the identification with NIR absorption spectra of the proposed molecules in the laboratory are necessary to reveal the carriers of NIR DIBs.

The NIR absorption spectra of 27 ionized PAHs (cations and anions) are obtained by \citet{mat05} using the matrix-isolation technique. \citet{cox14}, who explored the DIBs in the NIR wavelength range, compared their newly detected DIB candidates and the electronic transitions of ionized PAHs. They found that some of the NIR DIBs are detected within a few tens of angstroms from the wavelengths of the strong electronic transitions of ionized PAHs reported by \citet{mat05}. Among the eight NIR DIB candidates that are confirmed as DIBs in this study, they suggested that wavelengths of $\lambda \lambda 10393, 10438, 10504, 10780, \text{and }13027$ are near those of strong bands of singly ionized PAHs. Among the eight NIR DIBs newly detected in this study, the DIB $\lambda 10792$ is detected close to the wavelength of the strong electronic transition of C$_{42}$H$_{22}^+$ ($\lambda = 10790$\r{A}), but the wavelength of the other DIBs do not correspond to the strong electronic transitions of ionized PAHs. However, as \citet{cox14} pointed out, the pure gas-phase spectra of ionized PAHs are required for the firm identification of NIR DIBs with ionized PAHs.

\section{Summary}

In order to search for new DIBs in the NIR wavelength range and study their properties, we obtained NIR high-resolution spectra of 25 early-type stars with color excesses of $0.07 < E(B-V) < 3.4$ using the newly developed NIR Echelle spectrograph WINERED. Consequently, we identified 15 new DIBs in the wavelength range $0.91 < \lambda < 1.32$ $\mu$m in addition to five DIBs previously detected in this wavelength range. For the first time, we investigated the correlations among NIR DIBs, eight strong optical DIBs investigated by \citet{fri11} ($\lambda \lambda$5487.7, 5705.1, 5780.5, 5797.1, 6196.0, 6204.5, 6283.8, and 6613.6), and the reddening of stars with a large number of lines of sight. Our findings are summarized as follows:

\begin{enumerate}

\item All identified NIR DIBs are found to have correlation coefficients with $E(B-V)$ that are not as high as those of the optical DIBs. The lower correlation coefficients of NIR DIBs with $E(B-V)$ show that the abundance of the NIR DIB carriers depends on some physical properties of gas clouds, such as UV flux, as well as on the amount of gas in the line of sight. 

\item Three NIR DIBs, $\lambda \lambda 10780, 10792, \text{and } 11797$, are found to be well correlated with each other, suggesting that these three DIBs consist of a ``family,'' wherein the carrier molecules share similar physical and chemical properties. Because some data points do not quite fit the correlation, these DIBs would appear to originate from a different molecule.
 
\item The three strongest NIR DIBs, $\lambda \lambda 10780, 11797, \text{and }13175$, are found to be correlated better with $\lambda \lambda 5705.1, 5780.5, 6204.5, \text{and }6283.8$ ($r=0.8 - 0.9$) than with $\lambda \lambda 5797.1, 6196.0, \text{and }6613.6$ ($r=0.6-0.7$). Moreover, the NIR DIBs are found to be relatively weaker in $\zeta -$type clouds, where the molecules are shielded from UV flux, in the same manner as $\lambda 5780.5$, whose carrier is proposed to be a cation molecule with high ionization potential. Considering the similarity between the strongest NIR DIBs as well as some other NIR DIBs with $\lambda 5780.5$, we suggest that the carriers of these three NIR DIBs are also cation molecules. 

\end{enumerate}

We are grateful to the staff of Koyama Astronomical Observatory for their support during our observation. We thank Alan T. Tokunaga and Setsuko Wada for their useful comments. We would like to acknowledge the helpful comments made by the anonymous referee. This study has been financially supported by KAKENHI (16684001) Grant-in-Aid for Young Scientists (A), KAKENHI (20340042) Grant-in-Aid for Scientific Research (B), KAKENHI (26287028) Grant-in-Aid for Scientific Research (B), KAKENHI (21840052) Grant-in-Aid for Young Scientists (Start-up), and the Japan Society for the Promotion of Science, MEXT-Supported Program for the Strategic Research Foundation at Private Universities, 2008-2012 (No.S0801061) and 2014-2018 (No. S1411028). S.H. is supported by Grant-in-Aid for JSPS Fellows Grant No. 13J10504.

\begin{deluxetable*}{cccccccccc}
\tabletypesize{\scriptsize}
\tablecaption{Summary of Targets}
\tablewidth{0pt}
\tablehead{
 \colhead{Star} & \colhead{Spectral Type} & \colhead{$J$} & \colhead{$E(B-V)$} & \colhead{Integration Time} & \colhead{S/N\footnote{The signal-to-noise ratio of the spectrum at $\lambda \sim 10400$\r{A}.}} &\colhead{S/N\footnote{The signal-to-noise ratio of the spectrum at $\lambda \sim 10400$\r{A} after the correction of telluric absorption lines.}} & \colhead{Telluric \footnote{Telluric standard stars used for the correction of the telluric absorption lines.}} & \colhead{$v_\odot$ \footnote{The heliocentric velocity of the strongest velocity component of the interstellar clouds.}} & \colhead{Ref.\footnote{Left: References for DIBs in the optical wavelength range. Right: References for atomic absorption lines to define the velocities of interstellar clouds: (1) \citet{fri11}, (2) \citet{wel01} (\ion{K}{1}), (3) on-line catalog by Dan Welty (http://astro.uchicago.edu/~dwelty/gal.html), (4) \citet{wel03} (\ion{Ca}{1}), (5) \citet{wel94} (\ion{Na}{1}), (6) \citet{pri01} (\ion{Na}{1}, \ion{Ca}{2}) }} \\
 \colhead{} & \colhead{} & \colhead{(mag)} & \colhead{(mag)} & \colhead{(s)} & \colhead{Raw} & \colhead{Corrected} & \colhead{} & \colhead{(km s$^{-1}$)}
}
\startdata
HD2905 &  B1Iae &   4.141 &  0.33 & 720 & 600 & 600 & 50 Cas & -21 & 1, 2 \\
HD12953 &  A1Iae &   4.062 &  0.62 & 960 & 700 & 600 & 50 Cas & $-$ & $-$, $-$ \\
HD14489 &  A2Ia &  4.528 &  0.40 & 1,800 & 750 & 650 & 50 Cas & $-$ & $-$, $-$ \\
HD20041 &  A0Ia &   4.057 &  0.73 & 1,200  & 750 & 750 & 50 Cas & -5 & 1, 3 \\
HD21291 &  B9Ia & 2.712 &  0.42 & 1,200 & 400 & 300 & 50 Cas & 0 & $-$, 2 \\
HD21389 &  A0Iab & 3.050 &  0.55 & 1,500 & 500 & 500 & 50 Cas & -10 & 1, 3  \\
HD23180 &  B1III &   3.610 &  0.31 & 1,300 & 600 & 500 & 7 Cam & 13 & 1, 2 \\
HD24398 &  B1Ib & 2.650 &  0.32 & 400 & 450 & 300 & 50 Cas & 15 & 1, 2  \\
HD24912 &  O7.5IIIe & 3.987 &  0.33 & 1,200 & 400 & 300 & 50 Cas & 7 & 1, 2  \\
HD25204 &  B3V+A4IV &  3.572 &  0.30 & 720 & 650 & 400 & 7 Cam & $-$ & $-$, $-$ \\
HD30614 &  O9.5Ia &   4.243 &  0.30 & 960 & 550 & 400 & 7 Cam & -3 & 1, 4 \\
HD36371 &  B4Ib &  4.246 &  0.43 & 720 & 700 & 550 & 7 Cam & $-$ & 1, $-$ \\
HD36486 &  B0III+O9V &  2.744 &  0.07 & 240 & 750 & 400 & 21 Lyn & 24 & 1, 5  \\
HD36822 &  B0III &  4.807 &  0.11 & 1,200 & 600 & 450 & 7 Cam & 25 & 1, 6  \\
HD37043 &  O9III &  3.490 &  0.07 & 600 & 600 & 350 & 7 Cam & 8 & 1, 5  \\
HD37128 &  B0Iab &  2.191 &  0.08 & 240 & 700 & 450 & HR1483 & 25 & 1, 2  \\
HD37742 &  O9Iab &  2.210 &  0.08 & 240 & 850 & 400 & 7 Cam & 23 & 1, 6  \\
HD38771 &  B0Iab & 2.469 &  0.07 & 240 & 600 & 350 & HR1483 & 20 & 1, 2  \\
HD41117 &  B2Iaev &  4.186 &  0.44 & 720 & 500 & 400 &21 Lyn & 15 & 1, 2  \\
HD43384 & B3Iab &  5.187 &  0.57 & 1,440 & 500 & 400 & 21 Lyn & $-$ & 1, $-$  \\
HD50064 & B6:Ia &  6.175 &  0.85 & 1,200 & 300 & 300 & omi Aur & $-$ & 1, $-$  \\
HD190603 & B1.5Ia+ &  4.416 &  0.70 & 800 & 200 & 200 & HIP87108 & $-$ &  $-$, $-$ \\
HD202850 & B9Iab &  3.973 &  0.13 & 180  & 170 & 170 & 50 Cas & $-$ & 1, $-$ \\
HD223385 & A3Iae &  3.500 &  0.59 & 720 & 650 & 550 & 50 Cas & -28 & $-$, 2 \\
Cyg OB2 No.12 & B3-4Ia+ & 4.667 & 3.4 & 800 & 100 & 100 & HIP87108 &$-$ &  1, $-$ \\
Rigel &  B8Iab & 0.206 &  0.00 & 8  & 650 & 300 & HR1483 & $-$ & $-$, $-$
\enddata
\tablecomments{}
\label{targets}
\end{deluxetable*}

\begin{deluxetable*}{ccccccc}
\tabletypesize{\scriptsize}
\tablecaption{Summary of the Identified DIBs}
\tablewidth{0pt}
\tablehead{
 \colhead{DIB \footnote{The rest-frame wavelength in air.}} & \colhead{$\lambda _0$ (\r{A})\footnote{The center wavelength measured for the profile of DIBs toward HD20041. The 1 $\sigma$ uncertainties are shown.}} & \colhead{EW (m\r{A})\footnote{The equivalent widths of DIBs toward HD20041. The 1 $\sigma$ uncertainties are shown.}} & \colhead{FWHM\footnote{The 1 $\sigma$ uncertainties are shown.} (\r{A})} & \colhead{Ref. \footnote{References, in which the DIB was reported for the first time: (1) \citet{job90}, (2) \citet{cox14}, (3) \citet{gro07}}} & \colhead{Stellar line \footnote{Contaminated stellar absorption lines}} & \colhead{Telluric effect \footnote{How heavily the DIB is affected by the contamination of telluric absorption lines.}}
}
\startdata
$\lambda$9880 & 9880.34 $\pm 0.10$ & 18.2 $\pm$ 1.9 & 1.25 	$\pm	0.12 	$  &   This study & $-$ & slightly affected \\
$\lambda$10360 & 10360.69 $\pm 0.10$ & 17.5 $\pm$ 1.2 & 1.63 	$\pm	0.11 	$ &   2   & $-$ & none \\
$\lambda$10393 & 10393.17 $\pm 0.10$ & 25.6 $\pm$ 1.6 & 2.62 	$\pm	0.18 	$ & 2 & \ion{Mg}{2} & none \\
$\lambda$10438 & 10438.00 $\pm 0.21 $ & 21.6 $\pm$ 1.6 & 6.06 	$\pm	0.52 	$ &   2  & $-$ & none \\
$\lambda$10504 & 10504.52 $\pm 0.10$ & 36.2 $\pm$ 0.9 & 1.46 	$\pm	0.25 	$ &   2 & \ion{N}{1}, \ion{Fe}{2} & none \\
$\lambda$10697 & 10696.71 $\pm0.05$  & 120.3 $\pm$ 5.3 & 4.73 	$\pm	0.12 	$ &  2 & many \ion{C}{1}, \ion{Si}{1} & slightly affected \\
$\lambda$10780 & 10780.46 $\pm 0.10$ & 57.1 $\pm$ 2.1 & 1.64 	$\pm	0.07 $ &   3 & $-$ & slightly affected  \\
$\lambda$10792 & 10792.15 $\pm 0.18$ & 17.3 $\pm$ 2.0 & 2.52 	$\pm	0.43 $ &   3 & $-$ & slightly affected \\
$\lambda$11797 & 11797.53  $\pm 0.10$ & 61.0 $\pm$ 3.5 & 1.51 	$\pm	0.04 $ &   1 & $-$ & overlapped \\
$\lambda$12293\footnote{$\lambda _0$ and FWHM of this DIB are measured for HD21389. See \textsection 4.2 for detail.} & 12293.69   $\pm 0.10$ & $-$ & 1.76 	$\pm	0.24 	$ &   This study  & \ion{N}{1} & overlapped \\
$\lambda$12337 & 12337.32  $\pm 0.23$ & 25.5 $\pm$ 2.0 & 5.08 	$\pm	0.53 $  &   2   & $-$ & slightly affected \\
$\lambda$12518 & 12518.69 $\pm 0.18$ & 9.7 $\pm$ 1.6 & 2.32 	$\pm	0.42 $ &   This study   & $-$ & slightly affected \\
$\lambda$12536 & 12536.69 $\pm 0.13$ & 13.1 $\pm$ 1.8 & 1.72 	$\pm	0.31 $ &   This study   & $-$ & slightly affected \\
$\lambda$12623 & 12623.86 $\pm 0.11$ & 38.1 $\pm$ 4.1 & 2.16 	$\pm	0.25 $ &   This study   & $-$ &  overlapped \\
$\lambda$12799 & 12799.13 $\pm 0.09$ & 50.8 $\pm$ 3.1 & 2.22 	$\pm	0.20 $ &   This study   & \ion{He}{1}, \ion{H}{1}(Pa $\beta$) & slightly affected \\
$\lambda$12861 & 12861.62 $\pm 0.22$ & 24.4 $\pm$ 2.3 & 4.36 	$\pm	0.51 	$ &   This study   & $-$ & slightly affected \\
$\lambda$13027 & 13027.68 $\pm 0.35$  & 37.6 $\pm$ 2.2 & 3.36 $\pm	0.81 $ &   2   & $-$ &  overlapped \\
$\lambda$13175 & 13175.55 $\pm 0.10$ & 285.2 $\pm$ 5.6 & 4.07 	$\pm	0.19 	$ &   1   & $-$ & overlapped
\enddata
\tablecomments{}
\label{DIBs}
\end{deluxetable*}
\clearpage

\clearpage

\begin{deluxetable*}{ccccccccccc}
\tabletypesize{\scriptsize}
\tablecaption{EWs of DIBs}
\tablewidth{0pt}
\tablehead{
 \colhead{Stars} & \colhead{\tiny E(B-V)} & \colhead{$\lambda 9980$} & \colhead{$\lambda 10360$} & \colhead{$\lambda 10393$} &\colhead{$\lambda 10438$} &\colhead{$\lambda 10504$}  & \colhead{$\lambda 10697$} & \colhead{$\lambda 10780$} & \colhead{$\lambda 10792$}& \colhead{$\lambda 11797$}
}
\startdata
HD2905 & 0.33 & 10.9 $\pm$ 2.2 & 6.0 $\pm$ 0.9 & 5.8 $\pm$ 0.9 & 4.6 $\pm$ 0.9 & 8.3 $\pm$ 1.2 & 93.7 $\pm$ 5.3 & 47.8 $\pm$ 2.0 & 7.5 $\pm$ 1.5 & 38.3 $\pm$ 2.7\\ 
HD12953 & 0.62 & 13.0 $\pm$ 1.6 & 19.4 $\pm$ 1.6 & 11.4 $\pm$ 0.9 & $<$ 3.3 & $-$ & 90.5 $\pm$ 5.5 & 71.2 $\pm$ 2.6 & 25.3 $\pm$ 2.9 & 78.7 $\pm$ 3.2\\ 
HD14489 & 0.40 & 11.8 $\pm$ 1.8 & 14.1 $\pm$ 1.0 & $-$ & $<$ 3.0 & 30.8 $\pm$ 0.7 & 86.1 $\pm$ 4.9 & 66.4 $\pm$ 1.5 & 19.4 $\pm$ 2.0 & 70.1 $\pm$ 2.8\\ 
HD20041 & 0.73 & 18.2 $\pm$ 1.9 & 17.5 $\pm$ 1.2 & 25.6 $\pm$ 1.6 & 21.6 $\pm$ 1.6 & 36.2 $\pm$ 0.9 & 120.3 $\pm$ 5.3 & 57.1 $\pm$ 2.1 & 17.3 $\pm$ 2.0 & 61.0 $\pm$ 3.5\\ 
HD21291 & 0.42 & $<$ 8.4 & 8.3 $\pm$ 1.3 & $<$ 4.2 & 16.9 $\pm$ 2.2 & 18.4 $\pm$ 1.5 & 102.9 $\pm$ 9.2 & 22.7 $\pm$ 3.3 & 11.5 $\pm$ 3.0 & 18.0 $\pm$ 4.6\\ 
HD21389 & 0.55 & 16.8 $\pm$ 2.0 & 15.5 $\pm$ 1.2 & $-$ & 13.2 $\pm$ 1.5 & 45.9 $\pm$ 1.1 & 217.3 $\pm$ 9.2 & 123.9 $\pm$ 3.3 & 36.3 $\pm$ 2.8 & 103.6 $\pm$ 3.5\\ 
HD23180 & 0.31 & $<$ 5.3 & $<$ 2.5 & $<$ 2.8 & 10.8 $\pm$ 1.3 & $<$ 4.3 & $-$ & 11.8 $\pm$ 1.3 & $<$ 4.1 & 10.9 $\pm$ 1.4\\ 
HD24398 & 0.32 & $<$ 8.1 & $<$ 4.0 & $<$ 3.6 & $<$ 4.4 & $<$ 6.2 & $-$ & 9.2 $\pm$ 2.3 & $<$ 7.9 & $<$ 19.8\\ 
HD24912 & 0.33 & $<$ 7.5 & $<$ 4.9 & $<$ 4.2 & 10.7 $\pm$ 1.9 & 18.6 $\pm$ 2.0 & 55.4 $\pm$ 5.6 & 44.5 $\pm$ 3.4 & 17.7 $\pm$ 2.8 & 33.8 $\pm$ 4.6\\ 
HD25204 & 0.30 & $<$ 5.2 & $<$ 2.2 & $<$ 2.3 & $<$ 2.4 & $<$ 4.6 & $-$ & $<$ 6.9 & $<$ 6.5 & $<$ 9.6\\ 
HD30614 & 0.30 & $<$ 5.0 & $<$ 2.7 & 6.3 $\pm$ 0.9 & $<$ 3.1 & 10.4 $\pm$ 1.2 & $-$ & 6.7 $\pm$ 1.0 & $-$ & 24.9 $\pm$ 3.2\\ 
HD36371 & 0.43 & 9.3 $\pm$ 1.6 & 9.1 $\pm$ 1.1 & 9.2 $\pm$ 1.0 & 18.3 $\pm$ 1.4 & 29.1 $\pm$ 1.1 & 139.4 $\pm$ 7.4 & 58.0 $\pm$ 2.6 & 15.4 $\pm$ 1.0 & 67.4 $\pm$ 3.3\\ 
HD36486 & 0.07 & $<$ 5.1 & $<$ 2.3 & $<$ 2.2 & $<$ 2.7 & $<$ 3.7 & $-$ & 9.6 $\pm$ 1.6 & $<$ 5.4 & $<$ 7.2\\ 
HD36822 & 0.11 & $<$ 5.6 & $<$ 2.4 & $<$ 2.8 & $<$ 3.2 & $<$ 4.5 & $-$ & 11.9 $\pm$ 1.5 & $<$ 5.3 & 17.1 $\pm$ 3.5\\ 
HD37043 & 0.07 & $<$ 6.1 & $<$ 3.3 & $<$ 2.5 & $<$ 2.4 & $<$ 4.4 & $-$ & $<$ 6.1 & $<$ 6.1 & $<$ 14.4\\ 
HD37128 & 0.08 & $<$ 7.7 & $<$ 2.5 & $<$ 2.0 & $<$ 2.4 & $<$ 3.4 & $-$ & 9.5 $\pm$ 1.7 & $<$ 4.9 & $<$ 7.2\\ 
HD37742 & 0.08 & $<$ 6.2 & $<$ 2.0 & $<$ 1.9 & $<$ 2.2 & $<$ 3.3 & $-$ & $<$ 6.0 & $<$ 6.0 & $<$ 9.6\\ 
HD38771 & 0.07 & $<$ 8.4 & $<$ 2.5 & $<$ 2.3 & $<$ 2.6 & $<$ 3.8 & $-$ & 13.0 $\pm$ 2.0 & $<$ 6.2 & $-$\\ 
HD41117 & 0.44 & 16.0 $\pm$ 1.7 & 8.2 $\pm$ 1.0 & 6.5 $\pm$ 0.9 & 16.5 $\pm$ 1.4 & 18.1 $\pm$ 1.0 & 85.6 $\pm$ 5.8 & 47.2 $\pm$ 2.0 & 20.6 $\pm$ 2.0 & 47.2 $\pm$ 2.3\\ 
HD43384 & 0.57 & 15.2 $\pm$ 1.9 & 14.4 $\pm$ 1.5 & $<$ 3.9 & 11.5 $\pm$ 1.3 & 26.9 $\pm$ 1.1 & 117.7 $\pm$ 4.6 & 54.3 $\pm$ 2.3 & 14.5 $\pm$ 2.2 & 70.5 $\pm$ 3.1\\ 
HD50064 & 0.85 & $<$ 18.3 & $<$ 6.4 & $<$ 6.5 & 40.9 $\pm$ 2.9 & $-$ & 171.6 $\pm$ 9.6 & 96.3 $\pm$ 4.2 & 36.0 $\pm$ 4.3 & 105.1 $\pm$ 6.6\\ 
HD190603 & 0.70 & 18.2 $\pm$ 5.2 & 18.2 $\pm$ 2.7 & $<$ 15.8 & $<$ 13.0 & 45.5 $\pm$ 6.6 & 233.8 $\pm$ 12.8 & 136.2 $\pm$ 5.2 & 50.9 $\pm$ 3.7 & 139.2 $\pm$ 7.8\\ 
HD202850 & 0.13 & $<$ 14.1 & $<$ 10.9 & $<$ 10.7 & $<$ 12.8 & $-$ & $-$ & 43.2 $\pm$ 4.1 & $-$ & 46.2 $\pm$ 8.5\\ 
HD223385 & 0.59 & 22.1 $\pm$ 2.5 & 27.5 $\pm$ 1.4 & $-$ & 24.6 $\pm$ 1.5 & $-$ & 169.1 $\pm$ 6.3 & 156.3 $\pm$ 3.3 & 57.3 $\pm$ 3.4 & 170.4 $\pm$ 4.1\\ 
\tiny CygOB2No.12\scriptsize & 3.4 & 92.6 $\pm$ 10.2 & 72.7 $\pm$ 5.9 & 37.7 $\pm$ 5.3 & $<$ 27.8 & 99.0 $\pm$ 3.5 & 434.6 $\pm$ 16.5 & 322.9 $\pm$ 6.1 & 72.8 $\pm$ 6.5 & 300.5 $\pm$ 14.5\\ 
Rigel & 0.0 & $<$ 11.0 & $<$ 2.2 & $<$ 2.4 & $<$ 2.5 & $-$ & $-$ & $<$ 5.7 & $<$ 5.8 & $<$ 9.3
\enddata
\tablecomments{The EWs are given in units of m\r{A}. The bars mean that the EW or upper limit of the DIB could not be evaluated due to the overlapped stellar absorption lines and/or the residual features of the telluric correction.}
\label{DIBew1}
\end{deluxetable*}

\setcounter{table}{2}

\begin{deluxetable*}{ccccccccccc}
\tabletypesize{\scriptsize}
\tablecaption{Continued.}
\tablewidth{0pt}
\tablehead{
 \colhead{Stars} & \colhead{\tiny E(B-V)} & \colhead{$\lambda 12293$} & \colhead{$\lambda 12337$} & \colhead{$\lambda 12518$} & \colhead{$\lambda 12536$} & \colhead{$\lambda 12623$} & \colhead{$\lambda 12799$} & \colhead{$\lambda 12861$} & \colhead{$\lambda 13027$} & \colhead{$\lambda 13175$}
}
\startdata
HD2905 & 0.33 & 5.9 $\pm$ 1.1 & 7.4 $\pm$ 1.1 & $<$ 5.0 & $<$ 4.3 & 10.1 $\pm$ 2.1 & 12.3 $\pm$ 1.6 & $<$ 5.5 & 28.2 $\pm$ 2.2 & 99.2 $\pm$ 5.0\\ 
HD12953 & 0.62 & 7.9 $\pm$ 1.1 & 13.7 $\pm$ 1.5 & $<$ 3.2 & 10.9 $\pm$ 1.3 & 20.1 $\pm$ 2.5 & 22.3 $\pm$ 1.4 & 27.3 $\pm$ 2.3 & 18.2 $\pm$ 1.8 & 294.0 $\pm$ 5.3\\ 
HD14489 & 0.40 & 14.5 $\pm$ 1.4 & 16.1 $\pm$ 2.0 & 8.1 $\pm$ 1.4 & 27.0 $\pm$ 2.4 & 37.5 $\pm$ 3.8 & 30.1 $\pm$ 2.3 & 21.4 $\pm$ 1.9 & $<$ 5.7 & 307.0 $\pm$ 6.1\\ 
HD20041 & 0.73 & $-$ & 25.5 $\pm$ 2.0 & 9.7 $\pm$ 1.6 & 13.1 $\pm$ 1.8 & 38.1 $\pm$ 4.1 & 50.8 $\pm$ 3.1 & 24.4 $\pm$ 2.3 & 37.6 $\pm$ 2.2 & 285.2 $\pm$ 5.6\\ 
HD21291 & 0.42 & $<$ 9.0 & 17.9 $\pm$ 2.7 & $<$ 10.5 & $<$ 12.5 & $<$ 13.8 & 38.8 $\pm$ 4.8 & $<$ 9.2 & $<$ 12.5 & $-$\\ 
HD21389 & 0.55 & 19.4 $\pm$ 2.0 & 15.2 $\pm$ 1.7 & 7.4 $\pm$ 1.5 & 29.8 $\pm$ 2.3 & 53.9 $\pm$ 3.6 & 43.4 $\pm$ 2.7 & 58.4 $\pm$ 3.3 & 17.3 $\pm$ 2.5 & 479.4 $\pm$ 9.1\\ 
HD23180 & 0.31 & 7.9 $\pm$ 1.6 & $<$ 5.0 & $<$ 6.3 & $<$ 6.3 & $<$ 5.7 & $<$ 7.3 & $<$ 4.0 & $<$ 4.5 & 13.8 $\pm$ 3.1\\ 
HD24398 & 0.32 & $<$ 10.0 & $<$ 8.5 & $<$ 10.3 & $<$ 11.3 & $<$ 9.6 & $<$ 13.0 & $<$ 10.3 & $<$ 13.0 & $<$ 17.5\\ 
HD24912 & 0.33 & 11.9 $\pm$ 2.4 & 19.7 $\pm$ 2.9 & $<$ 9.2 & $<$ 11.5 & $<$ 15.1 & 25.1 $\pm$ 3.5 & $<$ 11.1 & $<$ 11.9 & 199.2 $\pm$ 13.7\\ 
HD25204 & 0.30 & $<$ 5.5 & $<$ 5.4 & $<$ 6.1 & $<$ 7.4 & $-$ & $<$ 9.1 & $<$ 7.0 & $<$ 6.9 & $<$ 16.7\\ 
HD30614 & 0.30 & $<$ 5.0 & $<$ 4.9 & $<$ 5.0 & $<$ 4.7 & $<$ 6.4 & $<$ 7.0 & $<$ 5.5 & $<$ 5.4 & $<$ 13.2\\ 
HD36371 & 0.43 & 12.1 $\pm$ 1.8 & $<$ 5.1 & 12.8 $\pm$ 1.9 & $<$ 7.0 & $-$ & 10.0 $\pm$ 2.8 & 16.2 $\pm$ 2.7 & 14.3 $\pm$ 2.5 & 219.0 $\pm$ 7.7\\ 
HD36486 & 0.07 & $<$ 8.3 & $<$ 7.0 & $<$ 7.0 & $<$ 6.7 & $<$ 7.8 & 16.9 $\pm$ 2.9 & 15.7 $\pm$ 2.2 & $<$ 6.9 & $<$ 13.7\\ 
HD36822 & 0.11 & $<$ 5.2 & $<$ 5.8 & $<$ 6.5 & $<$ 6.5 & $<$ 11.9 & $<$ 10.9 & $<$ 7.4 & $<$ 8.4 & $<$ 14.2\\ 
HD37043 & 0.07 & $<$ 8.0 & $<$ 7.5 & $<$ 8.9 & $<$ 11.6 & $<$ 13.3 & $-$ & $<$ 8.7 & $<$ 9.0 & $<$ 13.4\\ 
HD37128 & 0.08 & $<$ 5.3 & $<$ 4.4 & $<$ 6.1 & $<$ 7.3 & $<$ 7.1 & $-$ & 10.4 $\pm$ 2.0 & $<$ 7.4 & $-$\\ 
HD37742 & 0.08 & $<$ 6.2 & $<$ 6.1 & $<$ 6.6 & $<$ 6.5 & $<$ 8.6 & $<$ 8.4 & 11.0 $\pm$ 1.9 & $<$ 8.4 & $<$ 14.4\\ 
HD38771 & 0.07 & $<$ 5.6 & $-$ & $<$ 7.6 & $<$ 9.8 & $<$ 11.1 & $-$ & $<$ 7.7 & $<$ 8.0 & $-$\\ 
HD41117 & 0.44 & $<$ 5.1 & 11.9 $\pm$ 1.6 & $<$ 6.8 & 9.3 $\pm$ 1.8 & 18.6 $\pm$ 2.2 & 21.4 $\pm$ 1.9 & 18.7 $\pm$ 2.3 & 36.1 $\pm$ 2.7 & 155.3 $\pm$ 4.7\\ 
HD43384 & 0.57 & 8.2 $\pm$ 1.6 & 39.0 $\pm$ 3.0 & $<$ 7.4 & 11.1 $\pm$ 2.2 & 29.8 $\pm$ 2.9 & 32.4 $\pm$ 2.3 & 32.5 $\pm$ 3.2 & 19.0 $\pm$ 2.5 & 241.8 $\pm$ 6.6\\ 
HD50064 & 0.85 & 14.2 $\pm$ 3.4 & $<$ 9.6 & $<$ 10.3 & 55.7 $\pm$ 4.9 & 65.3 $\pm$ 5.9 & 56.0 $\pm$ 3.8 & $<$ 15.8 & 43.2 $\pm$ 4.4 & 386.5 $\pm$ 14.8\\ 
HD190603 & 0.70 & 21.5 $\pm$ 2.9 & 76.5 $\pm$ 5.6 & 31.5 $\pm$ 4.1 & $<$ 9.8 & 48.9 $\pm$ 3.7 & 21.6 $\pm$ 3.6 & 24.4 $\pm$ 4.0 & $-$ & 493.8 $\pm$ 11.6\\ 
HD202850 & 0.13 & $<$ 16.2 & $<$ 15.6 & $<$ 14.8 & $<$ 15.8 & $<$ 23.5 & $-$ & $<$ 14.8 & $<$ 16.2 & $-$\\ 
HD223385 & 0.59 & 12.5 $\pm$ 1.7 & 37.1 $\pm$ 2.8 & 24.3 $\pm$ 1.7 & 22.2 $\pm$ 1.6 & 41.3 $\pm$ 3.1 & 34.9 $\pm$ 2.2 & 27.4 $\pm$ 3.2 & $-$ & 436.6 $\pm$ 6.5\\ 
\tiny CygOB2No.12\scriptsize & 3.4 & $-$ & 162.3 $\pm$ 10.9 & 42.7 $\pm$ 5.5 & 48.7 $\pm$ 4.7 & 141.2 $\pm$ 6.5 & 87.7 $\pm$ 6.0 & 73.6 $\pm$ 4.2 & $-$ & 982.3 $\pm$ 12.5\\ 
Rigel & 0.0 & $<$ 7.5 & $-$ & $<$ 9.0 & $<$ 7.6 & $<$ 9.0 & $-$ & $<$ 7.6 & $<$ 9.4 & $-$
\enddata
\tablecomments{}
\label{DIBew2}
\end{deluxetable*}

\begin{deluxetable}{ccccc}
\tabletypesize{\scriptsize}
\tablecaption{Correlation Coefficients between EWs and $E(B-V)$ and the Results of Linear Least-square Fit}
\tablewidth{0pt}
\tablehead{
\colhead{DIB} & \colhead{$N$\footnote{Number of stars with DIB detection (5$\sigma$), which are used for the calculation of the correlation coefficients. The data of Cyg OB2 No.12 is not included in the calculations of correlation coefficients and the fitting of linear functions ($EW= a \times E(B-V) + b$). See \textsection{5.1} for detail.}} & \colhead{$r$ \footnote{Correlation coefficients ($E(B-V)$ - EW)}} & \colhead{$a$} & \colhead{$b$}
}
\startdata
$\lambda$9880 & 10 & 0.68 & 20$\pm$8 & 4$\pm$4 \\ 
$\lambda$10360 & 11 & 0.73 & 35$\pm$11 & -4$\pm$6 \\ 
$\lambda$10393 & 6 & 0.88 & 40$\pm$11 & -8$\pm$5 \\ 
$\lambda$10438 & 11 & 0.83 & 46$\pm$10 & -6$\pm$5 \\ 
$\lambda$10504 & 11 & 0.80 & 71$\pm$18 & -8$\pm$8 \\ 
$\lambda$10697 & 13 & 0.60 & 202$\pm$81 & 21$\pm$43 \\ 
$\lambda$10780 & 21 & 0.73 & 141$\pm$30 & -4$\pm$12 \\ 
$\lambda$10792 & 13 & 0.56 & 53$\pm$24 & -3$\pm$13 \\ 
$\lambda$11797 & 17 & 0.69 & 149$\pm$41 & -4$\pm$19 \\ 
$\lambda$12293 & 11 & 0.45 & 13$\pm$8 & 6$\pm$4 \\ 
$\lambda$12337 & 11 & 0.60 & 84$\pm$37 & -18$\pm$19 \\ 
$\lambda$12518 & 6 & 0.46 & 34$\pm$33 & -4$\pm$18 \\ 
$\lambda$12536 & 8 & 0.50 & 54$\pm$38 & -10$\pm$22 \\ 
$\lambda$12623 & 10 & 0.71 & 77$\pm$27 & -8$\pm$16 \\ 
$\lambda$12799 & 14 & 0.65 & 45$\pm$15 & 7$\pm$8 \\ 
$\lambda$12861 & 12 & 0.56 & 30$\pm$14 & 11$\pm$6 \\ 
$\lambda$13027 & 8 & 0.48 & 31$\pm$23 & 9$\pm$13 \\ 
$\lambda$13175 & 13 & 0.70 & 601$\pm$182 & -39$\pm$96 \\  \hline \\
$\lambda$5487.7 & 13 & 0.91 & 194$\pm$26 & -8$\pm$11 \\ 
$\lambda$5705.1 & 14 & 0.85 & 181$\pm$32 & 0$\pm$14 \\ 
$\lambda$5780.5 & 19 & 0.94 & 809$\pm$70 & -34$\pm$23 \\ 
$\lambda$5797.1 & 15 & 0.94 & 298$\pm$31 & -9$\pm$12 \\ 
$\lambda$6196.0 & 19 & 0.96 & 82$\pm$6 & -3$\pm$2 \\ 
$\lambda$6204.5 & 18 & 0.93 & 305$\pm$31 & -2$\pm$11 \\ 
$\lambda$6283.8 & 17 & 0.89 & 1662$\pm$221 & 19$\pm$81 \\ 
$\lambda$6613.6 & 18 & 0.98 & 358$\pm$18 & -22$\pm$6
\enddata
\tablecomments{}
\label{ebvcc}
\end{deluxetable}

\begin{deluxetable*}{cccccccccccc}
\tabletypesize{\scriptsize}
\tablecaption{Correlation Coefficients among NIR DIBs}
\tablewidth{0pt}
\tablehead{
\colhead{} & \colhead{$\lambda$9880} & \colhead{$\lambda$10360} & \colhead{$\lambda$10438} & \colhead{$\lambda$10504} & \colhead{$\lambda$10697} & \colhead{$\lambda$10780} & \colhead{$\lambda$10792} & \colhead{$\lambda$11797} & \colhead{$\lambda$12337} & \colhead{$\lambda$12623} & \colhead{$\lambda$13175}
}
\startdata
$\lambda$9880& $-$ & 0.75 (10)& 0.62 (7)& 0.57 (8)& 0.52 (10)& 0.70 (10)& 0.76 (10)& 0.71 (10)& 0.54 (9)& 0.61 (9)& 0.63 (10)\\ 
$\lambda$10360& $-$ & $-$ & 0.66 (8)& 0.89 (9)& 0.45 (11)& 0.78 (11)& 0.80 (11)& 0.85 (11)& 0.47 (10)& 0.57 (9)& 0.75 (10)\\ 
$\lambda$10438& $-$ & $-$ & $-$ & 0.47 (8)& 0.46 (10)& 0.46 (11)& 0.56 (10)& 0.55 (11)& 0.51 (8)& 0.77 (7)& 0.57 (10)\\ 
$\lambda$10504& $-$ & $-$ & $-$ & $-$ & 0.82 (10)& 0.88 (11)& 0.81 (10)& 0.90 (11)& 0.58 (9)& 0.99 (7)& 0.95 (9)\\ 
$\lambda$10697& $-$ & $-$ & $-$ & $-$ & $-$ & 0.82 (13)& 0.75 (13)& 0.77 (13)& 0.66 (11)& 0.77 (10)& 0.85 (12)\\ 
$\lambda$10780& $-$ & $-$ & $-$ & $-$ & $-$ & $-$ & 0.96 (13)& 0.97 (17)& 0.50 (12)& 0.65 (10)& 0.92 (13)\\ 
$\lambda$10792& $-$ & $-$ & $-$ & $-$ & $-$ & $-$ & $-$ & 0.94 (13)& 0.63 (11)& 0.63 (10)& 0.87 (12)\\ 
$\lambda$11797& $-$ & $-$ & $-$ & $-$ & $-$ & $-$ & $-$ & $-$ & 0.63 (11)& 0.63 (10)& 0.90 (13)\\ 
$\lambda$12337& $-$ & $-$ & $-$ & $-$ & $-$ & $-$ & $-$ & $-$ & $-$ & 0.54 (9)& 0.60 (10)\\ 
$\lambda$12623& $-$ & $-$ & $-$ & $-$ & $-$ & $-$ & $-$ & $-$ & $-$ & $-$ & 0.84 (10)\\ 
$\lambda$13175& $-$ & $-$ & $-$ & $-$ & $-$ & $-$ & $-$ & $-$ & $-$ & $-$ & $-$ 
\enddata
\tablecomments{The number in parentheses denotes the number of stars used for the calculation of the correlation coefficients. The data of Cyg OB2 No.12 is not included in the calculations of the correlation coefficients.}
\label{nirdibcc}
\end{deluxetable*}

\begin{deluxetable*}{ccccccccc}
\tabletypesize{\scriptsize}
\tablecaption{Correlations Coefficients among Optical DIBs and those between Optical and NIR DIBs}
\tablewidth{0pt}
\tablehead{
 \colhead{} & \colhead{$\lambda$5705.1 } & \colhead{$\lambda$5780.5 } & \colhead{$\lambda$6204.5 } & \colhead{$\lambda$6283.8 } & \colhead{$\lambda$5487.7 } & \colhead{$\lambda$5797.1 } & \colhead{$\lambda$6196.0 } & \colhead{$\lambda$6613.6}
}
\startdata
\\ $\lambda$5705.1& $-$ & 0.95 (14)& 0.94 (14)& 0.93 (14)& 0.94 (13)& 0.89 (14)& 0.80 (14)& 0.87 (14)\\ 
$\lambda$5780.5& $-$ & $-$ & 0.99 (18)& 0.97 (17)& 0.97 (13)& 0.94 (15)& 0.95 (19)& 0.96 (18)\\ 
$\lambda$6204.5& $-$ & $-$ & $-$ & 0.99 (16)& 0.95 (13)& 0.93 (15)& 0.93 (18)& 0.95 (18)\\ 
$\lambda$6283.8& $-$ & $-$ & $-$ & $-$ & 0.88 (13)& 0.89 (15)& 0.89 (17)& 0.91 (16)\\ 
$\lambda$5487.7& $-$ & $-$ & $-$ & $-$ & $-$ & 0.97 (13)& 0.96 (13)& 0.94 (13)\\ 
$\lambda$5797.1& $-$ & $-$ & $-$ & $-$ & $-$ & $-$ & 0.94 (15)& 0.94 (15)\\ 
$\lambda$6196.0& $-$ & $-$ & $-$ & $-$ & $-$ & $-$ & $-$ & 0.98 (18)\\ 
$\lambda$6613.6& $-$ & $-$ & $-$ & $-$ & $-$ & $-$ & $-$ & $-$ \\ 
\hline \\
$\lambda$10438& 0.39 (8)& 0.61 (9)& 0.64 (9)& 0.60 (9)& 0.54 (8)& 0.58 (9)& 0.40 (9)& 0.61 (9) \\
$\lambda$10780& 0.75 (13)& 0.81 (15)& 0.86 (14)& 0.91 (15)& 0.61 (12)& 0.71 (13)& 0.74 (15)& 0.68 (14) \\
$\lambda$11797& 0.86 (12)& 0.83 (11)& 0.90 (11)& 0.95 (11)& 0.73 (11)& 0.77 (11)& 0.74 (11)& 0.70 (11) \\
$\lambda$13175& 0.68 (9)& 0.78 (9)& 0.87 (9)& 0.92 (9)& 0.49 (8)& 0.63 (9)& 0.48 (9)& 0.60 (9) 
\enddata
\tablecomments{The number in parentheses denotes the number of stars used for the calculation of the correlation coefficients. The data of Cyg OB2 No.12 is not included in the calculations of the correlation coefficients.}
\label{optcc}
\end{deluxetable*}

\begin{figure*}[!ht]
 \centering
 \includegraphics[width=5cm,clip]{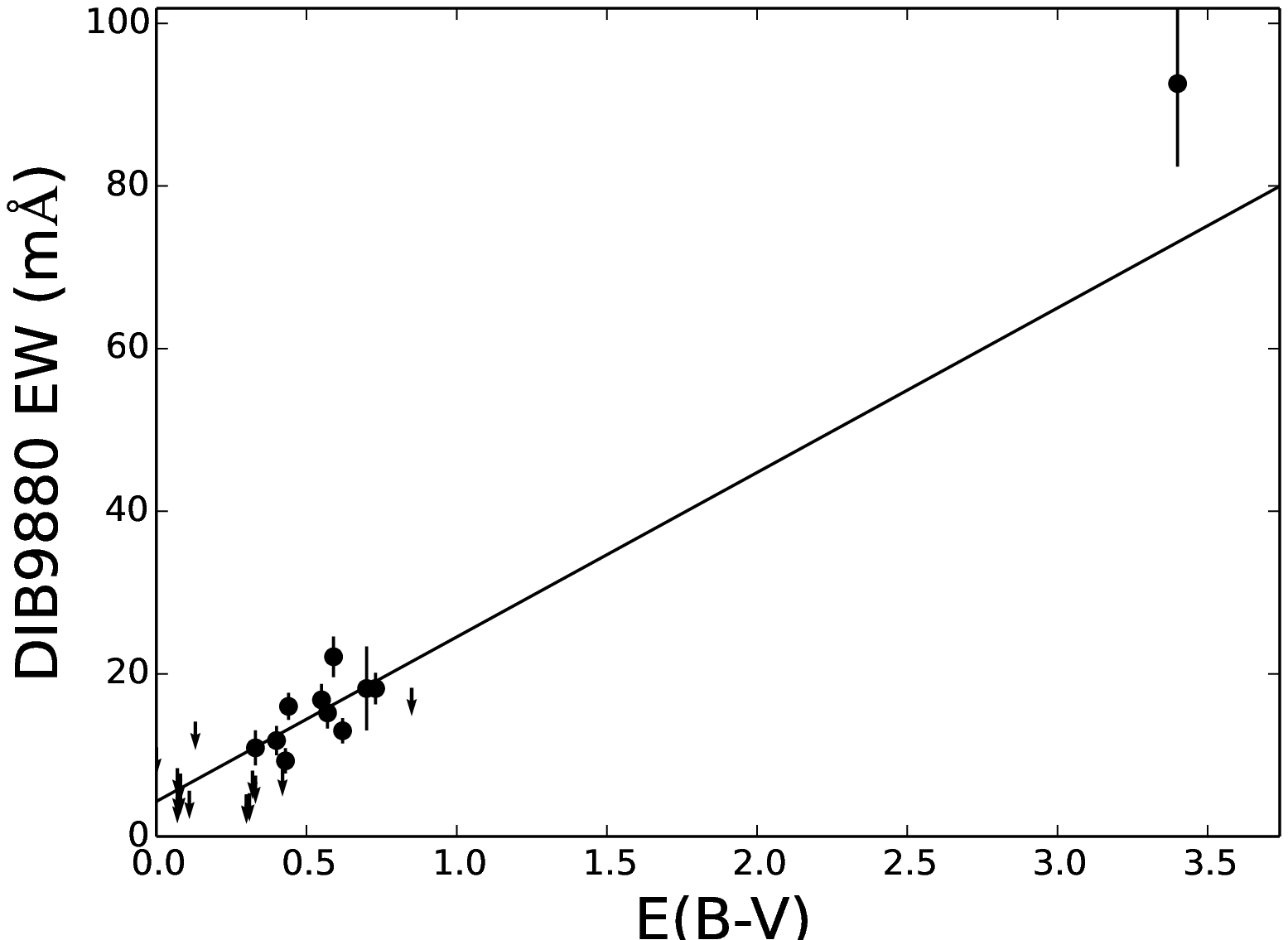}
 \includegraphics[width=5cm,clip]{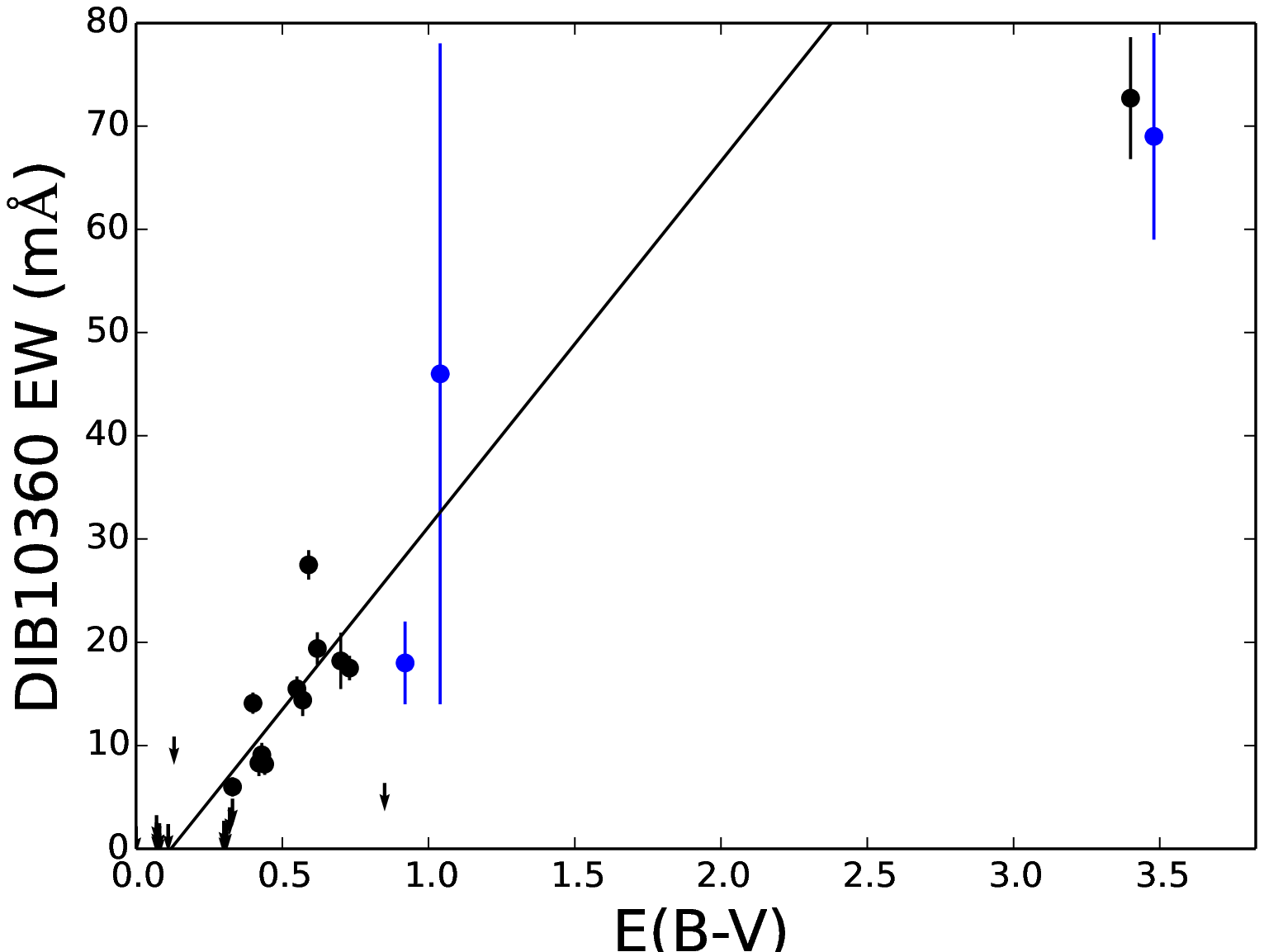}
 \includegraphics[width=5cm,clip]{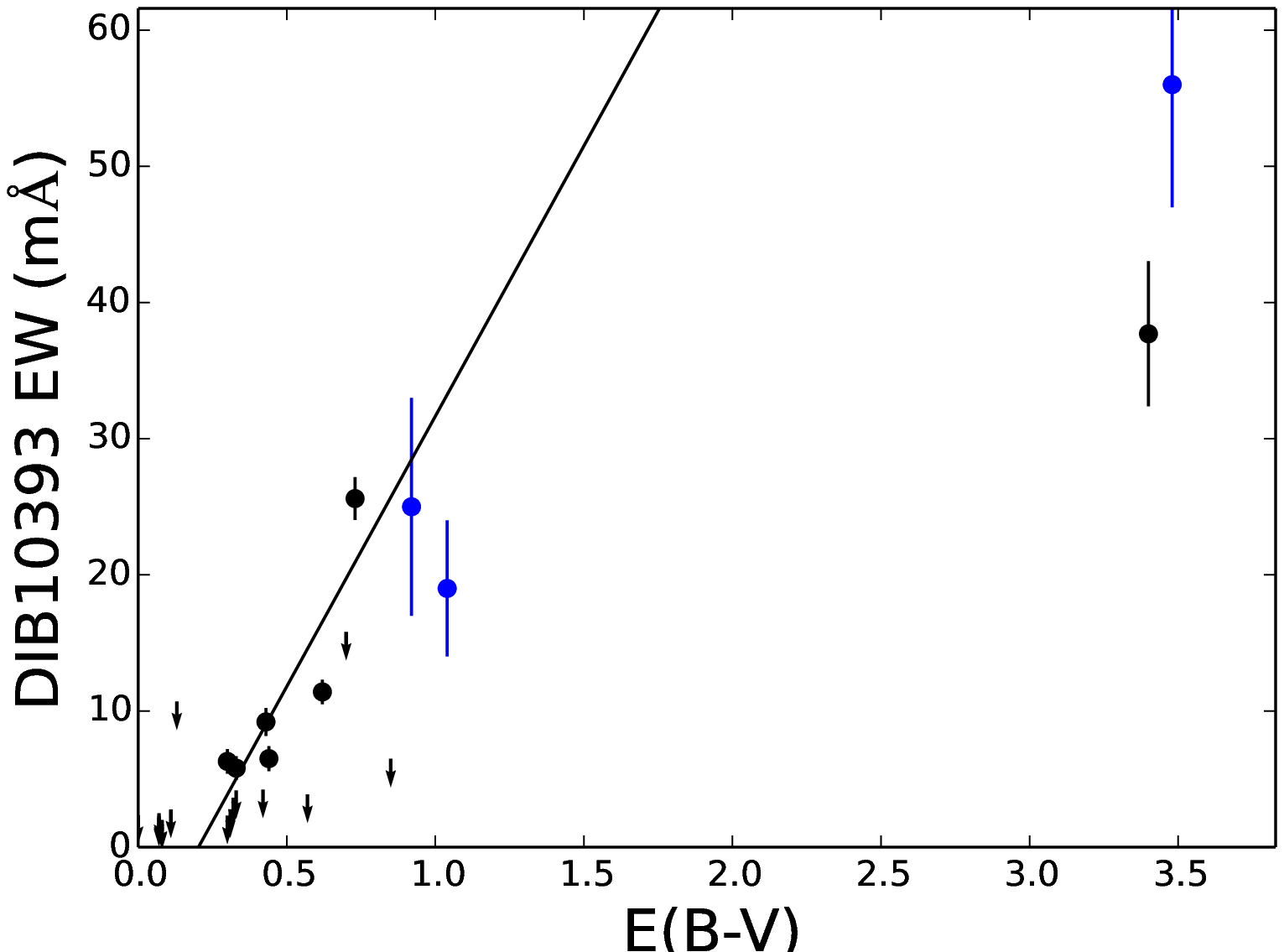}
 \includegraphics[width=5cm,clip]{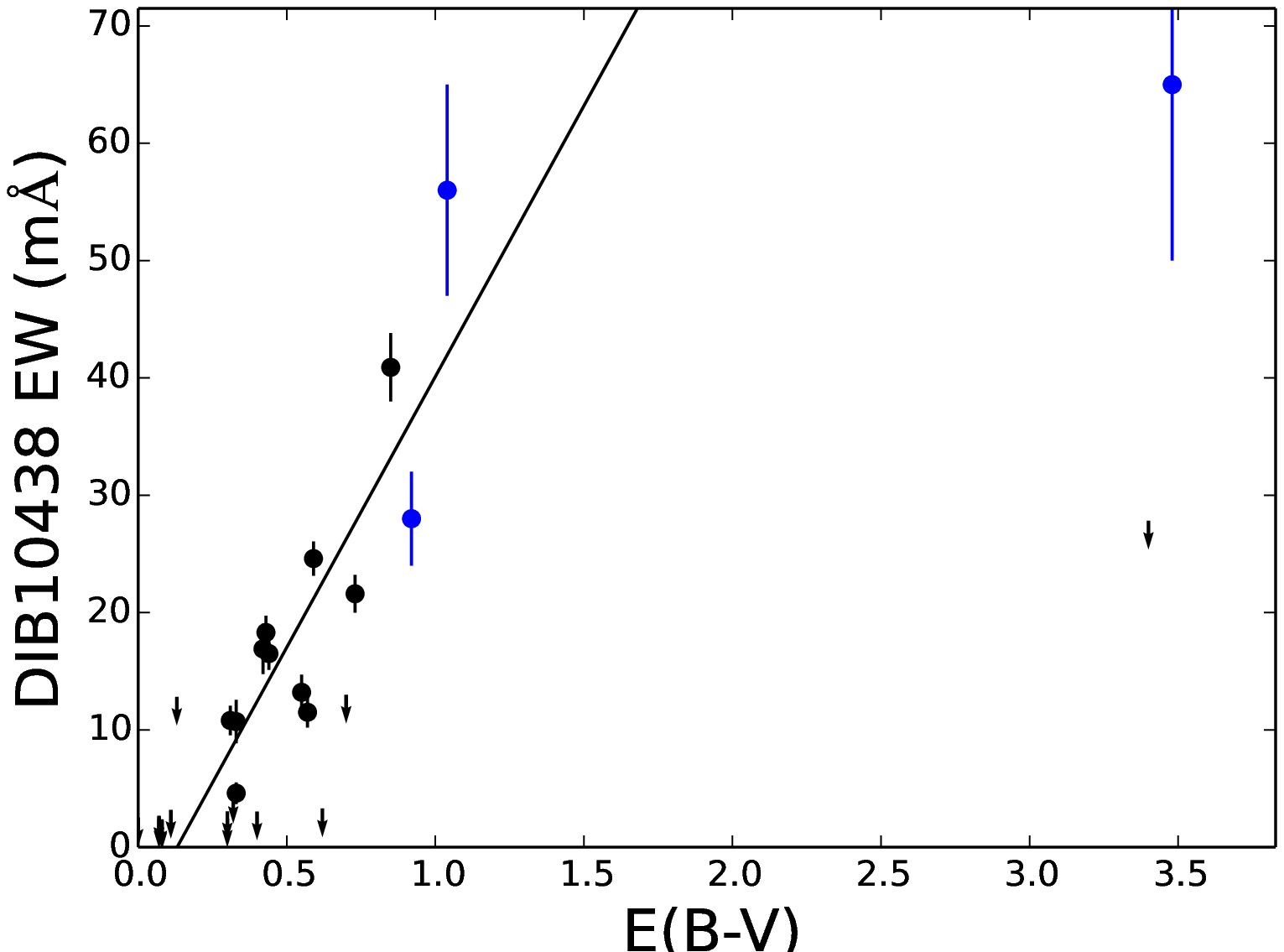}
 \includegraphics[width=5cm,clip]{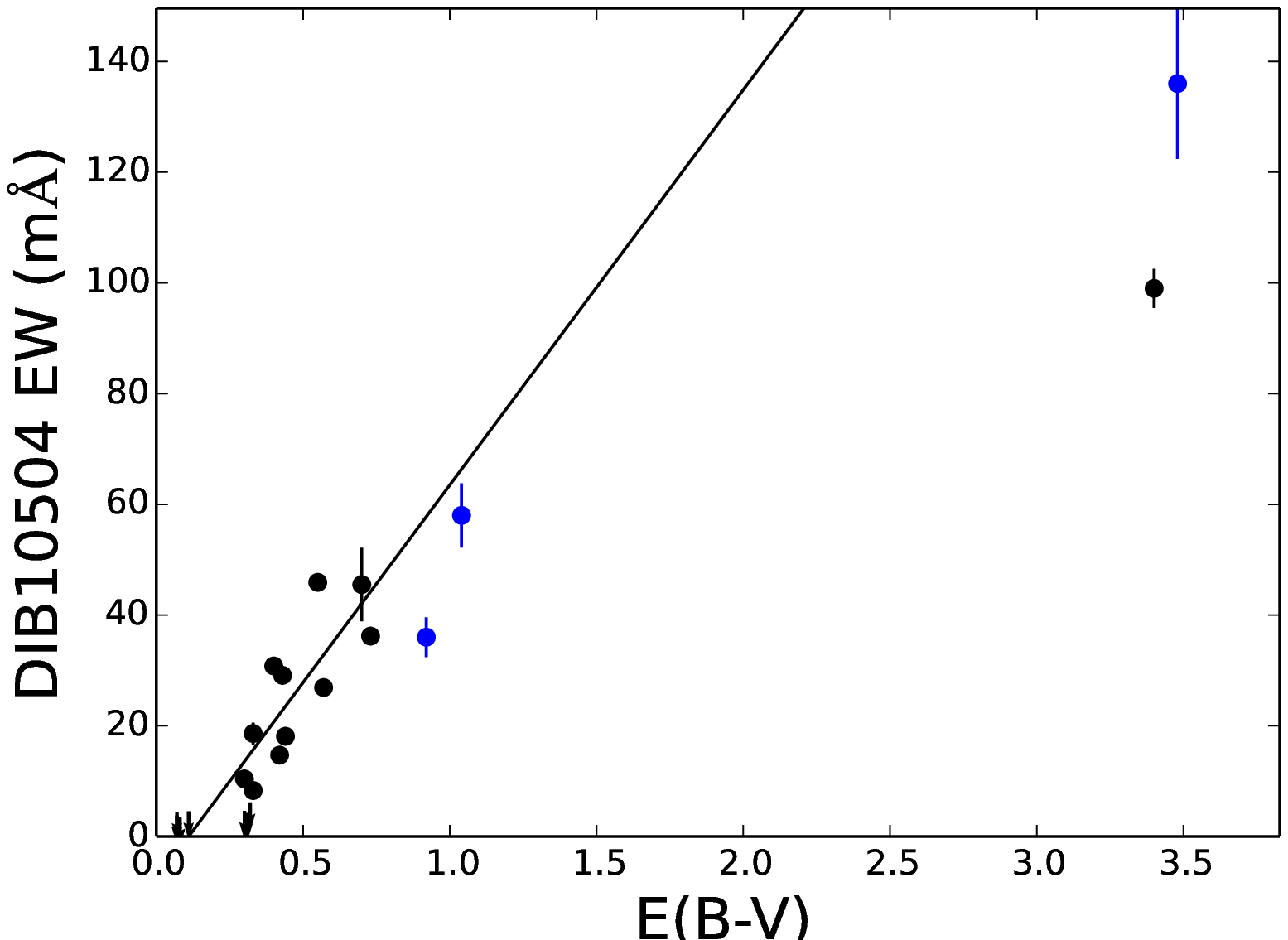}
 \includegraphics[width=5cm,clip]{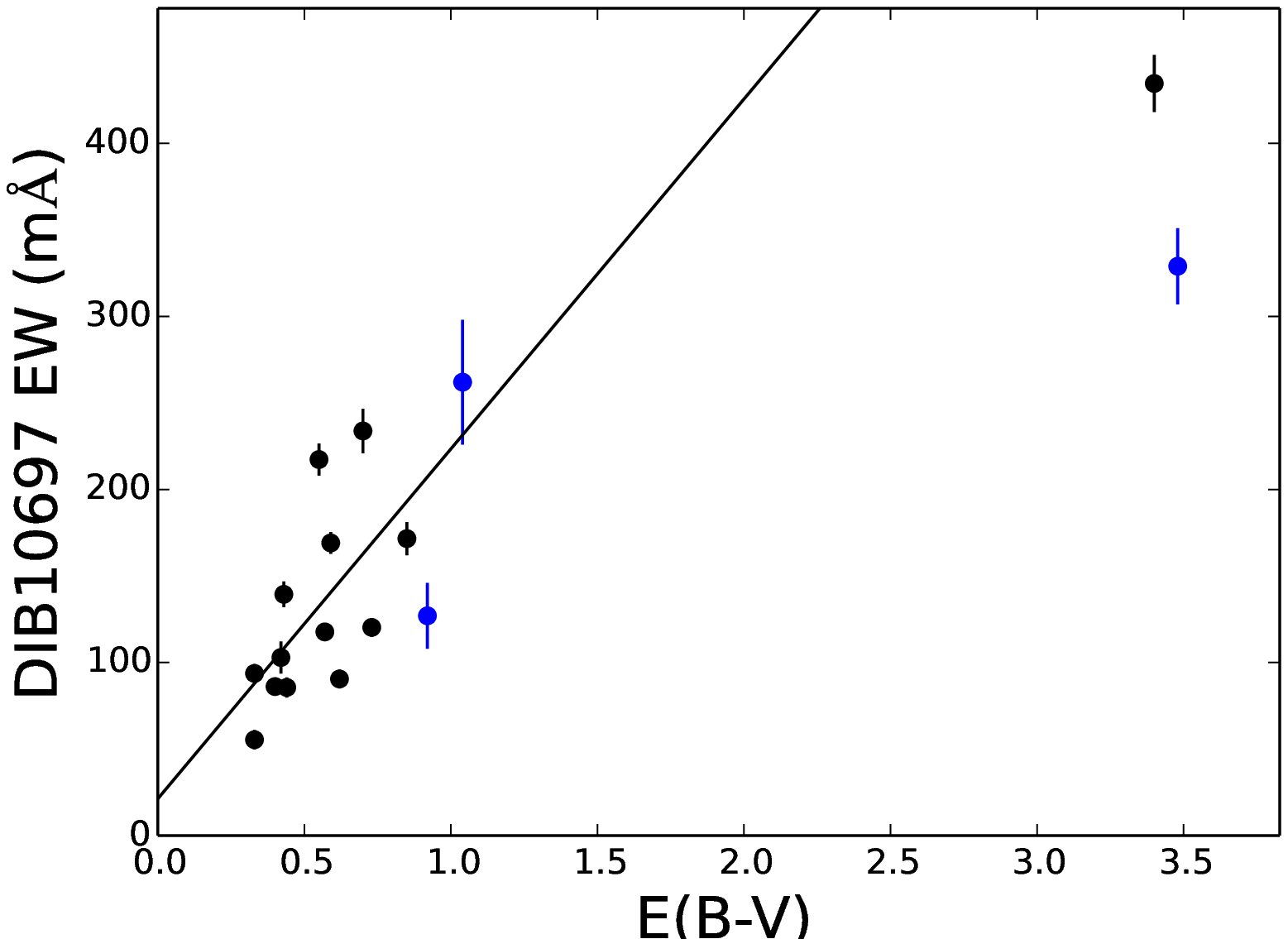}
 \includegraphics[width=5cm,clip]{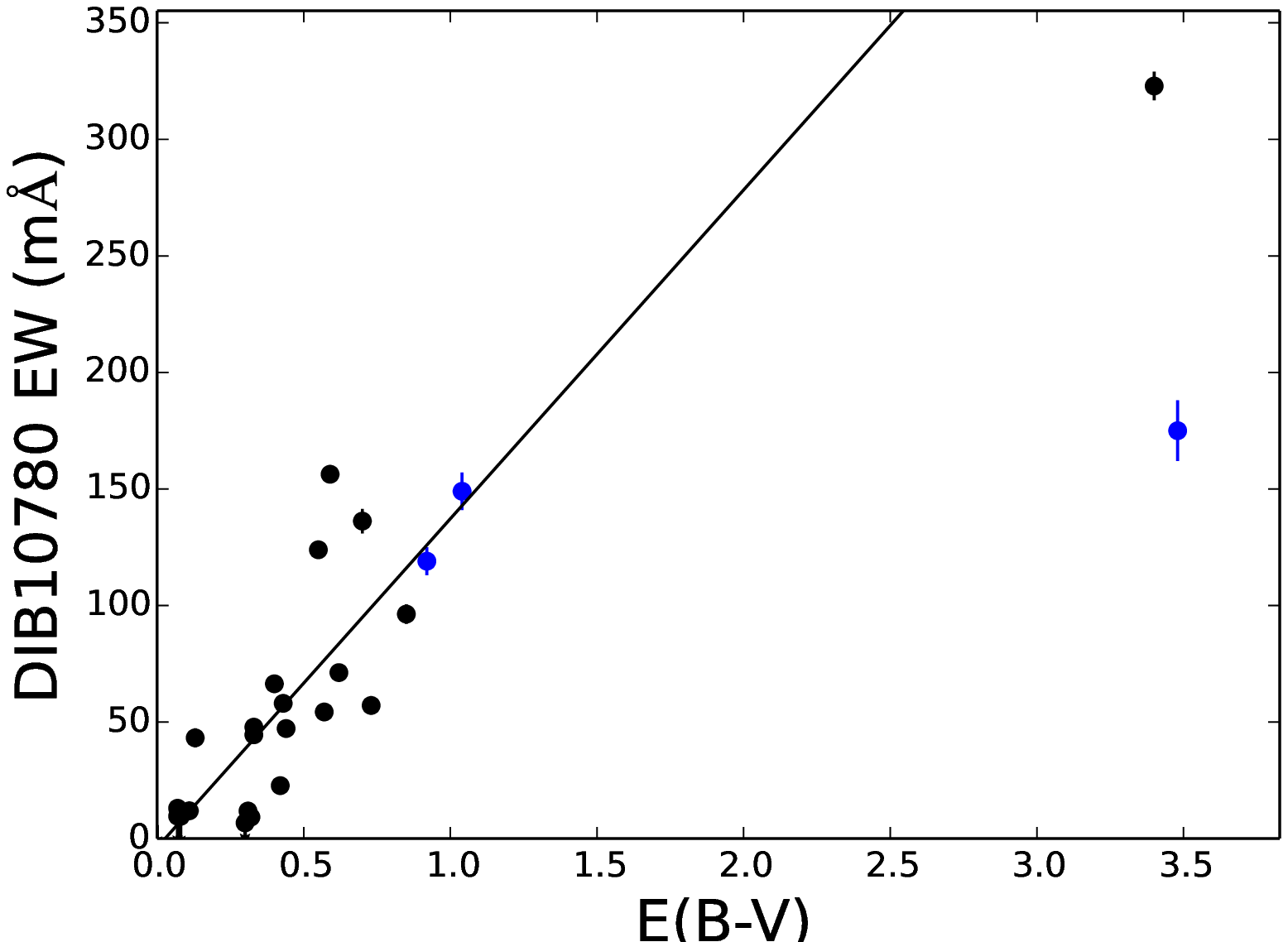}
 \includegraphics[width=5cm,clip]{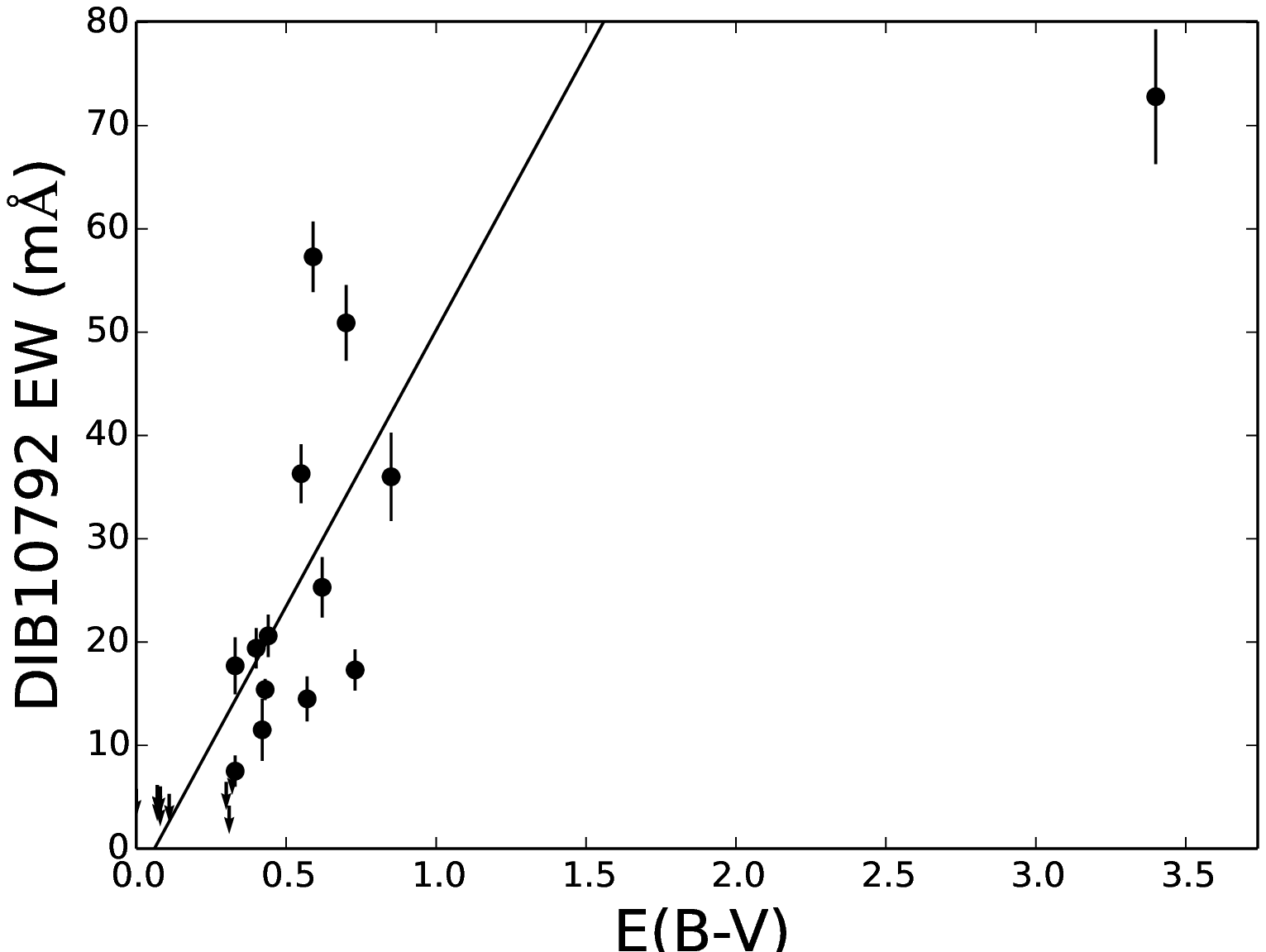}
 \includegraphics[width=5cm,clip]{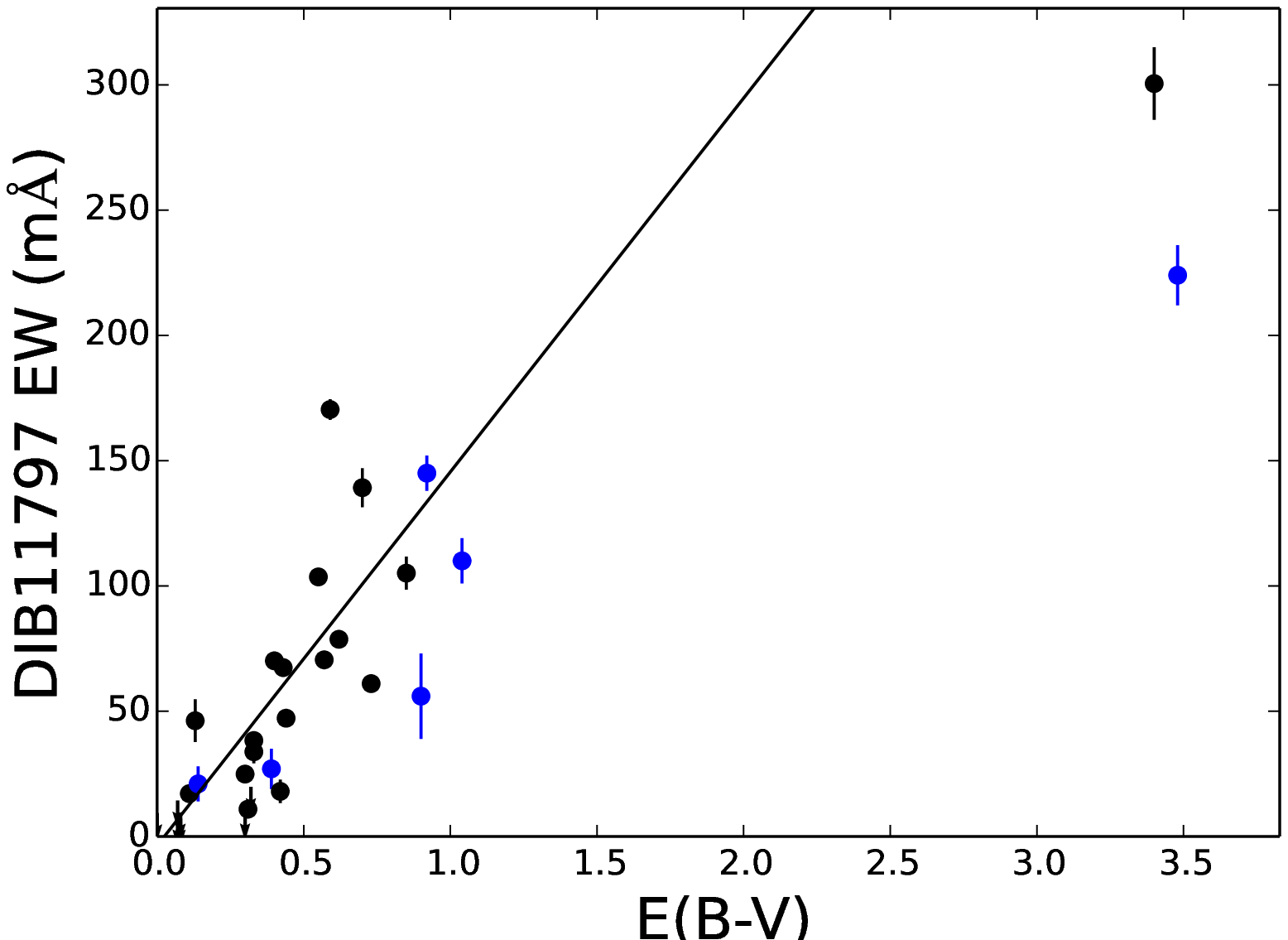}
  \caption{Correlations of NIR DIBs with $E(B-V)$. The black points show the EWs obtained by us, while the blue points show the EWs obtained by \citet{cox14}. The lines in each panel show the linear functions fitted to each plot. The blue points and the points of Cyg OB2 No.12 at $E(B-V)=3.4$ are not included in the fitting.}
 \label{ebvcc1}
\end{figure*}

\setcounter{figure}{7}

\begin{figure*}[!ht]
 \centering
 \includegraphics[width=5cm,clip]{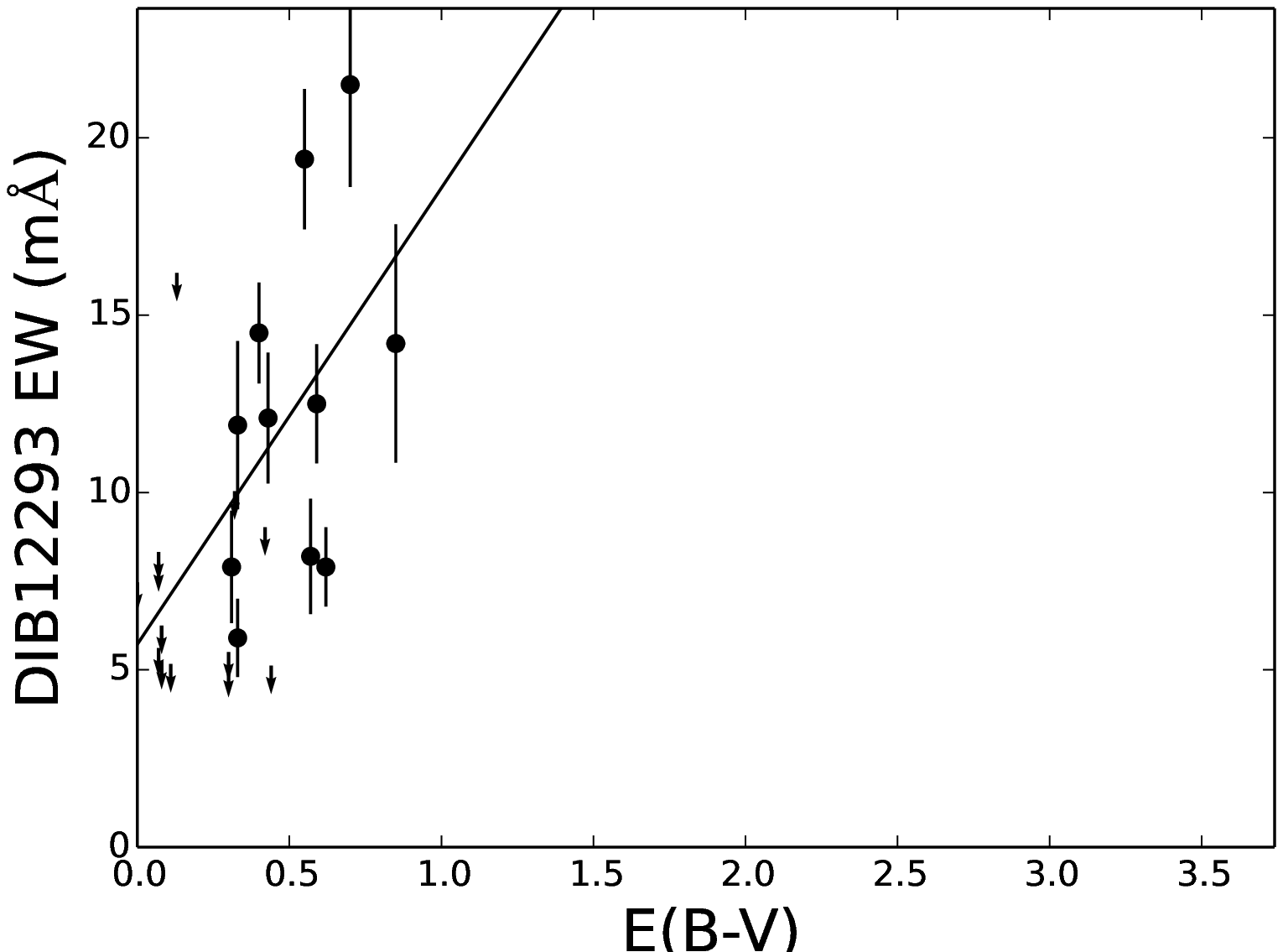}
 \includegraphics[width=5cm,clip]{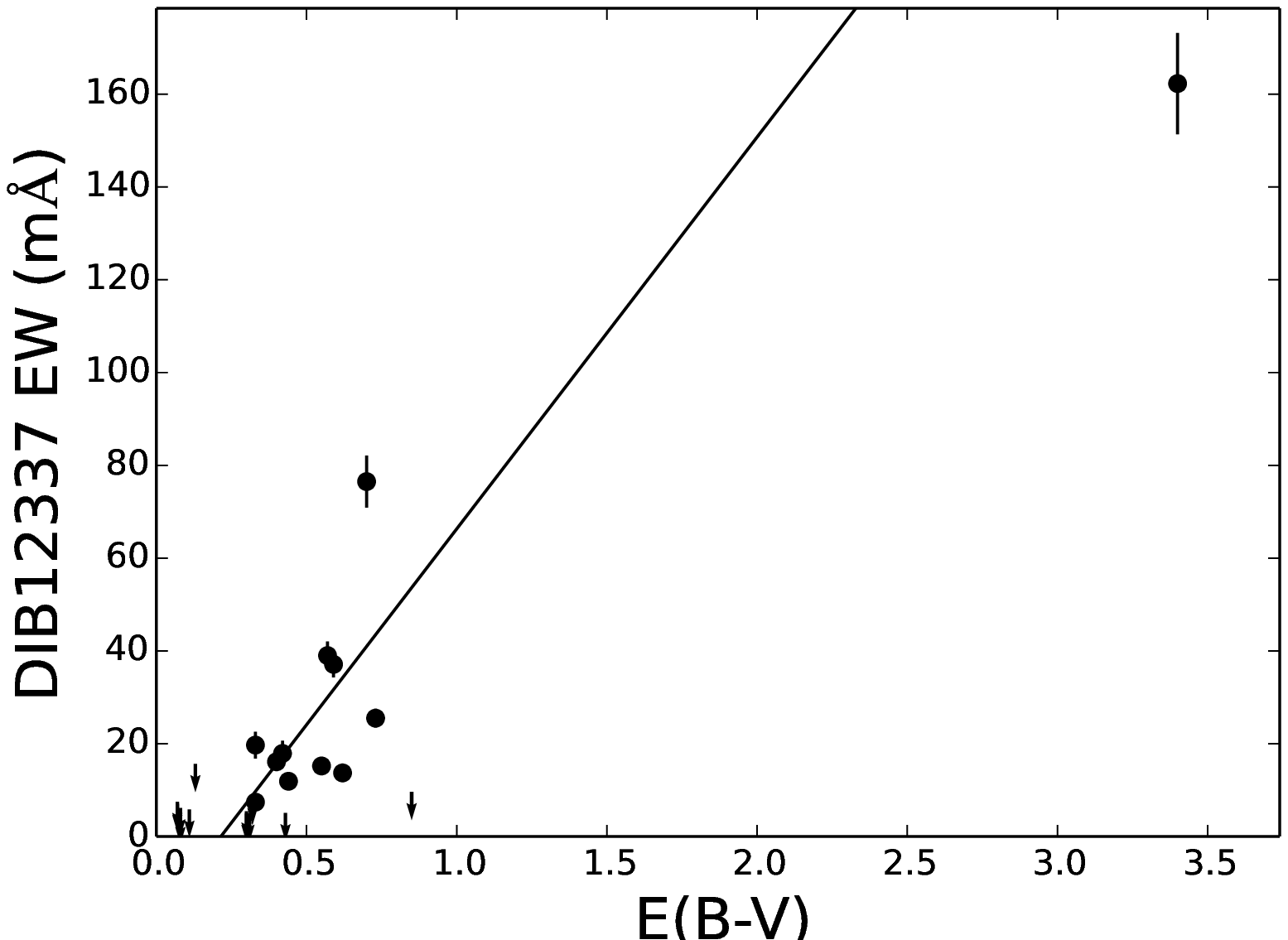}
 \includegraphics[width=5cm,clip]{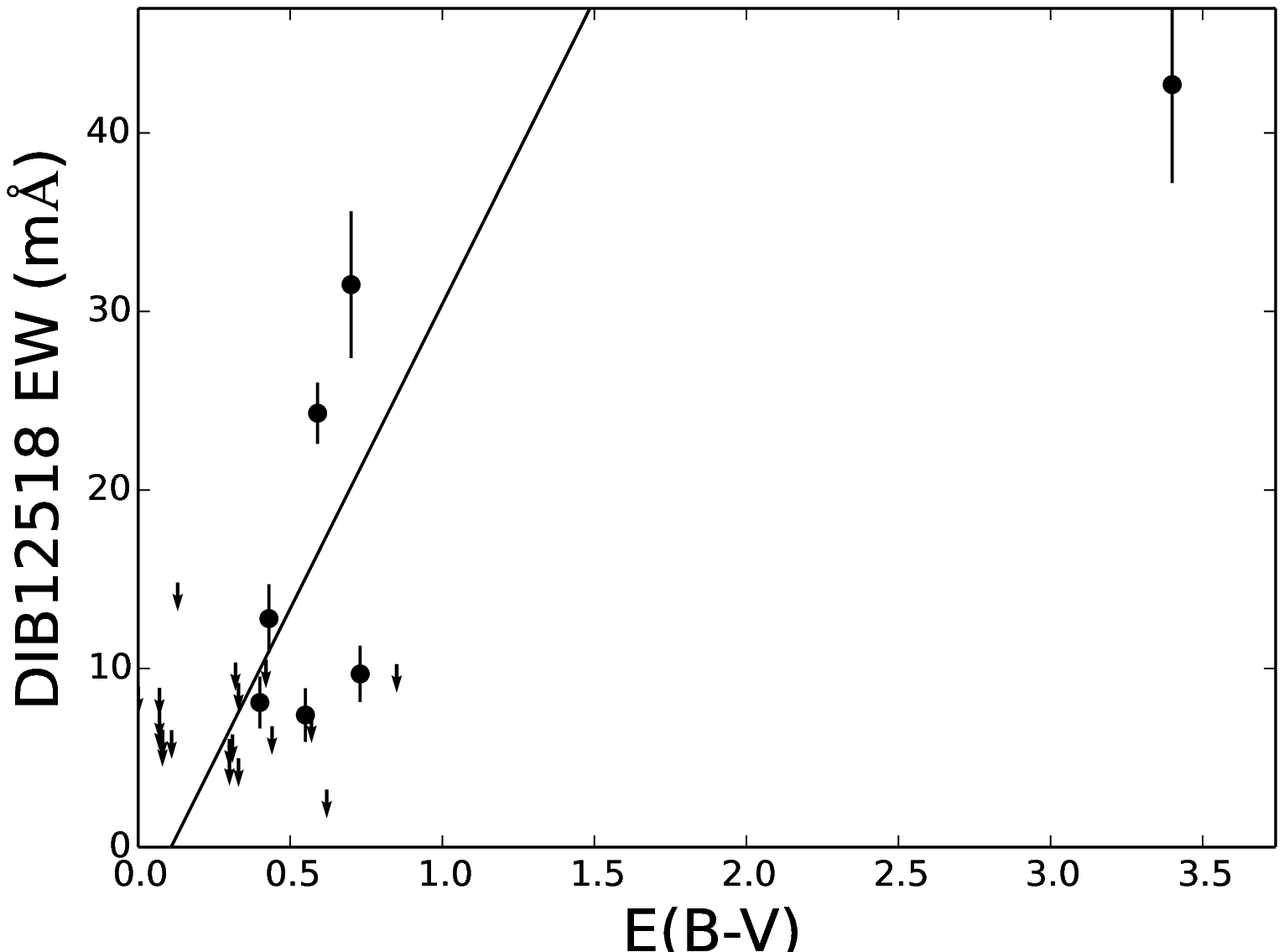}
 \includegraphics[width=5cm,clip]{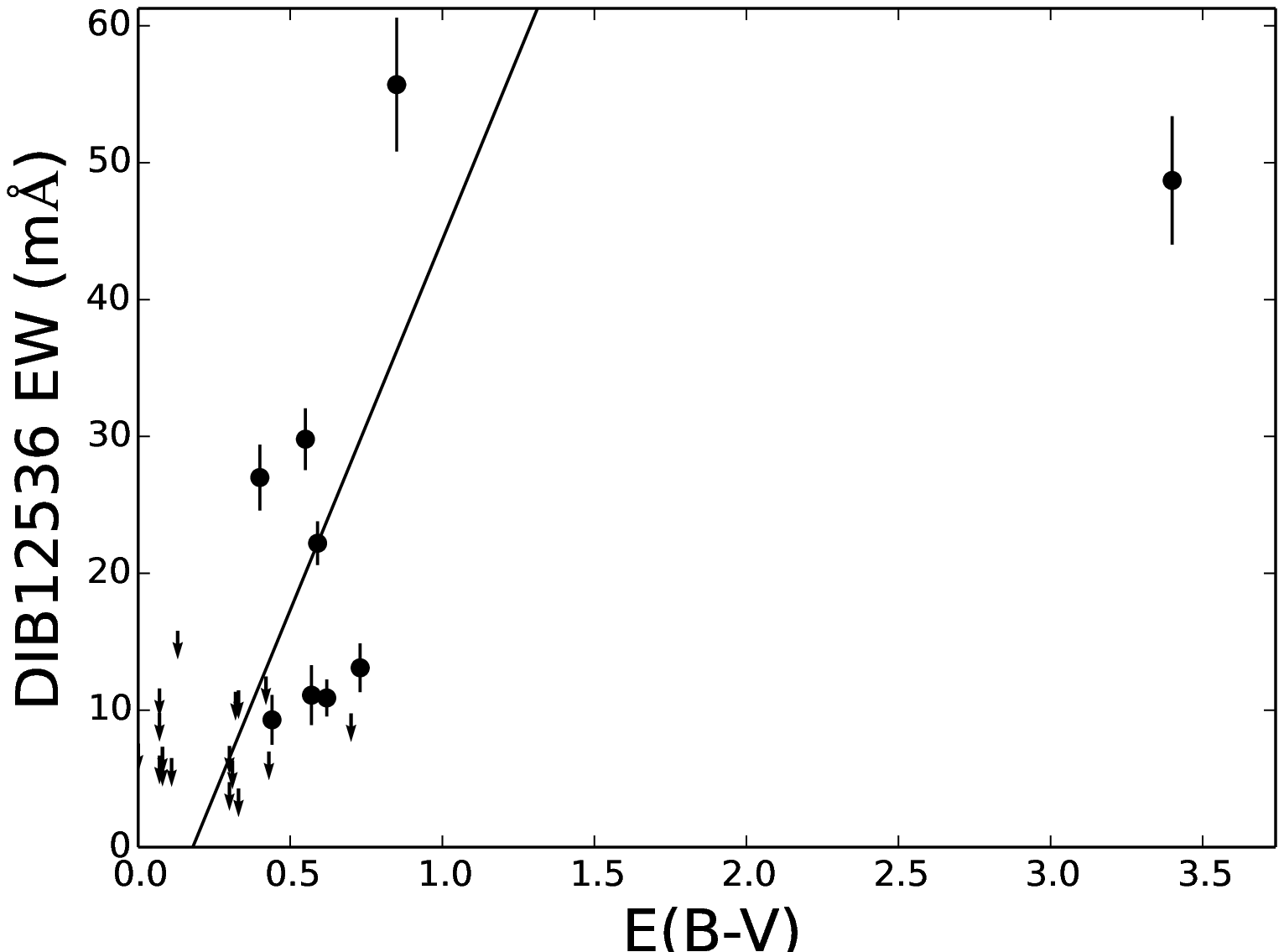}
 \includegraphics[width=5cm,clip]{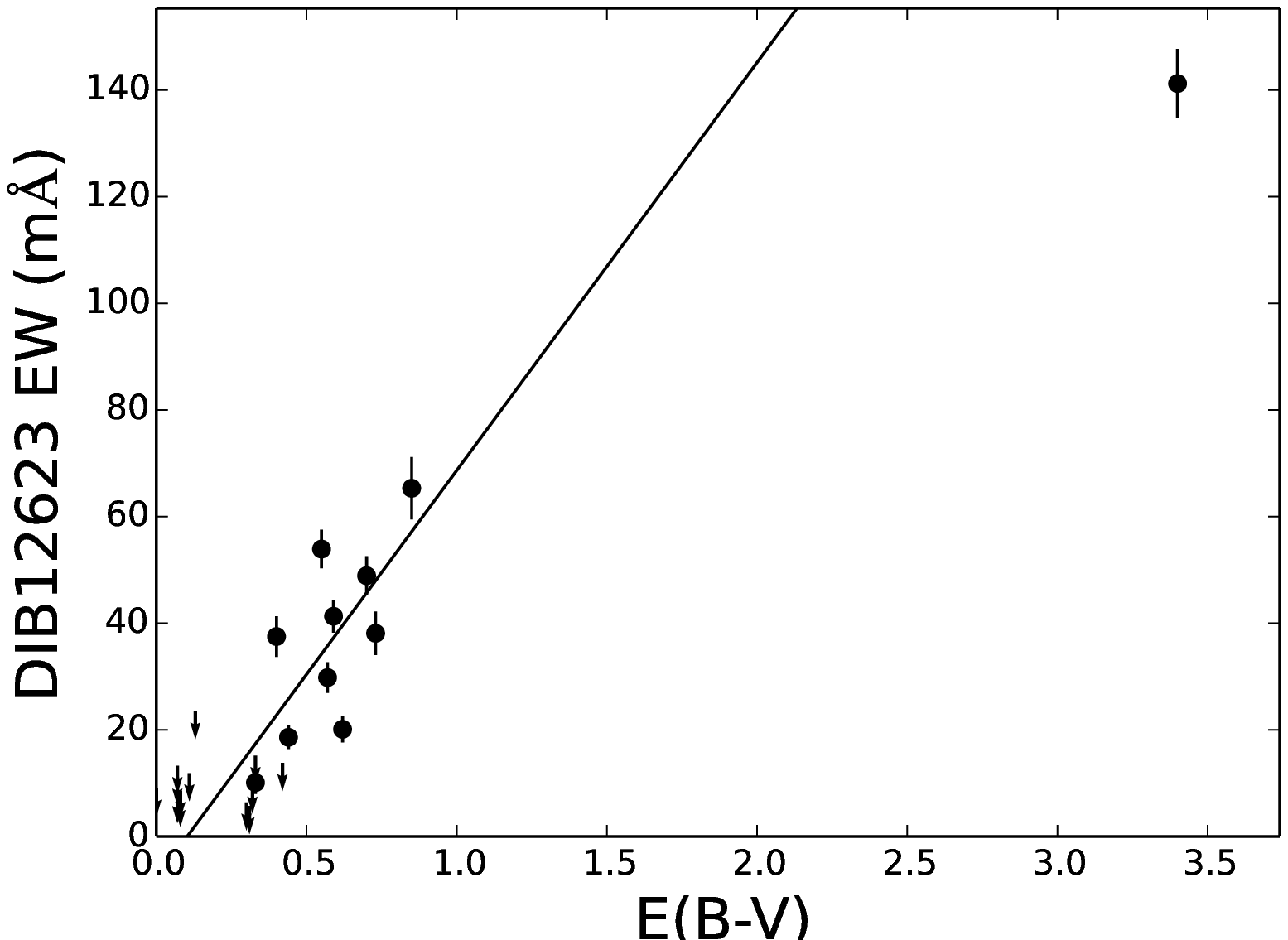}
 \includegraphics[width=5cm,clip]{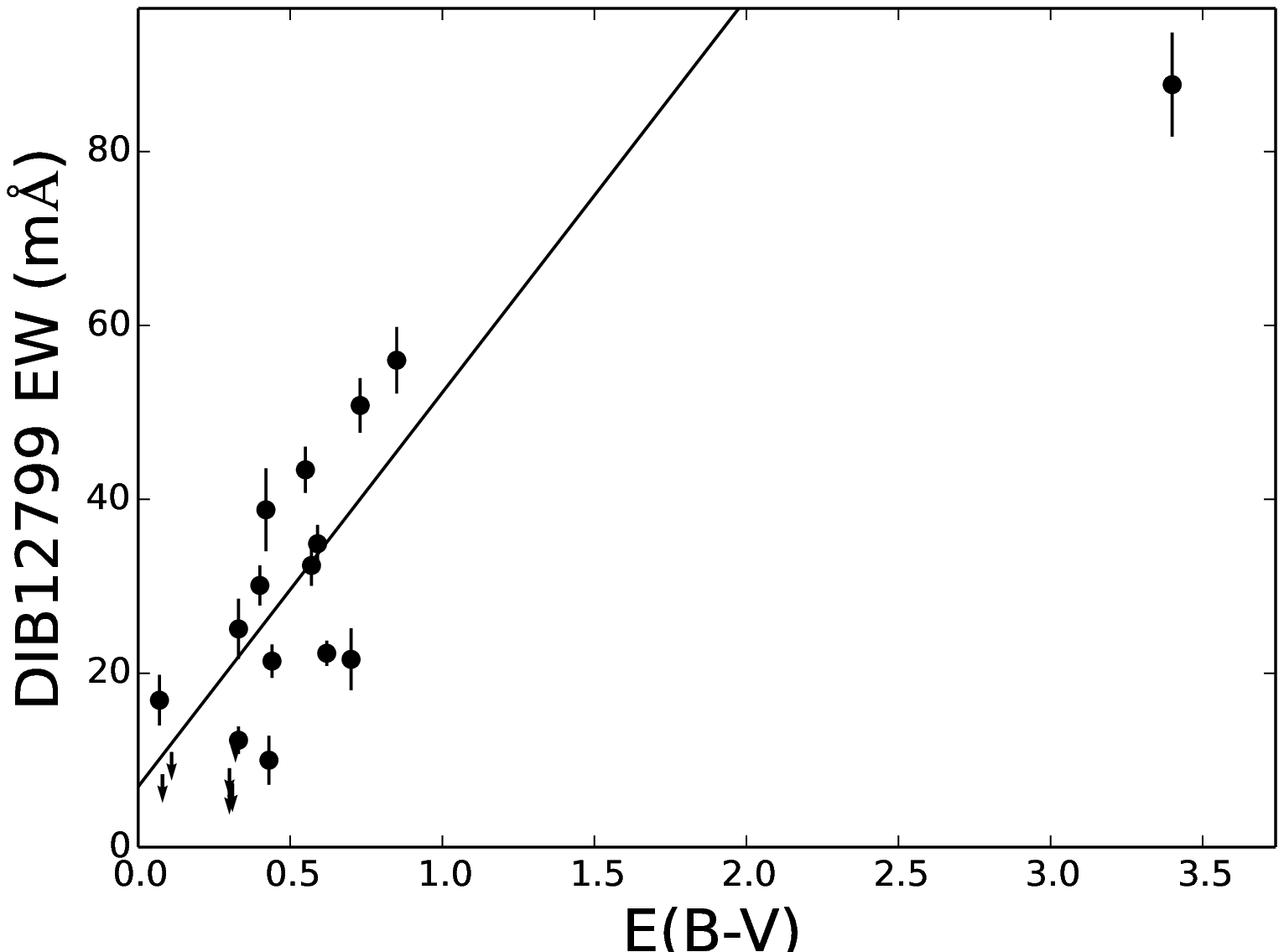}
 \includegraphics[width=5cm,clip]{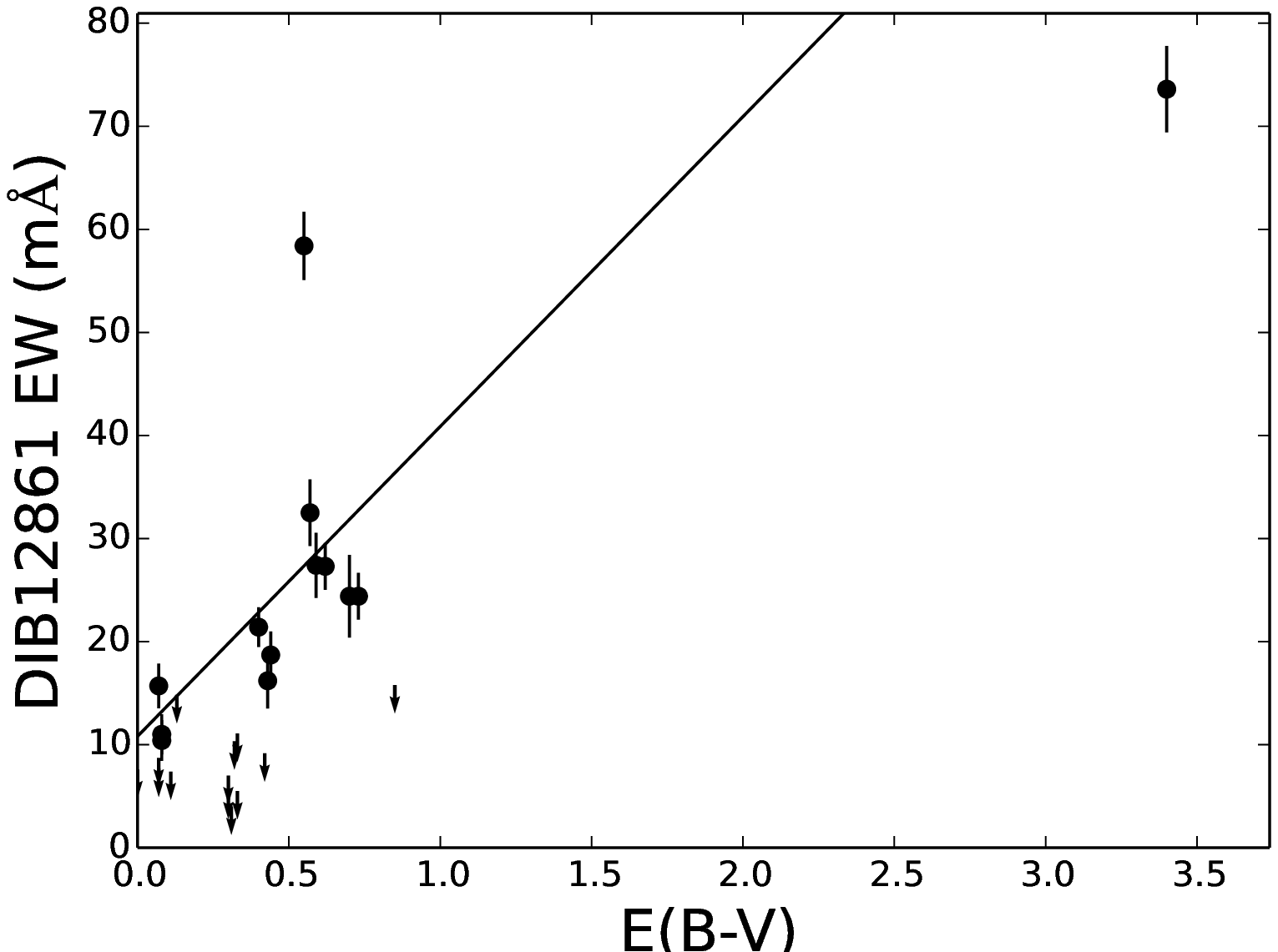}
 \includegraphics[width=5cm,clip]{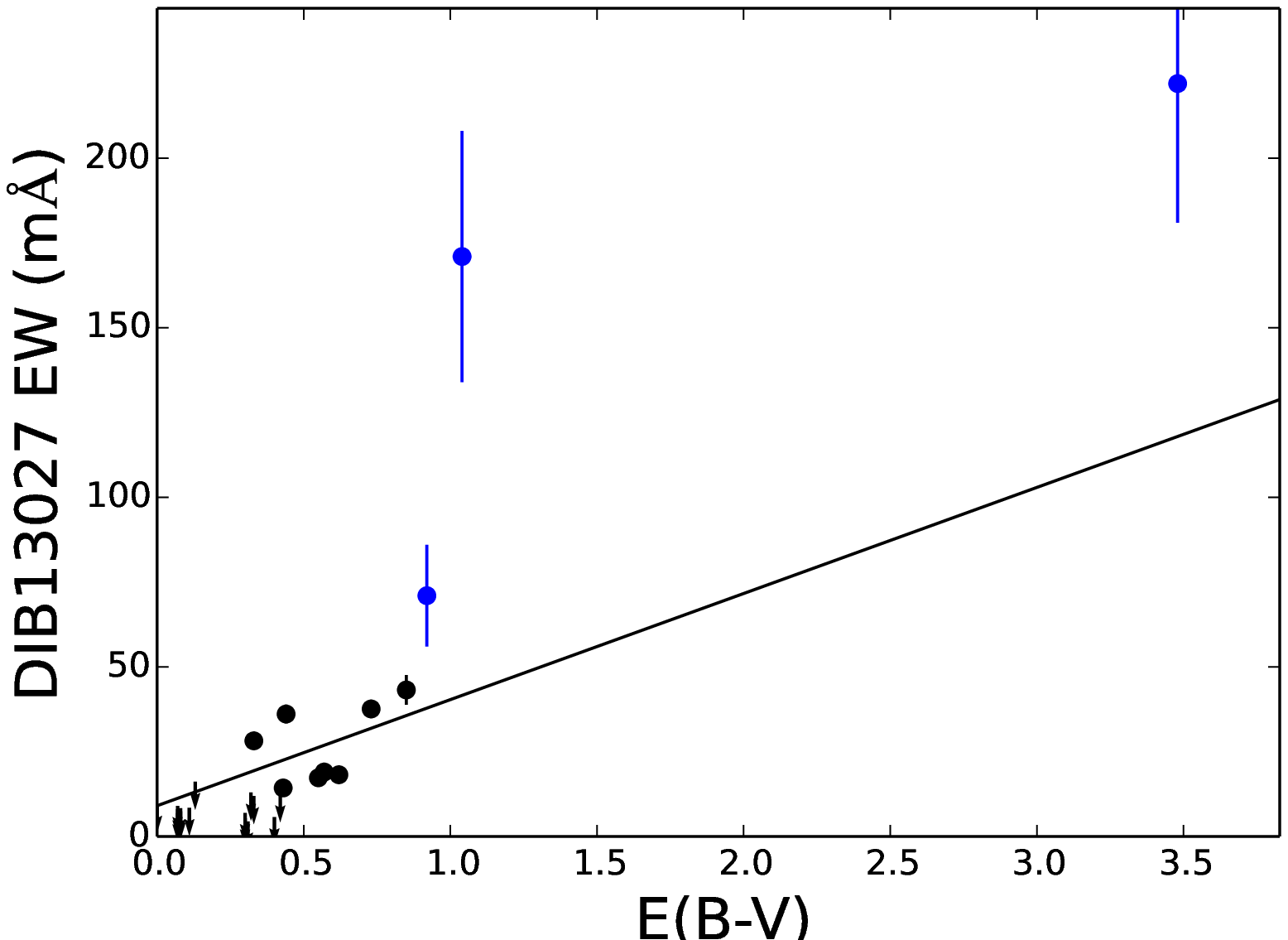}
 \includegraphics[width=5cm,clip]{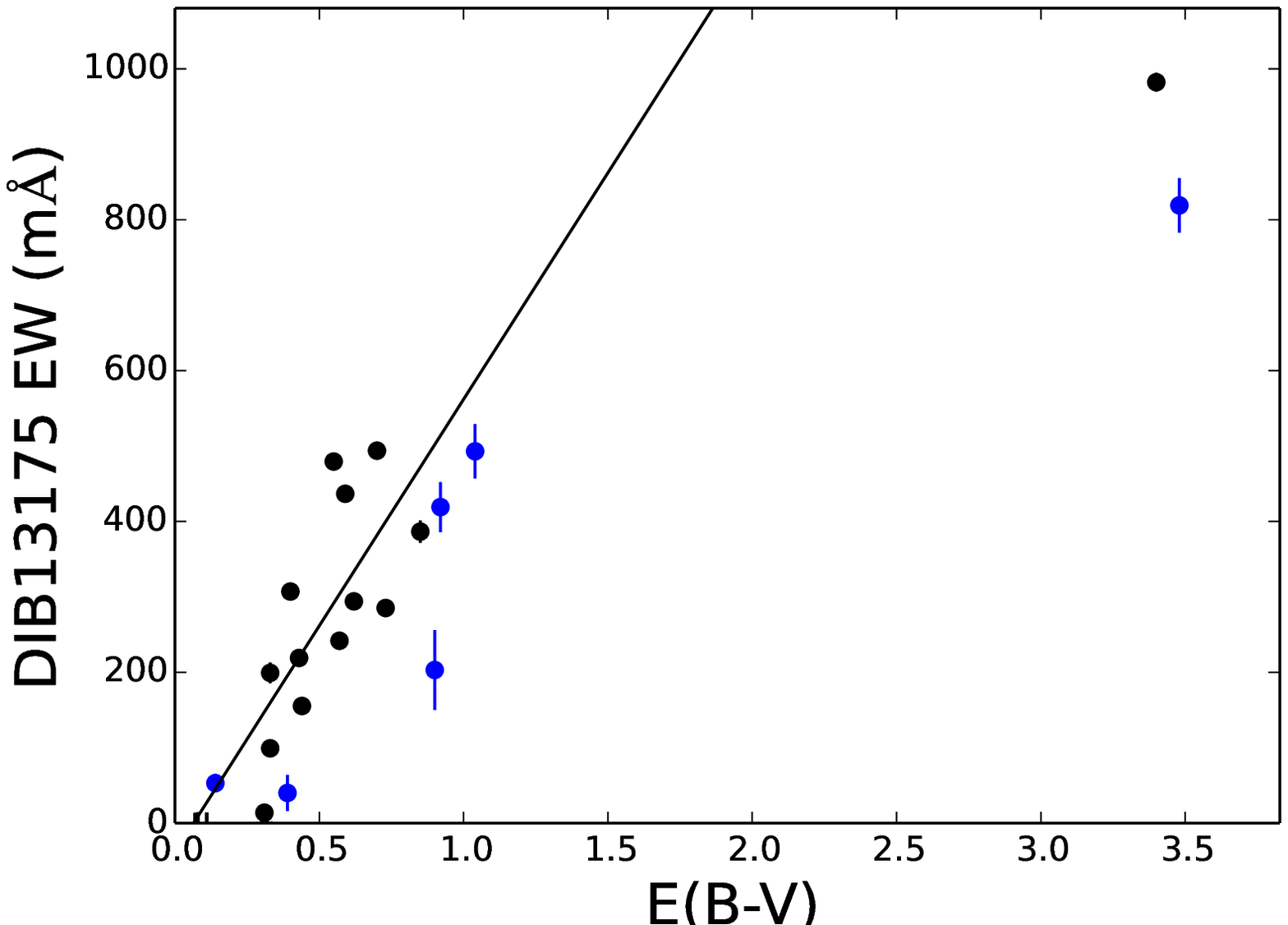}
  \caption{\textit{Continued.}}
 \label{ebvcc2}
\end{figure*}

\begin{figure}[!ht]
 \centering
 \includegraphics[width=7cm,clip]{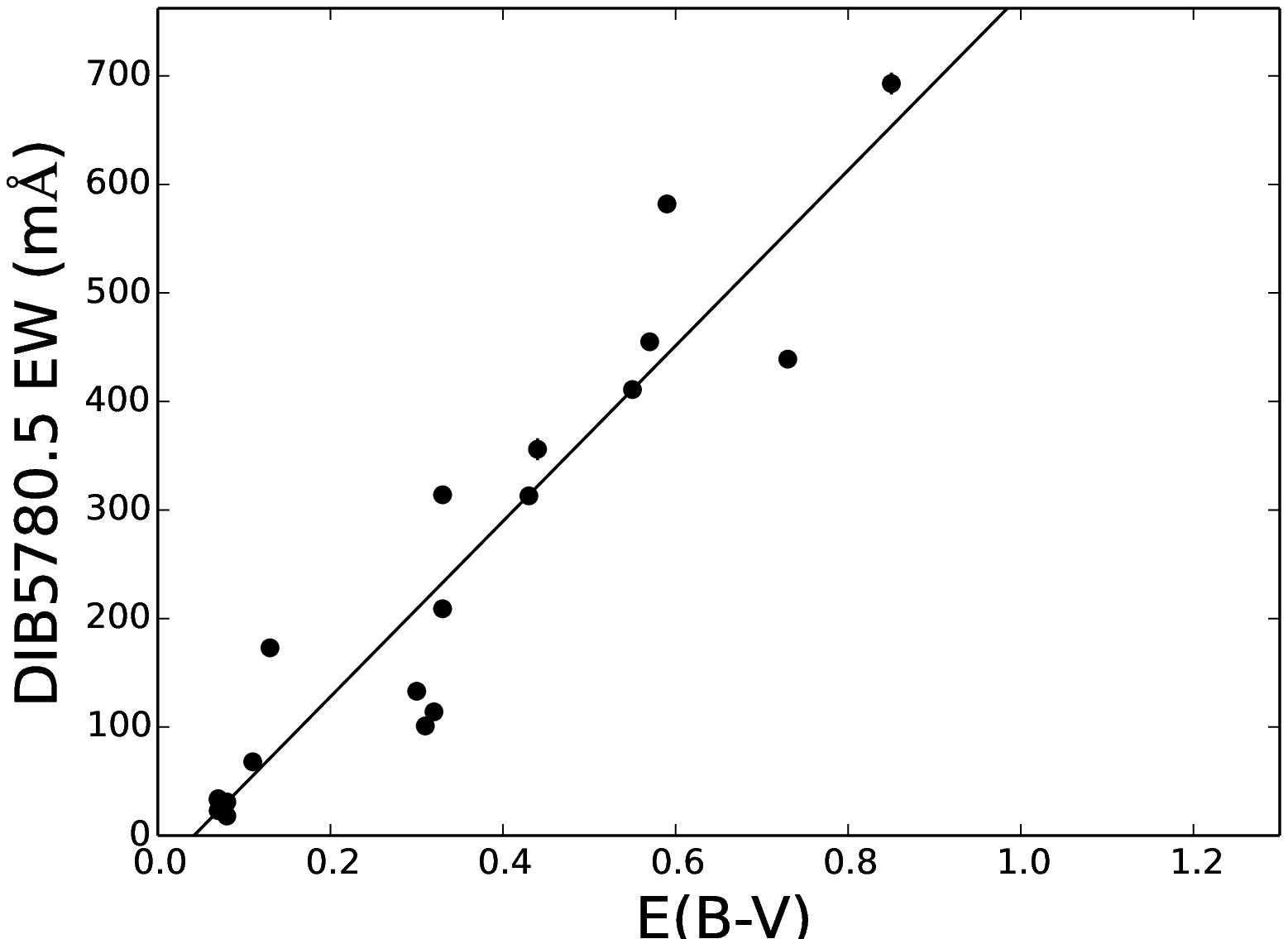}
 \includegraphics[width=7cm,clip]{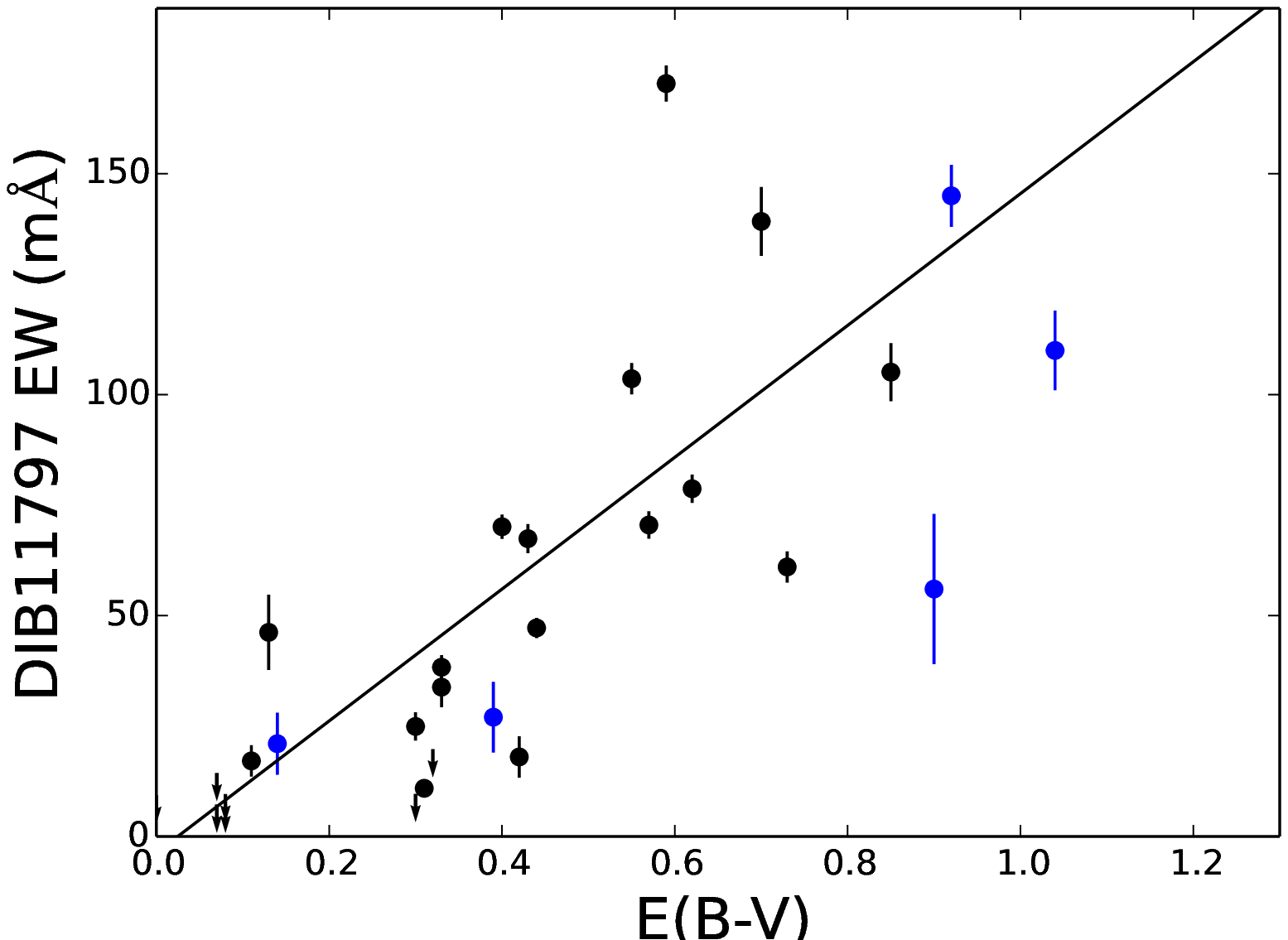}
\caption{Comparison of correlations of optical DIB $\lambda 5780.5$ (left panel) and NIR DIB $\lambda 11797$ (right panel) with $E(B-V)$ for the range of $E(B-V) < 1.1$. For the right panel, the black points show the EWs obtained by us, while the blue points show the EWs obtained by \citet{cox14}. For the left panel, the black points are from \citet{fri11}. The lines in each panel show the linear functions fitted to each plot. The blue points are not included in the fitting.}
 \label{ebvcorr_rep}
\end{figure}

\begin{figure*}[!ht]
 \centering
 \includegraphics[width=18cm,clip]{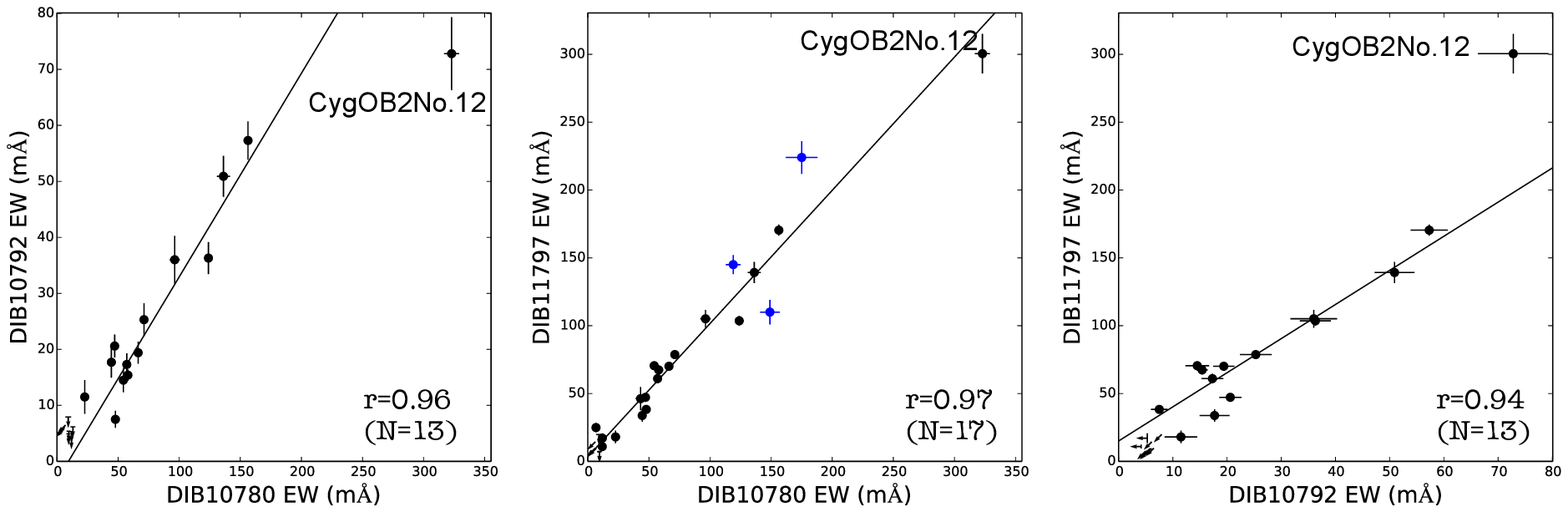}
 \caption{Correlations among three NIR DIBs, $\lambda \lambda 10780, 10792, \text{and }11797$. The black points show our data, while the blue points are from \citet{cox14}. The lines in each panel show the linear functions fitted to each plot. The correlation coefficients $(r)$ and the number of stars used in the calculation $(N)$ are shown in each panel. The blue points and the points of Cyg OB2 No.12 are not included in the fitting.}
 \label{NIRcorrelation}
\end{figure*}

\begin{figure*}[!ht]
 \centering
 \includegraphics[width=18cm,clip]{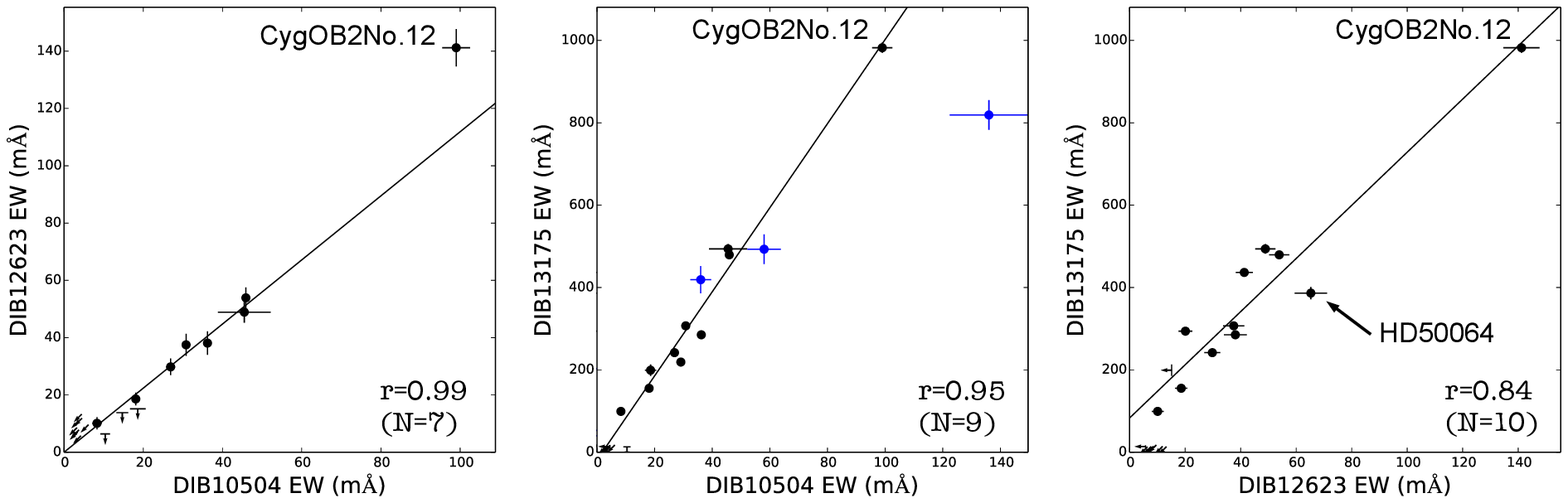}
 \caption{Correlations among three NIR DIBs, $\lambda \lambda 10504, 12623, \text{and }13175$. The notations are the same as those in Figure \ref{NIRcorrelation}.}
 \label{NIRcorrelation2}
\end{figure*}

\begin{figure*}[!ht]
 \centering
 \includegraphics[width=7cm,clip]{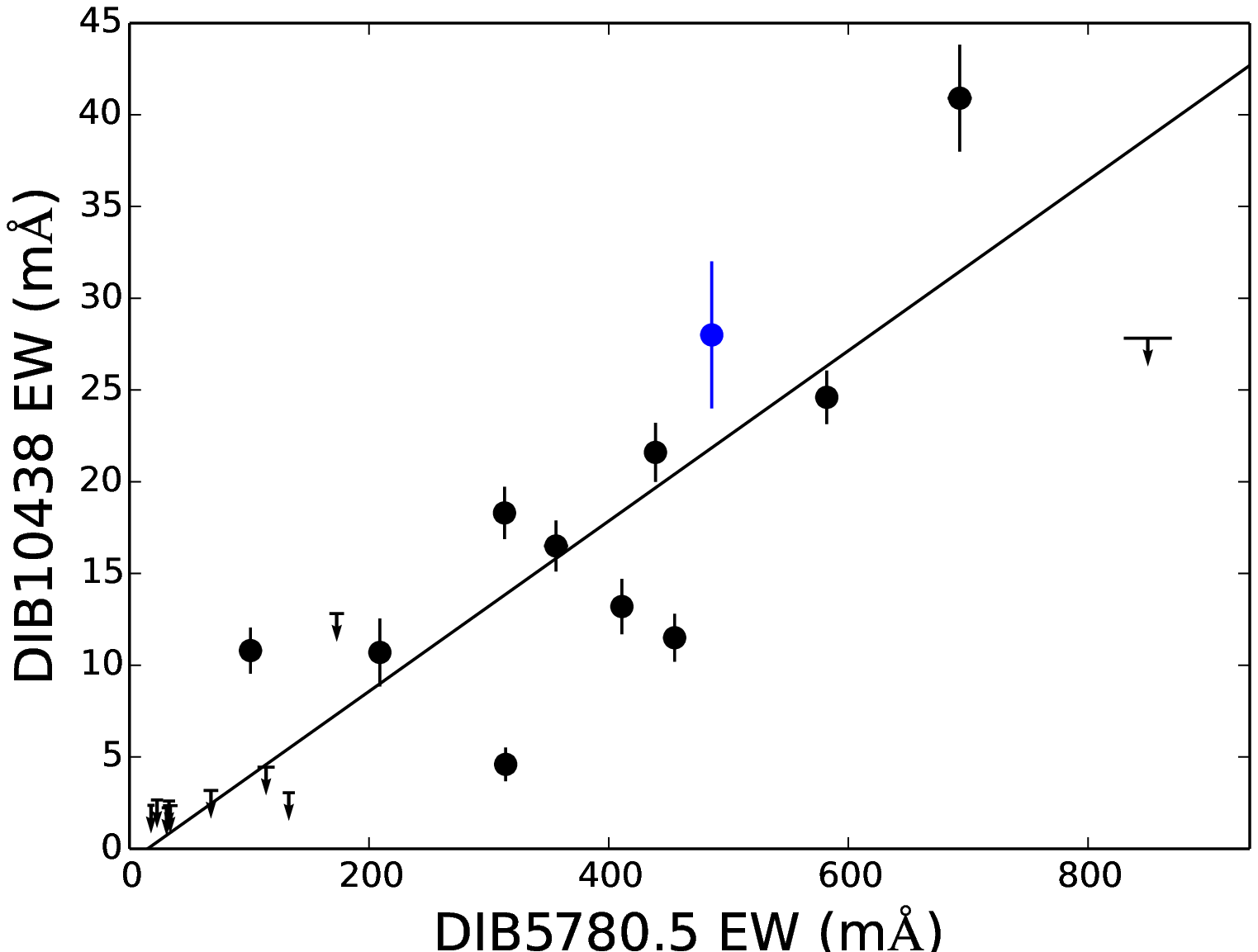}
 \includegraphics[width=7cm,clip]{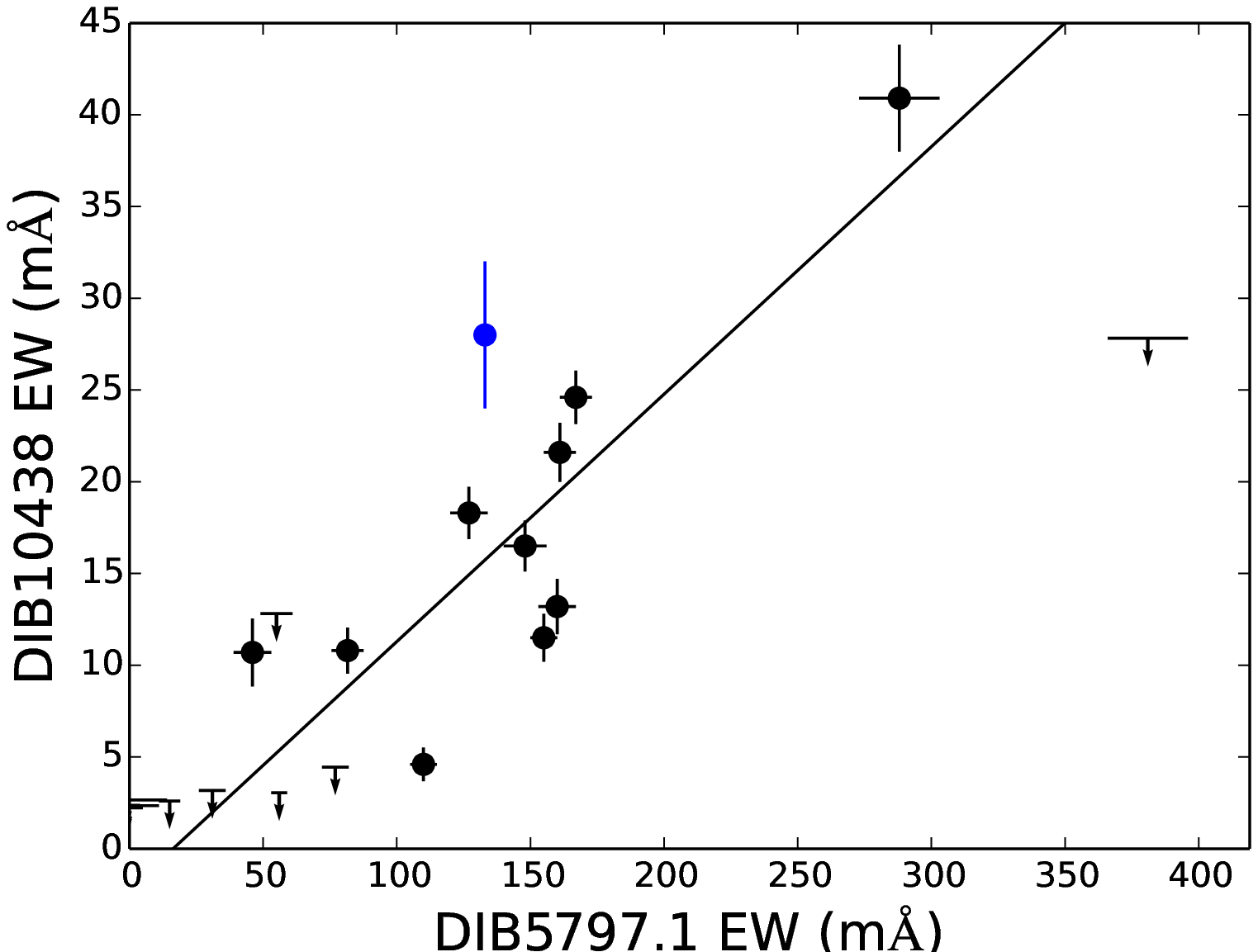}
 \includegraphics[width=7cm,clip]{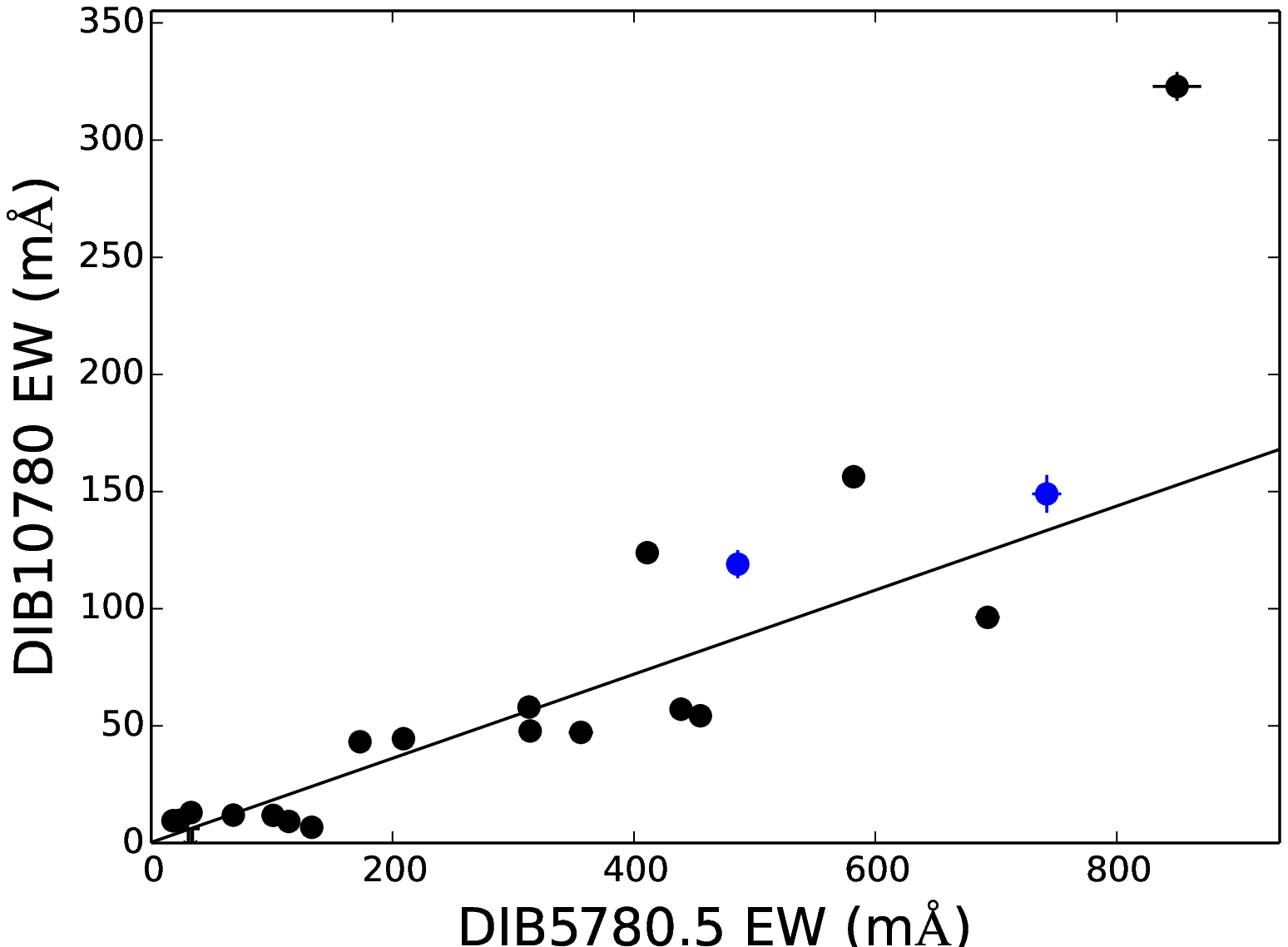}
 \includegraphics[width=7cm,clip]{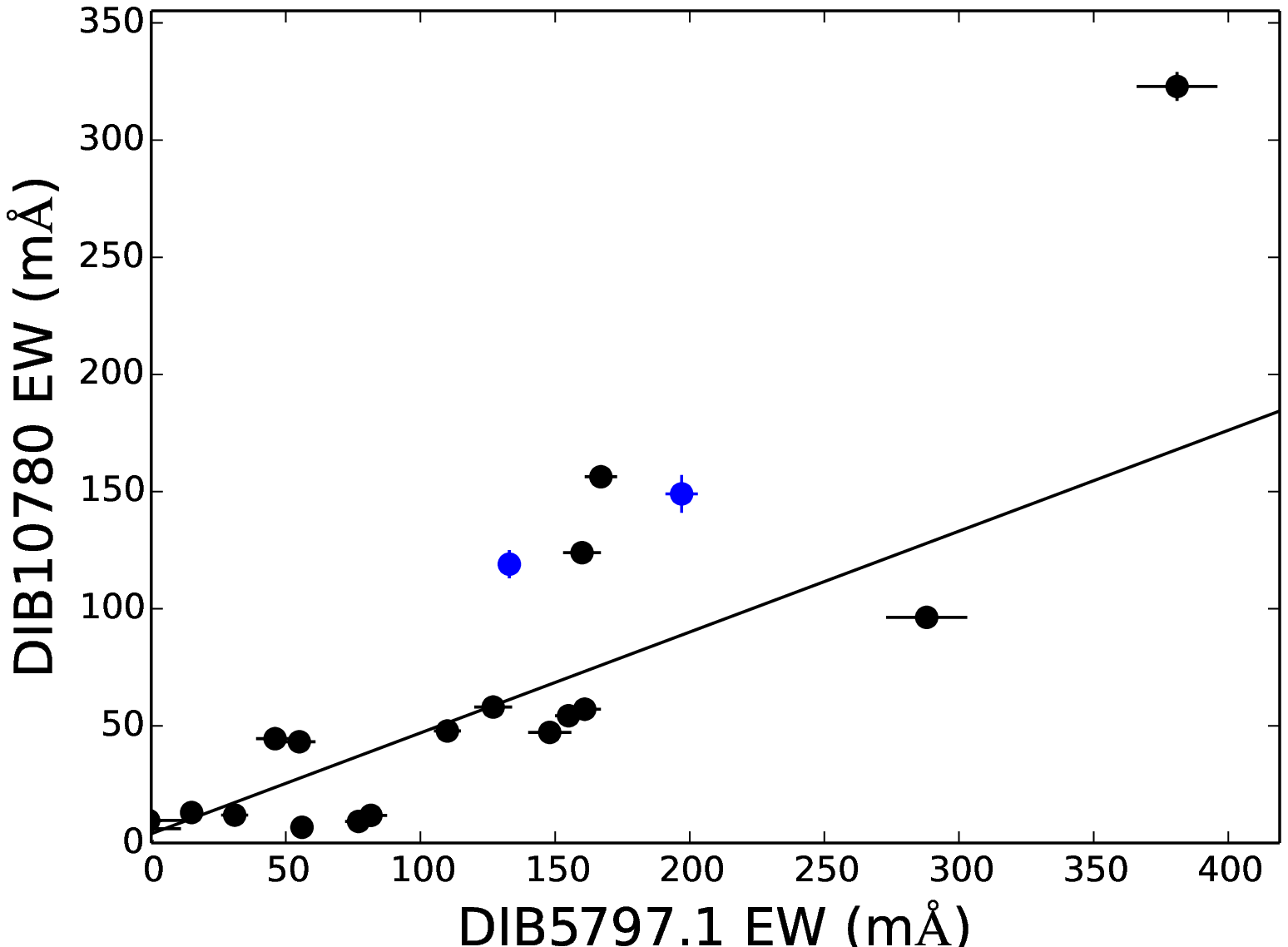}
 \includegraphics[width=7cm,clip]{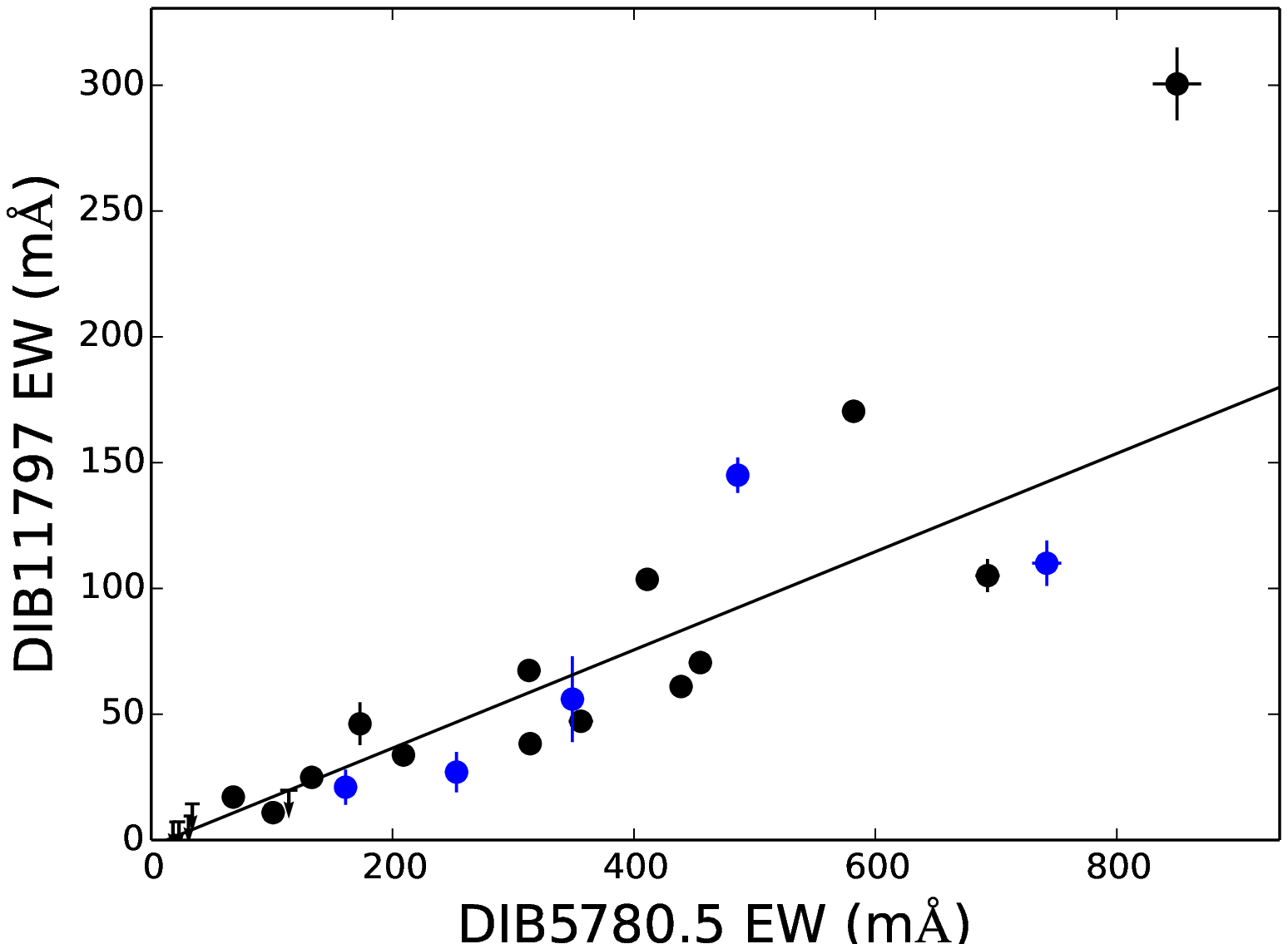}
 \includegraphics[width=7cm,clip]{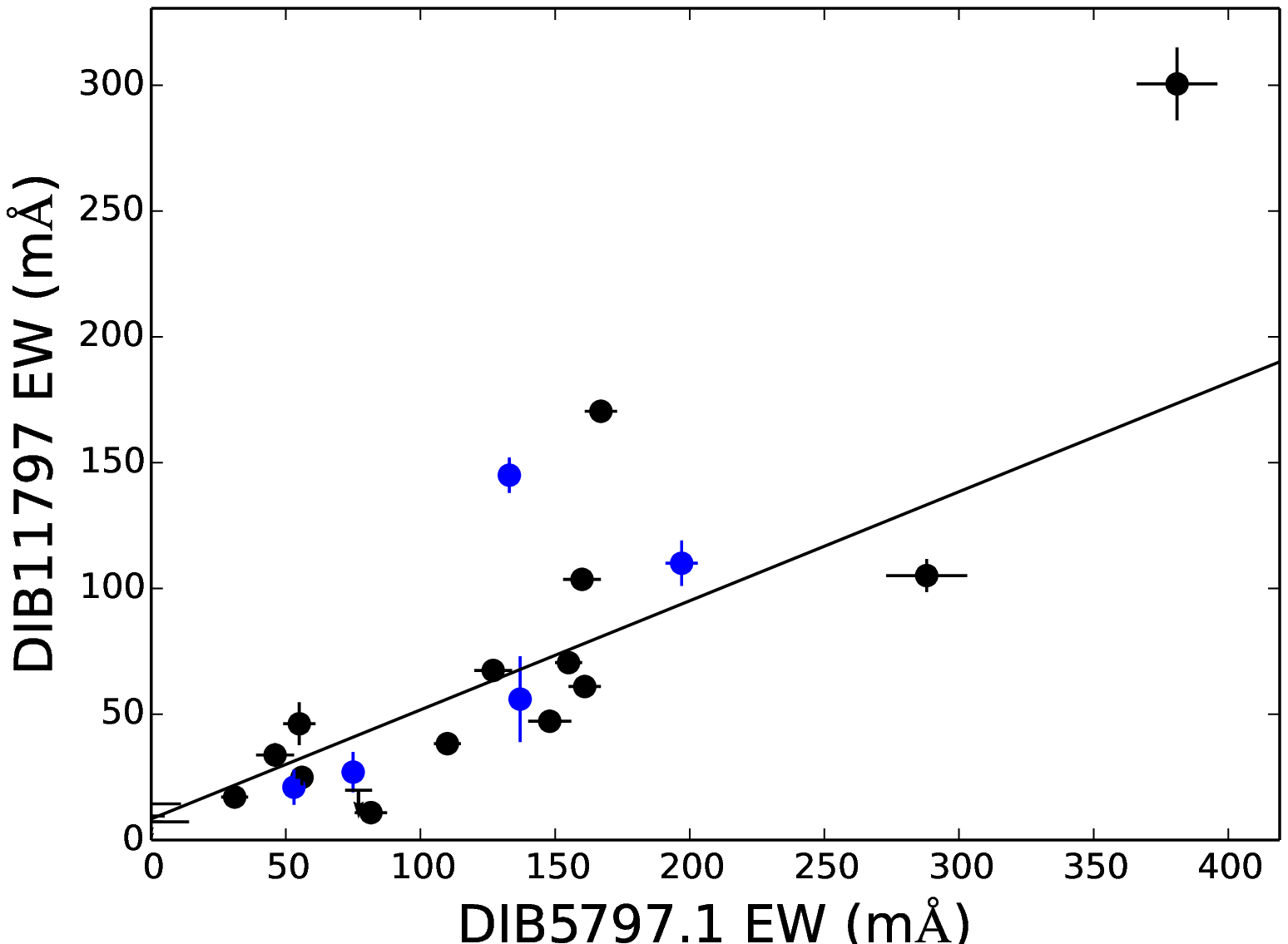}
 \includegraphics[width=7cm,clip]{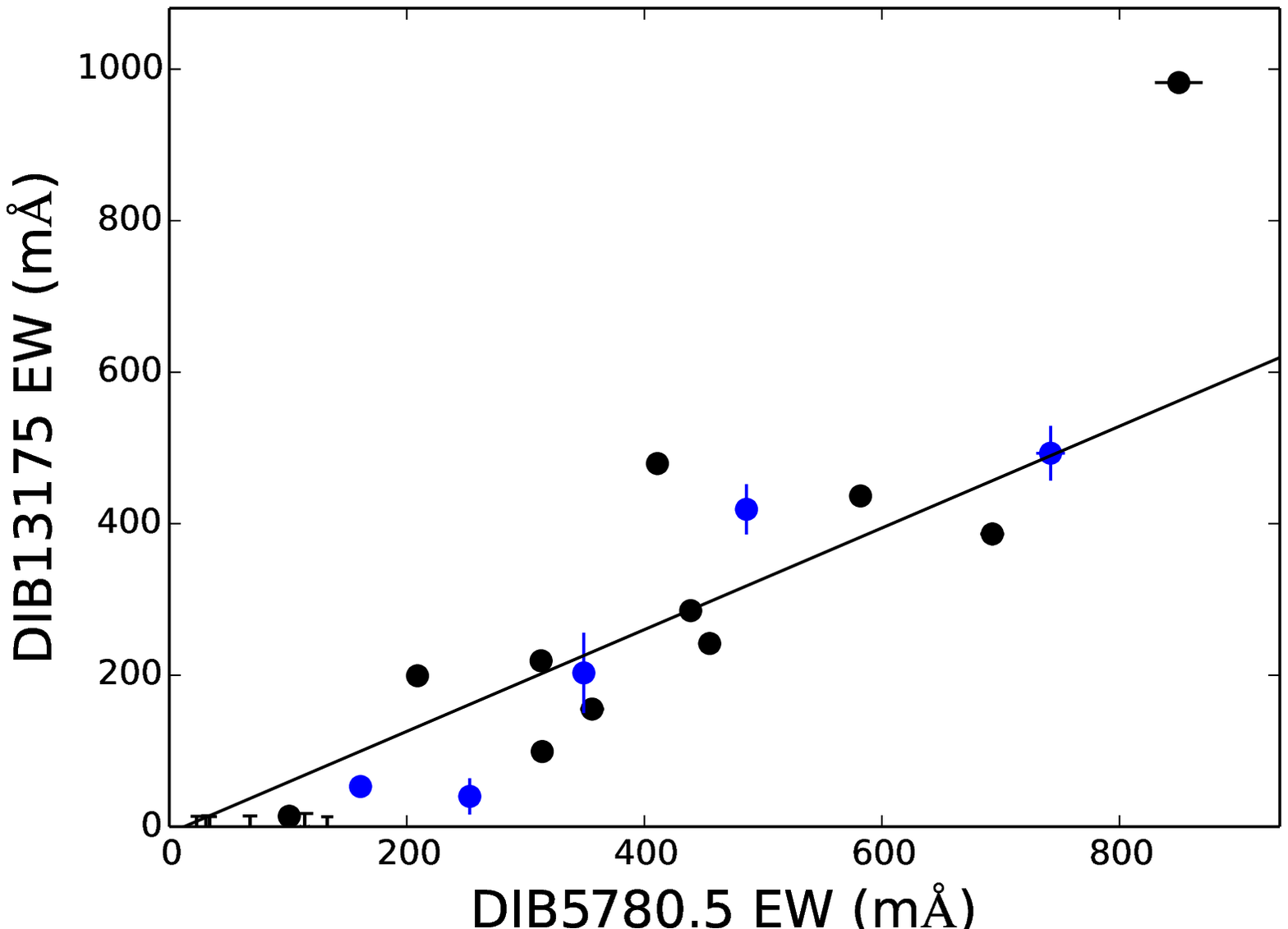}
 \includegraphics[width=7cm,clip]{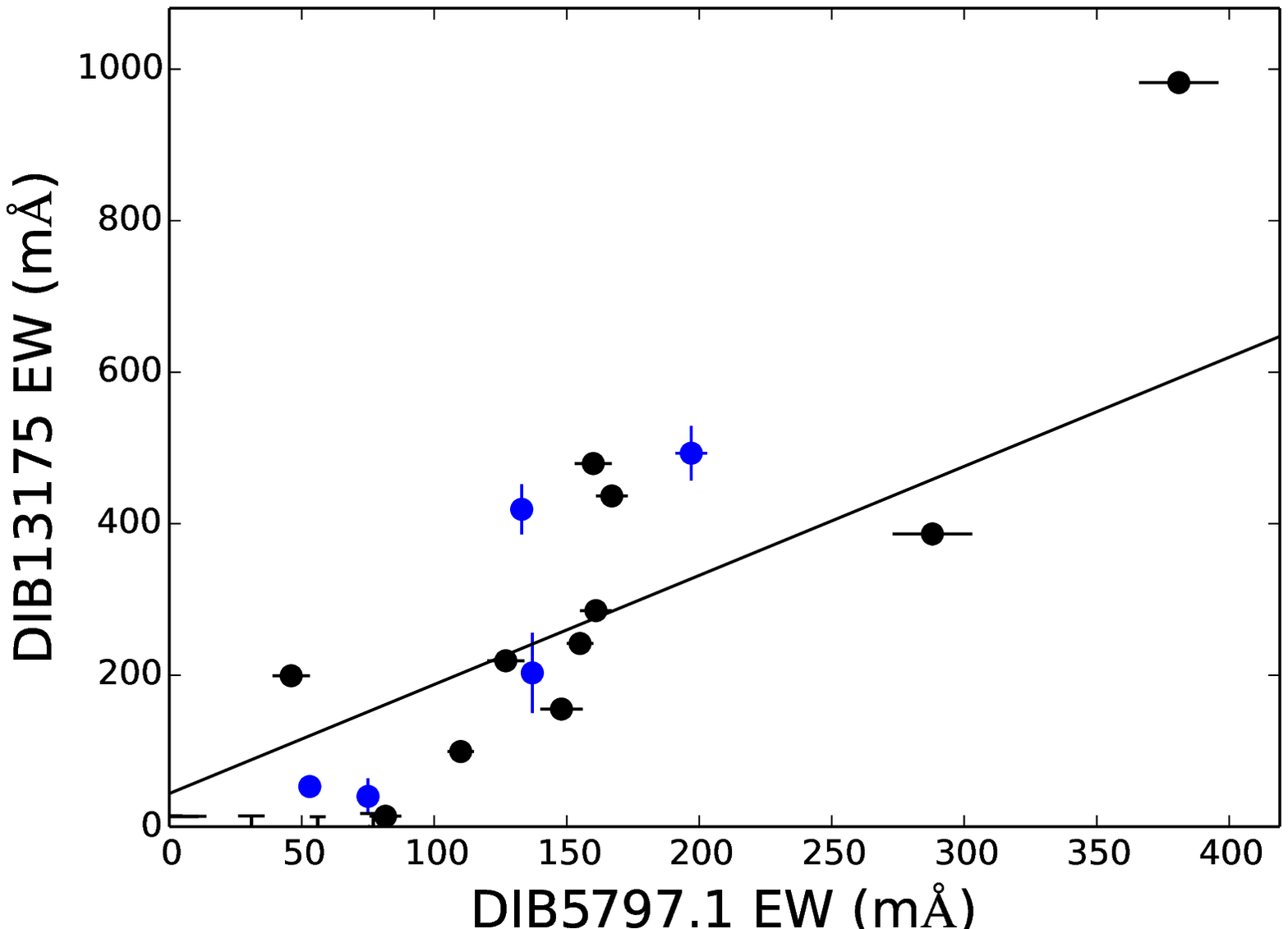}
 \caption{Correlations of four NIR DIBs, $\lambda \lambda 10438, 10780, 11797,\text{and } 13175$, with two optical DIBs, $\lambda \lambda 5780.5, 5797.1$. The black points show our data, while the blue points are from \citet{cox14}. The lines in each panel show the linear functions fitted to each plot. The blue points and the points of Cyg OB2 No.12 are not included in the fitting.}
 \label{NIROPTcorrelation}
\end{figure*}

\end{document}